\documentclass[12pt,openright]{extarticle}
\usepackage[dvips]{epsfig}
\usepackage{amsmath}
\usepackage{color}
\usepackage{amssymb}
\usepackage{changepage}
\usepackage{mdframed}
\usepackage{graphics}
\usepackage{wrapfig}
\usepackage[margin=1.1in]{geometry}
\usepackage{enumerate}

\begin{document}

\title{Ultracold collisions of molecules}
\author{Goulven Qu{\'e}m{\'e}ner \\
Laboratoire Aim{\'e} Cotton, CNRS,  \\
Universit{\'e} Paris-Sud, ENS Paris-Saclay, Universit{\'e} Paris-Saclay, \\
91405 Orsay, France \\ e-mail: goulven.quemener@u-psud.fr}

\maketitle

\abstract{This paper deals with the theory of collisions between two ultracold particles with a special focus on molecules. It describes the general features of the scattering theory of two particles with internal structure, using a time-independent quantum formalism. It starts from the Schr{\"o}dinger equation and introduces the experimental observables such as the differential or integral cross sections, and rate coefficients. Using a partial-wave expansion of the scattering wavefunction, the radial motion of the collision is described through a linear system of coupled equations, which is solved numerically. Using a matching procedure of the scattering wavefunction with its asymptotic form, the observables such as cross sections and rate coefficients are obtained from the extraction of the reactance, scattering and transition matrices. The example of the collision of two dipolar molecules in the presence of an electric field is presented, showing how dipolar interactions and collisions can be controlled.

\clearpage

\tableofcontents{}

\clearpage

\section{Introduction}

The achievement of slowing, cooling and trapping atoms~\cite{Chu_RMP_70_685_1998,Cohen-Tannoudji_RMP_70_707_1998,Phillips_RMP_70_721_1998} to quantum degeneracy in Bose--Einstein condensates~\cite{Cornell_RMP_74_875_2002,Ketterle_RMP_74_1131_2002} or degenerate Fermi gases has tremendously impacted the Atomic, Molecular, and Optical scientific community. The world of ultracold matter is governed by Quantum Mechanics. The particles move so slowly that one has enough time in an experiment to precisely control their internal structure and external motion, with for example electric or magnetic fields, electromagnetic waves and optical lattices. Ultracold atomic physics has been extensively investigated since those achievements and has led to the exploration of new quantum phenomena \cite{Lewenstein_AP_56_243_2007,Bloch_RMP_80_885_2008,Baranov_PRep_464_71_2008}.

Other types of particles, such as ultracold ions, ultracold atoms in Rydberg states and ultracold molecules, are also of specific interest. In this paper we focus mainly on ultracold molecules. Compared to atoms, molecules have a much richer structure, including rotation and vibration in addition to the electronic and spin structure. In contrast to atoms which are directly cooled with lasers, it is harder to cool the molecules with the same procedure due to the lack of closed cycles of absorption and spontaneous emission, even if it can work in certain cases \cite{Shuman_N_467_820_2010}. 
Other techniques are then employed \cite{Schnell_ACIE_48_6010_2009,Dulieu_RPP_72_086401_2009}: buffer gas cooling \cite{Hutzler_CR_112_4803_2012}, deceleration of molecules \cite{VanDeMeerakker_CR_112_4828_2012,Narevicius_CR_112_4879_2012}, Sisyphus cooling \cite{Zeppenfeld_N_491_570_2012}, association of ultracold atoms via photo-association \cite{Thorsheim_PRL_58_2420_1987,Fioretti_PRL_80_4402_1998,Weiner_RMP_71_1_1999,Jones_RMP_78_483_2006, Ulmanis_CR_112_4890_2012}, magneto-association \cite{Kohler_RMP_78_1311_2006,Chin_RMP_82_1225_2010}, and coherent transfer driven by lasers  \cite{Bergmann_RMP_70_1003_1998,Ni_S_322_231_2008,Danzl_S_321_1062_2008, Koch_CR_112_4928_2012,Bergmann_JCP_142_170901_2015}.

If the molecules possess permanent electric or magnetic dipole moments, they can be manipulated  by electric or magnetic fields \cite{Krems_IRPC_24_99_2005,Krems_PCCP_10_4079_2008, Quemener_CR_112_4949_2012,Lemeshko_MP_111_1648_2013}. In addition to the individual energies of the molecules the strength and orientation of the molecule-molecule interaction can also be controlled, leading to promising applications \cite{Carr_NJP_11_055049_2009}. The precise control over the initial ultracold particles and their interactions can be used to engineer different quantum edifices such as dipolar particles in optical lattices. Such controlled and tunable set-ups can be used for quantum simulation to mimic the Hamiltonian of more complicated systems of condensed matter, quantum magnetism and many-body physics \cite{Micheli_PRA_76_043604_2007,Gorshkov_PRL_107_115301_2011,Baranov_CR_112_5012_2012,Wall_BookChapter_2014} or to design schemes of quantum information \cite{DeMille_PRL_88_067901_2002,Yelin_PRA_74_050301_2006,Karra_JCP_144_094301_2016}. Dipolar molecules can also be used for testing fundamental theories \cite{Hinds_PS_1997_34_1997,Tarbutt_BookChapter_2009}, or to explore a novel ultracold chemistry in a fully determined way \cite{Carr_NJP_11_055049_2009,Gonzalez-Martinez_PRA_90_052716_2014,Tscherbul_PRL_115_023201_2015}.

Once the molecules are cooled, collisions between molecules and/or atoms can then occur \cite{Krems_IRPC_24_99_2005,Weck_IRPC_25_283_2006,Hutson_IRPC_26_1_2007,Quemener_BookChapter_2009,Krems_PCCP_10_4079_2008, Quemener_CR_112_4949_2012}. In all cases, collisions play an important role for understanding the stability, the lifetime and the dynamics of an ultracold gas. This paper is devoted to the time-independent quantum description of collisions between two atoms or molecules with internal structure, therefore allowing for changes of the internal state during the collision. The proposed approach is general enough to describe atom-atom, atom-molecule, and molecule-molecule collisions, and we will emphasize on the latter case. As it is based on an angular expansion of the scattering wavefuntion in partial waves, the formalism is specially suited for ultralow collision energies. 
Section \ref{sec:SE} starts with a reminder on the Schr{\"o}dinger equation for one and two particles, on the system of coordinates, and on the types of collisions. 
Two different parts of the colliding motion are tackled. Section \ref{sec:far} describes the region beyond the range of interactions where the particles hardly feel each other. The relevant observables are introduced there. Section \ref{sec:short} is devoted to the zone where the particles interact. This is where the partial wave expansion of the scattering wavefunction is introduced, leading to a system of coupled equations for the radial motion. The coupled system is solved using the method of the log-derivative matrix propagation. Symmetry considerations are also invoked, linked to the isotropy of space, to the symmetrization of identical particles, or to the presence of an external field. 
Section \ref{sec:matching} proceeds to the matching between the two latter regions. The  reactance, scattering and transition matrices are defined and their relations with the observables are established.
Section \ref{sec:ultracold} describes certain properties of collisions in the ultracold regime.
In Section \ref{sec:krb}, as an application, we use this formalism to study the dipolar collisions between two ultracold KRb molecules in an electric field.  We show how we can simplify the full problem to restrict the physical process to its main relevant element. Two cases are explored: (i) collisions of molecules in the ground rotational state and (ii) collisions of molecules in the first excited rotational state. It is found that collision rates can be enhanced or suppressed. We conclude and give some perspectives in Section \ref{sec:conclusion}.

For readers that desire additional information we refer for instance to references
\cite{Brandsen_Joachain_Book_2003,Cohen-Tannoudji_Book_1997,Friedrich_Book_2005,Landau_Book_1958,Child_Book_1996,
Atkins_Friedman_Book_2005,Launay_Book_2000} among many others.

\section{The Schr\"odinger equation}
\label{sec:SE}

\subsection{The Schr\"odinger equation for one particle}

The dynamics of a quantum particle of mass $m$ moving in a potential characterized by the operator $\widehat{V}$ is described by the time-dependent Schr\"odinger equation in the $\langle \vec{r}|$ representation:
\begin{eqnarray} \label{TISE1PLE}
 i\hbar \, \frac{\partial \Psi(\vec{r},t)}{\partial t} = \widehat{H} \, \Psi(\vec{r},t) 
\end{eqnarray}
where $ \langle \vec{r}| \Psi(t) \rangle = \Psi(\vec{r},t)$ is the wavefunction of the particle  at the position $\vec{r}$ and time $t$. The operator $\widehat{H}=\widehat{T} + \widehat{V}$ is the Hamiltonian of the particle, where $\widehat{T}$ and $\widehat{V}$ are the kinetic and potential energy operators, respectively, defined as:
\begin{eqnarray}
\widehat{T} = \frac{\widehat{\vec{p}}^{\ 2}}{2m} \equiv \frac{1}{2m} \, \bigg(\frac{\hbar}{i} \,\vec{\nabla}\bigg)^2 = - \frac{\hbar^2}{2m} \, \vec{\nabla}^2 \qquad ; \qquad \widehat{V} \equiv V(\vec{r},t).
\end{eqnarray}
Wide hats will be used to represent the quantum operators in this paper.
We define the presence probability density of a particle as 
$\rho(\vec{r},t) = |\Psi(\vec{r},t)|^2$ which has unit of a volume density. 
This quantity determines the probability $dP(\vec{r},t)=\rho(\vec{r},t) \, d\vec{r}$ to find the particle at time $t$ at position $\vec{r}$ in the volume element $d\vec{r}$.
The presence probability of the particle in a finite volume $\cal V$ is $P (t) = \int_{\cal V} \, dP(\vec{r},t) = \int_{\cal V} \, \rho(\vec{r},t) \, d\vec{r} $. 
Since the probability of finding the particle over all space must be unity,
the wavefunction has to be normalized using 
$ \int_{-\infty}^{+\infty} \, |\Psi(\vec{r},t)|^2 \, d\vec{r}  = 1 $.
This normalization is possible 
if the wavefunction is square integrable, typically when the wavefunction 
represents a bound state of a particle.
For a continuum state of a particle, that is when the particle 
is not bound in a specific space, 
this normalization is not possible. Several methods are used to normalize 
such wavefunctions for example using a Dirac delta function in the normalization
(see for example \cite{Brandsen_Joachain_Book_2003,Cohen-Tannoudji_Book_1997,Friedrich_Book_2005}).
We define the probability current of a particle as:
\begin{eqnarray} \label{current}
\widehat{\vec{j}}(\vec{r},t) &=& - \frac{\hbar}{2mi} \bigg[  \Psi^*(\vec{r},t) \, \vec{\nabla} \Psi(\vec{r},t) - \Psi(\vec{r},t) \, \vec{\nabla} \Psi^*(\vec{r},t)  \bigg] \nonumber \\
&=& \operatorname{Re} \left\{ \Psi^*(\vec{r},t) \, \left( \frac{\widehat{\vec{p}}}{m} \, \Psi(\vec{r},t) \right) \right\} .
\end{eqnarray}
The probability density and the probability current are related by:
\begin{eqnarray} \label{continuity}
\frac{\partial \rho(\vec{r},t)}{\partial t} + \vec{\nabla} . \widehat{\vec{j}} = 0.
\end{eqnarray}
Eq. \eqref{continuity} is the continuity equation showing that the probability is conserved locally, just like a charge is conserved in electrostatics. Indeed we have, from the divergence theorem,
\begin{eqnarray} \label{flux}
\int_{\cal V} \, \frac{\partial \rho(\vec{r},t)}{\partial t} \, d\vec{r} = 
\frac{\partial P(t)}{\partial t} = - \int_{\cal V}  \, \vec{\nabla} . \widehat{\vec{j}} \, d\vec{r} 
= - \oint_{\cal S} \, \widehat{\vec{j}} . d\vec{{\cal S}} .
\end{eqnarray}
The decrease (increase) in time of $P(t)$ inside the volume ${\cal V}$ at time t is equal to an outgoing (incoming) flux of $\widehat{\vec{j}}$ through the surface ${\cal S}$ enclosing the volume ${\cal V}$.
Note that $\widehat{\vec{j}}$ has unit of a surface density per unit of time.  \\

\noindent If the potential energy is independent of time $\widehat{V} = V(\vec{r})$, we can find a stationary solution $\Psi(\vec{r},t)$ with a well defined energy $E_{tot}$:
\begin{eqnarray} \label{stationarywf}
\Psi(\vec{r},t) = \psi^{E_{tot}}(\vec{r}) \, e^{-\frac{i E_{tot} t}{\hbar}},
\end{eqnarray}
with a separation of space and time in the wavefunction. The solution is stationary since
$|\Psi(\vec{r},t)|^2 = |\psi^{E_{tot}}(\vec{r})|^2$ is independent of time. Putting Eq.~\eqref{stationarywf} into Eq.~\eqref{TISE1PLE} gives the time-independent Schr\"odinger equation for $\psi^{E_{tot}}(\vec{r})$:
\begin{eqnarray}
\widehat{H} \, \psi^{E_{tot}}(\vec{r})=  \bigg[- \frac{\hbar^2}{2m} \, \vec{\nabla}^2 + V(\vec{r}) \bigg] \, \psi^{E_{tot}}(\vec{r}) = E_{tot} \, \psi^{E_{tot}}(\vec{r}).
\end{eqnarray}
It is often the case that in collisions the potential energy $V$ is independent of time and the total energy $E_{tot}$ is conserved. The time-independent formalism still applies when static electric or magnetic fields are present but not anymore when the fields vary in time. Similarly, the time-independent probability density and probability current are given by 
$\rho(\vec{r}) = |\psi^{E_{tot}}(\vec{r})|^2$ and $\widehat{\vec{j}}(\vec{r}) = - \frac{\hbar}{2mi} [ [\psi^{E_{tot}}(\vec{r})]^* \, \vec{\nabla} \psi^{E_{tot}}(\vec{r}) - \psi^{E_{tot}}(\vec{r}) \, \vec{\nabla} [\psi^{E_{tot}}(\vec{r})]^*]$. 

\subsection{The Schr\"odinger equation for two colliding particles}

\subsubsection{Coordinate systems}

\begin{figure}[h]
\begin{center}
\includegraphics*[width=6cm,keepaspectratio=true,angle=-90,trim=110 130 300 200]{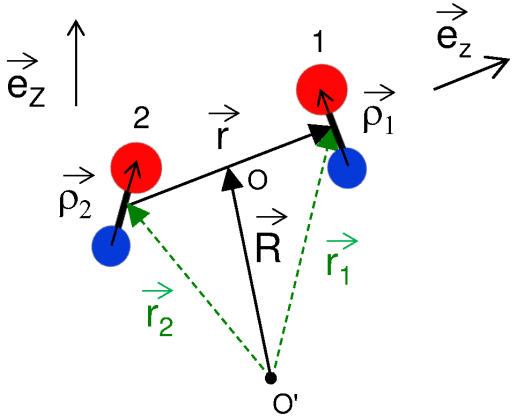}
\caption{System of two composite particles 1 and 2 (here two diatomic molecules)
described by two individual 
external vectors $\vec{r}_1, \vec{r}_2$ from an arbitrary point $O'$
and two individual internal vectors $\vec{\rho}_1, \vec{\rho}_2$. 
The relative and center-of-mass vectors 
$\vec{r}, \vec{R}$ are also shown. 
The point $O$ defines the center-of-mass of the system.
The unit vector $\vec{e}_{Z}$
orients the axis $OZ$ of the space-fixed frame while the unit vector $\vec{e}_{z}$
orients the axis $Oz$ of the body-fixed frame of the system (see text for details).}
\label{FIG1}
\end{center}
\end{figure}

We consider a time-independent collision problem of a system 
of two composite particles $i=1,2$ (for example two molecules) 
of mass $m_1$, $m_2$ (see Fig.~\ref{FIG1}), described with individual 
external coordinates $\vec{r}_1, \vec{r}_2$ from an arbitrary point $O'$ and internal 
coordinates $\vec{\rho}_1, \vec{\rho}_2$ (not to mistake with the 
presence probability density).
It is also useful to introduce the center-of-mass (CM) coordinates $\vec{R}$ and the relative (rel) coordinates $\vec{r}$ (see Fig.~\ref{FIG1}) defined by:
\begin{align}
\vec{R} &= \frac{m_1 \, \vec{r}_1 + m_2 \, \vec{r}_2}{m_{tot}} & \vec{r} & = \vec{r}_1 - \vec{r}_2,
\end{align}
where $m_{tot} = m_1 + m_2$ is the total mass and $m_{red} = m_1 \, m_2 / (m_1 + m_2)$ 
is the reduced mass. If we define the point $O$ so that $\overrightarrow{O'O} = \vec{R}$, then we
define a {\bf space-fixed frame} by the axes $OXYZ$. This is the space-fixed frame of the center-of-mass of the system, we could have defined any other arbitrary 
space-fixed frame $O'XYZ$. Space-fixed frames are also often called laboratory frames.
The axis $OZ$ is oriented along a unit vector $\vec{e}_Z$ as shown in Fig.~\ref{FIG1}.
We did not show the other unit vectors $\vec{e}_X$ and $\vec{e}_Y$ 
that orient the axes $OX$ and $OY$.
We also define a {\bf body-fixed frame} of the system 
by the axes $Oxyz$ when now the $Oz$ axis 
is oriented along a unit vector $\vec{e}_z$ also shown in Fig.~\ref{FIG1}, following
the orientation of the vector $\vec{r}$.
In this paper, we will choose the $OZ$ axis as the quantization 
axis in the space-fixed frame and the $Oz$ axis in the body-fixed frame of the system.
The two particles interact in general via a potential energy $V(\vec{\rho}_1, \vec{\rho}_2, \vec{r}_1, \vec{r}_2)$. The particles are initially located at large distances and they start to interact as they approach from each other. They are scattered in a given direction, reflecting the strength and the anisotropy of the potential energy.
In general the potential energy $V$ of the system can be separated in two terms: a potential energy $V_{int}$ which describes the internal interactions of the particles between themselves, and a potential energy $V_{ext}$ which describes eventual external potentials.
The former term $V_{int}$, contains all electrostatic Coulombic interactions between the electrons and the nuclei of the atoms composing the system, and do not depend on the absolute position of the charges but rather on their relative separation. Using extensive ab initio calculations, the full electronic problem is solved for parametric positions of the nuclei within the so called Born--Oppenheimer approximation. It results in a potential energy term $V_{int}(\vec{\rho}_1, \vec{\rho}_2, \vec{r}_1 - \vec{r}_2)$ = $V_{int}(\vec{\rho}_1, \vec{\rho}_2, \vec{r})$ and is usually called the {\bf potential energy surface} of the system, as it represents an energy as a function of multi-coordinates in space. 
The set of vectors $(\vec{\rho}_1, \vec{\rho}_2, \vec{r})$ are often called the Jacobi coordinates.
As it does not depend on the individual positions 
$\vec{r}_1, \vec{r}_2$ but only on the relative position of the molecules, 
it is separable in $\vec{R}$ and $\vec{r}$.
The latter term $V_{ext}$, can describe the interaction of the molecules with external fields for example a static electric or magnetic field. 
If these fields are uniform throughout space, these potentials do not depend on the individual positions of the molecules $\vec{r}_1, \vec{r}_2$. An external potential can also depends on the individual positions of the molecules. Depending on the case, the potential can be separable in 
$\vec{R}$ and $\vec{r}$, for example if the external potential is described by an harmonic oscillator \cite{Quemener_PRA_83_012705_2011}, or not, for example if the potential is described by an optical lattice \cite{Grishkevich_PRA_84_062710_2011}. 
In the following we will consider a system described by an arbitrary potential energy surface
$V_{int}(\vec{\rho}_1, \vec{\rho}_2, \vec{r})$ and an eventual external potential $V_{ext_1}(\vec{\rho}_1) + V_{ext_2}(\vec{\rho}_2)$ that does not depend on the individual position of the molecules. 
The total potential energy is then separable in $\vec{R}$ and $\vec{r}$. The case of non-separable potentials is not treated here as it is beyond the scope of this paper.
When the particles are far apart $|\vec{r}| = |\vec{r}_1 - \vec{r}_2| \to \infty$, $V_{int}(\vec{\rho}_1, \vec{\rho}_2, \vec{r}) \to V_{int_1}(\vec{\rho}_1) + V_{int_2}(\vec{\rho}_2)$, the internal potential energy of the two separated molecules 1 and 2. We define the  interaction potential energy by:
\begin{eqnarray}
U_{int}(\vec{\rho}_1, \vec{\rho}_2, \vec{r}) = V_{int}(\vec{\rho}_1, \vec{\rho}_2, \vec{r}) - V_{int_1}(\vec{\rho}_1) - V_{int_2}(\vec{\rho}_2),
\end{eqnarray}
where $U_{int} \to 0$ if $|\vec{r}| \to \infty$. 
Then the time-independent Schr\"odinger equation gives:
\begin{multline} \label{TISE2PLE}
\bigg[- \frac{\hbar^2}{2m_1} \, \vec{\nabla}^2_1 - \frac{\hbar^2}{2m_2} \, \vec{\nabla}^2_2 + U_{int}(\vec{\rho}_1, \vec{\rho}_2, \vec{r}) + \widehat{h}_1(\vec{\rho}_1)
+ \widehat{h}_2(\vec{\rho}_2) \bigg] \, \psi(\vec{\rho}_1, \vec{\rho}_2, \vec{r}_1, \vec{r}_2)  \\
= E_{tot} \, \psi(\vec{\rho}_1, \vec{\rho}_2, \vec{r}_1, \vec{r}_2).
\end{multline}
The operators $\widehat{h}(\vec{\rho}_i)$ are the internal Hamiltonians of particles $i=1,2$:
\begin{eqnarray} \label{internalham}
\widehat{h}_i(\vec{\rho}_i) \, \phi_{\alpha_i}(\vec{\rho}_i) &=& \left\{ \widehat{T}_{i} + \widehat{V}_i(\vec{\rho}_i) \right\} 
\, \phi_{\alpha_i}(\vec{\rho}_i)\nonumber \\
&=& \varepsilon_{\alpha_i} \, \phi_{\alpha_i}(\vec{\rho}_i),
\end{eqnarray}
where $\widehat{V}_i(\vec{\rho}_i) = V_{int_i}(\vec{\rho}_i)  + V_{ext_i}(\vec{\rho}_i)$. 
The index $\alpha_i$ represents the quantum numbers describing the internal eigenfunctions $\phi_{\alpha_i}$ and eigenenergies $\varepsilon_{\alpha_i}$ of the hamiltonian 
$\widehat{h}_i$ of the individual particle $i$.
As an example, if we consider a diatomic molecule with no spin structure where $V_{int_i}(\vec{\rho}_i)$ represents the vibrational and rotational internal potential energy and if we consider no external potential energy $V_{ext_i}=0$, then $\phi_{\alpha_i}(\vec{\rho}_i) =$ $\frac{\chi_{v_i,n_i}(\rho_i)}{\rho_i} \, $ $Y_{n_i}^{m_{n_i}}(\hat{\rho}_i)$.
We note $\widehat{\vec{n}}_i$ the rotational angular momentum operator of the molecule $i$ characterized by the quantum number $n_i$, and $\widehat{n}_{Z_i}$ represents the projection operator of $\widehat{\vec{n}}_i$ onto the $OZ$ space-fixed frame axis characterized by the quantum numbers $m_{n_i}$. Then
$Y_{n_i}^{m_{n_i}}(\hat{\rho}_i)$ represents the rotational wavefunction where $\hat{\rho}_i$ represents the spherical angles of $\vec{\rho}_i$. Small hats corresponds to angles here, not to mistake with the wide hats of the quantum operators. 
$\chi_{v_i,n_i}$ represents the radial vibrational wavefunction characterized by the vibrational and rotational quantum numbers $v_i, n_i$. 
The quantum numbers describing the internal state are ${\alpha_i} \equiv v_i, n_i, m_{n_i}$.
We note $\varepsilon_{\alpha} = \varepsilon_{\alpha_1} + \varepsilon_{\alpha_2}$, $\phi_{\alpha} = \phi_{\alpha_1} \, \phi_{\alpha_2}$, with $\alpha \equiv \alpha_1 \, \alpha_2$. 
The total energy of the system $E_{tot} = E_{k \, 1} + E_{k \, 2} + \varepsilon_{\alpha}$, where $E_{k \, i}$ is the kinetic energy of particle $i$ and is conserved during the collision.
Because we consider potentials that do not depend on the individual position of the molecules, one can separate the center-of-mass with the relative coordinates, and we can write the wavefunction as a product
$\psi(\vec{\rho}_1, \vec{\rho}_2, \vec{r}_1, \vec{r}_2)$ = $\psi_{CM}(\vec{R})$
$\psi_{rel}(\vec{\rho}_1, \vec{\rho}_2, \vec{r})$.
One can show that Eq. \eqref{TISE2PLE} can be decoupled into an equation for the center-of-mass motion:
\begin{eqnarray} \label{HamiltonianCM}
\bigg[- \frac{\hbar^2}{2m_{tot}} \, \vec{\nabla}^2_{\vec{R}} \bigg] \, \psi_{CM}(\vec{R}) 
= E_{k \, CM} \, \psi_{CM}(\vec{R})
\end{eqnarray}
and one for the relative motion:
\begin{eqnarray} \label{Hamiltonianrel}
\bigg[ - \frac{\hbar^2}{2 m_{red}} \, \vec{\nabla}^2_{\vec{r}} + U_{int}(\vec{\rho}_1, \vec{\rho}_2, \vec{r}) + \widehat{h}_1(\vec{\rho}_1)
+ \widehat{h}_2(\vec{\rho}_2) \bigg] \, \psi_{rel}(\vec{\rho}_1, \vec{\rho}_2, \vec{r}) \nonumber \\
= (E_{tot} - E_{k \, CM}) \, \psi_{rel}(\vec{\rho}_1, \vec{\rho}_2, \vec{r}).
\end{eqnarray}
with $E_{k \, 1} + E_{k \, 2}$ = $E_{k \, CM} + E_{k \, rel}$.
For the type of separable interaction potential energy $U_{int}(\vec{\rho}_1, \vec{\rho}_2, \vec{r})$, the solution in Eq. \eqref{HamiltonianCM} for the CM motion is a free motion unaffected by the internal interactions, and is represented as a plane wave. It can be separated from the collision problem. 
In the following, we consider the collision in the space-fixed frame $OXYZ$ 
of the center-of-mass
so that $\vec{R}=0$, $E_{k \, CM} = 0$, $E_{k \, rel} = E_k$, and $E_{tot}=E_k+\varepsilon_{\alpha}$.
Then Eq. \eqref{Hamiltonianrel} describes the motion of a fictitious particle of mass $m_{red}$, of internal state $\phi_{\alpha} = \phi_{\alpha_1} \, \phi_{\alpha_2}$ moving in the interacting potential $U_{int}(\vec{\rho}_1, \vec{\rho}_2, \vec{r})$. For simplicity we will omit the subscript ``rel'' in the wavefunction.

\subsubsection{Types of collisions}

\begin{figure}[h]
\begin{center}
\includegraphics*[width=10cm,keepaspectratio=true,angle=-90,trim=30 0 50 0]{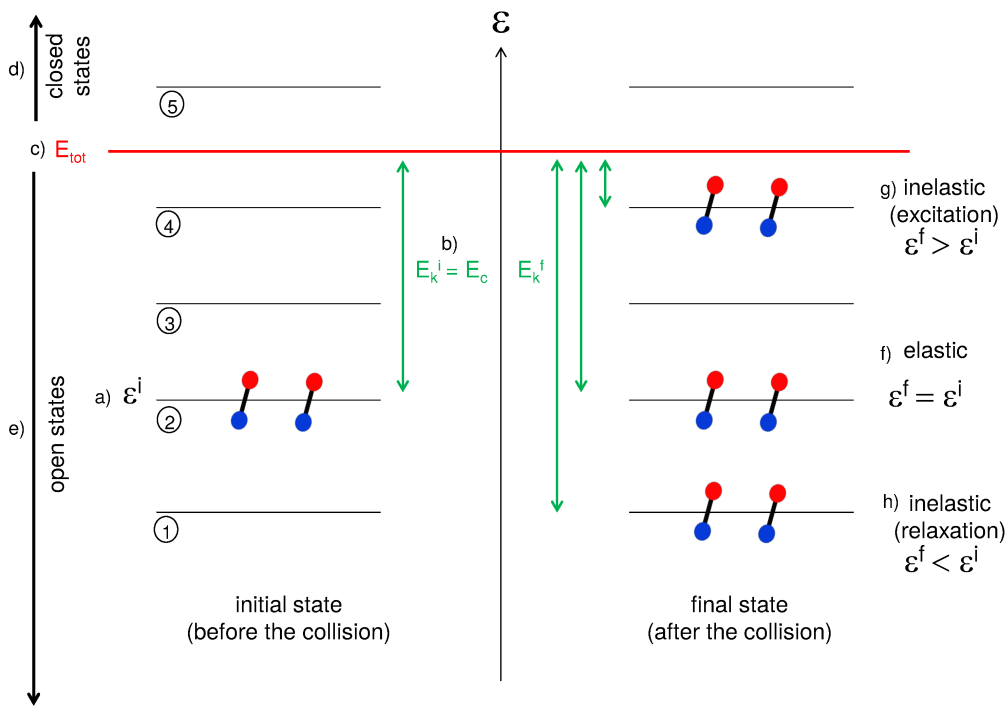}
\caption{Sketch of the energy thresholds of two particles (here two diatomic molecules)
when they are far from each other. The total energy $E_{tot}$ is indicated as a red line
and is conserved during a colliison. It separates the open states from the closed states.
The result of different types of collision 
is shown. For elastic collisions, the energy of the molecules is the same before and after the collision while for inelastic collisions, the energy is different. 
For excitation (relaxation) processes, the energy of the molecules 
has increased (decreased) so that 
their relative kinetic energy should decreased (increased) 
to conserve the total energy (see text for details). }
\label{FIG2}
\end{center}
\end{figure}

Fig.~\ref{FIG2} shows different types of collisions, according to the different internal energy levels $\varepsilon_{\alpha}$ of the pair of particles, before the collision (left) and after the collision (right). For this example, there are five different possible states labeled $\alpha=1,2,3,4,5$, with corresponding energies.
\begin{enumerate}[(a)]
\item On the left-hand side of the figure, the intial internal level is $\alpha=2$ and the initial internal 
energy $\varepsilon^i$  is $\varepsilon_2$ (the superscript `i' stands for `initial' here).
\item The initial kinetic energy $E_k^i$ is fixed and is called the {\bf collision energy} $E_c$.
\item The total energy is $E_{tot} = \varepsilon^i + E_k^i$.
\item The states with an internal energy larger than the total energy are called the {\bf closed states}, which are not energetically accessible after the collision.
\item The states with an internal energy smaller than the total energy are called the {\bf open states} which are energetically accessible after the collision.
\item An {\bf elastic collision} occurs when the final state is the same than the initial one,  $\varepsilon^f =  \varepsilon^i$ (the superscript `f' stands for `final' here). As the total energy is conserved, the final kinetic energy $E_k^f = E_k^i$ is also conserved. When the particles have the same internal energy after and before the collision, they also have the same kinetic energy.
\item If the final state is different than the initial one ($\varepsilon^f \neq  \varepsilon^i$), an {\bf inelastic collision} takes place. The case $\varepsilon^f > \varepsilon^i$  corresponds to an {\bf excitation} where $E_k^f < E_k^i$, leading to $E_k^f = E_{tot} - \varepsilon^f = \varepsilon^i - \varepsilon^f + E_k^i$. In this case, the particles have gained internal energy and lost kinetic energy.
\item The case $\varepsilon^f < \varepsilon^i$ refers to a {\bf relaxation} with $E_k^f > E_k^i$, and $E_k^f = \varepsilon^i - \varepsilon^f + E_k^i$. In this case, the particles have lost internal energy and gained kinetic energy.
\end{enumerate}

\noindent When the chemical identity of the products is different from the one of the reactants, various kinds of {\bf reactive collisions} can occur providing that the states of the products are open: 
\begin{eqnarray}
AB + CD & \to & AC + BD, AD + BC   \nonumber \\
        & \to & A + BCD, B + CDA, C + DAB, D + ABC \nonumber \\
        & \to & A + B + CD, B + C + DA, \nonumber \\
        & & C + D + AB, D + A + BC \nonumber \\
        & \to & A + B + C + D   \nonumber .
\end{eqnarray}
In such cases, a set of collective coordinates like the so-called hyperspherical coordinates \cite{Whitten_JMP_9_1103_1968,Johnson_JCP_79_1916_1983} should be employed as they are more appropriate than the Jacobi coordinates to treat the different arrangements or the four particles in a more symmetric way. The resulting collisional formalism is more complicated \cite{Pack_JCP_87_3888_1987,Launay_CPL_163_178_1989,Rittenhouse_JPBAMOP_44_172001_2011} and beyond the scope of this paper. Therefore we will not treat the case of reactive collisions
in the following, only the case of elastic and inelastic collisions.

\section{In the region far from collision}
\label{sec:far}

\subsection{Asymptotic form of the wavefunction}

\begin{figure}[h]
\begin{center}
\includegraphics*[width=10cm,keepaspectratio=true,angle=-90,trim=10 0 50 0]{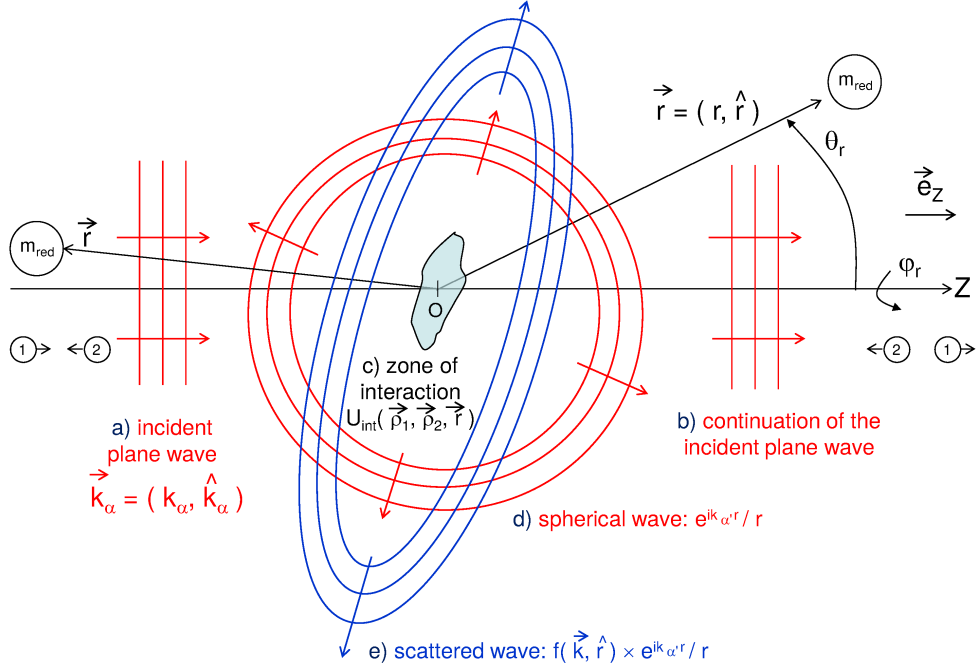}
\caption{Asymptotic form of the total wavefunction which decomposes
into a incident plane wave and a scattered wave due to the effect 
of the interaction $U_{int}$ (see text for details).}
\label{FIG3}
\end{center}
\end{figure}

\noindent As mentioned above, the motion of the center-of-mass can be separated from the collision problem. We then don't considered it anymore and we focus now on the relative motion described by the fictitious particle of mass $m_{red}$. The relative vector $\vec{r}$ can be written in spherical coordinates $\vec{r}=\{r,\hat{r}=(\theta_r,\varphi_r)\}$ with respect 
to the $OZ$ space-fixed frame axis.
The stationary scattering state for the relative motion in the CM frame for a given total energy $E_{tot}$ and for an initial state $\alpha, \vec{k}_{\alpha}$ behaves asymptotically as:
\begin{eqnarray} \label{asymexpansion}
\psi^{E_{tot}}_{\alpha, \vec{k}_{\alpha}}(\vec{\rho}_1, \vec{\rho}_2, \vec{r}) &\underset{r \to \infty}{=}& {\cal A} \bigg[ e^{i \vec{k}_{\alpha} . \vec{r} } \, \phi_{\alpha}(\vec{\rho}_1, \vec{\rho}_2) \nonumber \\
& & \quad \quad + \sum_{\alpha'} \, f^+_{\alpha \to \alpha'}(\vec{k}_\alpha,\hat{r}) \, \frac{ e^{i k_{\alpha'} r} }{r} \, \phi_{\alpha'}(\vec{\rho}_1, \vec{\rho}_2) \bigg] \\
 &=& \psi_{inc} + \psi_{scat} .  \nonumber
\end{eqnarray}
${\cal A}$ is a normalization factor which does not play a role for the result of the collision as we will see later. One could set ${\cal A}=1$ for simplicity. Fig.~\ref{FIG3} represents schematically the asymptotic form of the wavefunction and is separated in different parts:

\begin{enumerate}[(a)]
\item The incident wavefunction $\psi_{inc}$ is composed of an initial incident plane wave 
$e^{i \vec{k}_{\alpha} . \vec{r} }$ and the internal structure of the particles 
$\phi_{\alpha}(\vec{\rho}_1, \vec{\rho}_2)$. $\psi_{inc}$ is a solution of 
Eq.~\eqref{Hamiltonianrel} when $U_{int} \to 0$ at $r \to \infty$. 
The plane wave is characterized by a wavevector $\vec{k}_{\alpha}$ of magnitude $k_{\alpha}$ 
and incident direction $\hat{k}_{\alpha}$. The initial kinetic energy is 
$E_{k,\alpha} = \hbar^2 k_{\alpha}^2 / 2m_{red} = E_c$. In general $\vec{k}_{\alpha}$ can take 
any orientation with respect to $\vec{e}_Z$. 
In Fig.\ref{FIG3}, $\vec{k}_{\alpha}$ has been chosen with the same orientation 
than $\vec{e}_Z$.
The incident wavefunction $\psi_{inc}$ is expressed as a plane wave describing the particles at $Z \to -\infty$.
A part of the plane wave may continue to propagate towards $Z \to +\infty$ without interacting in the potential range.
\item In the zone of interaction around $Z \sim 0$, the particles interacts via the interaction potential energy $U_{int}(\vec{\rho}_1, \vec{\rho}_2, \vec{r})$.
\item Due to this interaction, the plane wave can also be scattered in a spherical manner. This is represented by a spherical wave $e^{i k_{\alpha'} r}/r$.
\item Due to the specific shape of the interaction potential, the plane wave is scattered  with an amplitude $f^+_{\alpha \to \alpha'}(\vec{k}_\alpha,\hat{r})$, referred to as
the {\bf scattering amplitude}. It represents the probability amplitude of the two particles for being scattered in the direction $\hat{r}$ from the initial state $\alpha$ with wavevector $\vec{k}_\alpha$ into the final state $\alpha'$. $\psi_{scat}$ represents the overall scattered wavefunction including the internal structure.
\end{enumerate}

\subsection{Observables}

\begin{figure}[h]
\begin{center}
\includegraphics*[width=6cm,keepaspectratio=true,angle=-90,trim=50 0 250 0]{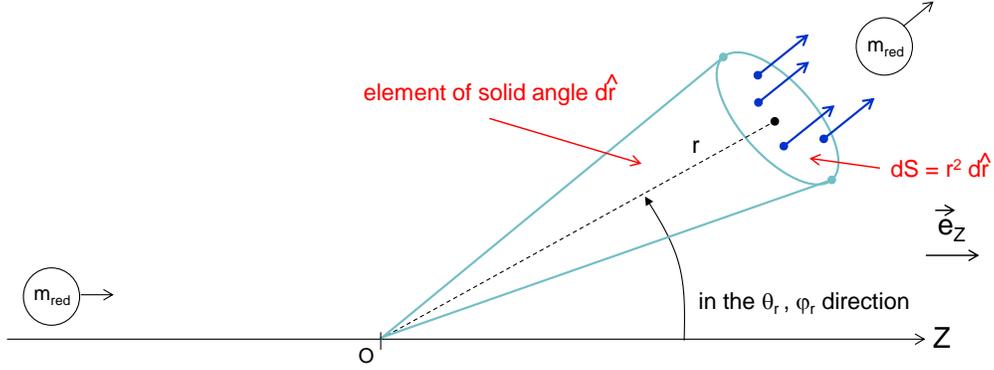}
\caption{Flux of fictive particles of mass $m_{red}$ into a detector far from the collision region.}
\label{FIG4}
\end{center}
\end{figure}

\noindent We relate now the scattering amplitude to the observables. Considering a typical beam/target collision experiment, the observable is the number of the beam particles, say particles 1, scattered out of the target particles, say particles 2, per unit of time and solid angle and detected by a detector in the laboratory frame somewhere far from the region of collision. In the CM frame it translates into the the number of fictitious particles of mass $m_{red}$ scattered out of the potential per unit of time and of solid angle $d\hat{r} = \sin\theta_r \, d\theta_r \, d\varphi_r$, detected by the detector in the direction $\hat{r} = (\theta_r, \varphi_r)$ for a transition $\alpha \to \alpha'$, and for a given incident direction $\vec{k}_{\alpha}$ (see Fig.~\ref{FIG4}). This number is proportional to the incident probability current $J_{inc} = N_{inc} \, j_{inc}$ of the $N_{inc}$ incoming particles of mass $m_{red}$. This is given by:
\begin{eqnarray} \label{diffXS1}
\frac{\partial N_{scat}}{\partial t \, \partial \hat{r} \, \partial \hat{k}_{\alpha}}(\vec{k}_{\alpha}, \hat{r}) \, \bigg|_{\alpha \to \alpha'} = J_{inc} \, \frac{\partial \sigma_{\alpha \to \alpha'}(\vec{k}_{\alpha},\hat{r})}{\partial \hat{r} \, \partial \hat{k}_{\alpha}}.
\end{eqnarray}
The quantity $\frac{\partial \sigma(\vec{k}_{\alpha},\hat{r})}{\partial \hat{r} \, \partial \hat{k}_{\alpha}}$ is called the {\bf differential cross section}. The flux of $J_{inc}$ in the differential cross section gives the number of particles scattered per unit of time and solid angle. By expressing $j_{inc}$ and $j_{scat}$ from $\psi_{inc}$ and $\psi_{scat}$ in Eq. \eqref{asymexpansion}, using the first line of the time-independent version of Eq. \eqref{current}, one can show that:
\begin{eqnarray} \label{diffXS2}
\frac{\partial \sigma_{\alpha \to \alpha'}(\vec{k}_{\alpha},\hat{r})}{\partial \hat{r} \, \partial \hat{k}_{\alpha}} = \frac{k_{\alpha'}}{k_\alpha} \, |f^+_{\alpha \to \alpha'}(\vec{k}_\alpha,\hat{r})|^2.
\end{eqnarray}
The {\bf integral cross section} for a given direction $\vec{k}_{\alpha}$ of collision is given by integrating the differential cross section over all scattering directions:
\begin{eqnarray} \label{intXS1}
\sigma_{\alpha \to \alpha'}(\vec{k}_{\alpha}) = \int  \frac{\partial \sigma(\vec{k}_{\alpha},\hat{r})}{\partial \hat{r} \, \partial \hat{k}_{\alpha}} \, d\hat{r} = 
\frac{k_\alpha'}{k_\alpha} \, \int  |f_{\alpha \to \alpha'}(\vec{k}_\alpha,\hat{r})|^2 \, d\hat{r}.
\end{eqnarray}
If the direction of collision is not specified (for example in a gas-cell experiment in contrast to a beam experiment), one also has to average over the value of the incident directions to obtain the averaged integral cross section for a given collision energy $E_{c} = \hbar^2 k_{\alpha}^2 / 2m_{red} $:
\begin{eqnarray} \label{intXS2}
\sigma_{\alpha \to \alpha'}({k}_{\alpha}) = \sigma_{\alpha \to \alpha'}(E_{c}) = \frac{\int  \sigma_{\alpha \to \alpha'}(\vec{k}_{\alpha}) \, d\hat{k}_{\alpha}}{\int d\hat{k}_{\alpha}} \, = \frac{1}{4\pi} \, \int  \sigma_{\alpha \to \alpha'}(\vec{k}_{\alpha}) \, d\hat{k}_{\alpha} .
\end{eqnarray}
Because:
\begin{eqnarray} \label{Nscatovertime1}
\frac{\partial N_{scat}}{\partial t} \, \bigg|_{\alpha \to \alpha'}  = \frac{1}{4\pi} \, \int d\hat{k}_{\alpha} \, \int d\hat{r} \, \frac{\partial N_{scat}}{\partial t \, \partial \hat{r} \, \partial \hat{k}_{\alpha}}(\vec{k}_{\alpha}, \hat{r}) \, \bigg|_{\alpha \to \alpha'}  = J_{inc} \times \sigma_{\alpha \to \alpha'}(E_{c}),
\end{eqnarray}
we see that the number of scattered particles per unit of time, summed over all incident directions is the flux of the incident probability current $J_{inc}$ through the averaged integral cross section $\sigma_{\alpha \to \alpha'}(E_{c})$. 
Then because $J_{inc}= N_{inc}/ \Delta S \Delta t$ is the number of incident particles crossing a given surface $\Delta S$ in the time interval $\Delta t$,  $N_{scat}/N_{inc} = \sigma / \Delta S$. Choosing a unit surface $\Delta S = 1$~cm$^2$, $\sigma$ expressed in cm$^2$ represents the number of scattered particles relative to the number of incident particles.
For gas-cell experiments, one usually has access to the initial volumic density $\rho_{gas}$ of the gas, not to the initial current $J_{inc}$. Using the second line of the time-independent version of Eq. \eqref{current} and applying it to $\psi_{inc}$ in Eq. \eqref{asymexpansion}, one can notice that $j_{inc} = |\psi_{inc}|^2 \, v = \rho_{inc} \, v$, where $v = \hbar k_\alpha / m_{red} = \sqrt{2 E_c / m_{red}}$ is the initial velocity of the fictitious particle, that is the relative initial velocity of the two colliding particles. Then $J_{inc} = (N_{inc} \, \rho_{inc}) \, v = \rho_{gas} \, v$. If we define another observable:
\begin{eqnarray} \label{rate}
\beta_{\alpha \to \alpha'}(E_{c}) = \sigma_{\alpha \to \alpha'}(E_{c}) \times v ,
\end{eqnarray}
called the {\bf rate coefficient},
then Eq. \eqref{Nscatovertime1} becomes now:
\begin{eqnarray} \label{Nscatovertime2}
\frac{\partial N_{scat}}{\partial t} \, \bigg|_{\alpha \to \alpha'}  =  \rho_{gas} \times 
\beta_{\alpha \to \alpha'}(E_{c}).
\end{eqnarray}
Therefore by knowing the initial volumic density instead of the the initial current, one extracts directly the rate coefficients instead of the cross sections, from the number of particles scattered per unit of time. If the cross section has unit of cm$^2$ and the velocity has unit of cm/s, the rate coefficient has unit of cm$^3$/s.

\section{In the region of collision}
\label{sec:short}

So far we have just defined some relevant quantities for the collisional properties of a system. We are now interested in how to calculate the cross section and the rate coefficient from a given potential energy. 

\subsection{Partial wave expansion}

In Eq.~\eqref{asymexpansion}, the plane wave appearing in $\psi_{inc}$ is a function of the vector $\vec{r}$. When $\vec{r}$ is represented by spherical coordinates 
$\vec{r}=\{r,\hat{r}=(\theta_r,\varphi_r)\}$, 
the kinetic energy operator in Eq.~\eqref{Hamiltonianrel} can be 
expressed in spherical coordinates by:
\begin{eqnarray} \label{kineticoperatorspherical}
 - \frac{\hbar^2}{2 m_{red} } \, \vec{\nabla}^2_{\vec{r}} \equiv
 - \frac{\hbar^2}{2 m_{red} }  
 \bigg[ \frac{1}{r^2} \, \frac{\partial}{\partial r} \, \left( r^2 \, \frac{\partial}{\partial r} \right) \bigg] + \frac{\widehat{\vec{l}}^{\ 2}}{2 m_{red}  r^2},
 \end{eqnarray}
The spherical harmonics $Y_{l}^{m_l}(\hat{r})$
are eigenfunctions of the square of the angular momentum operator $\widehat{\vec{l}}$ so that
$\widehat{\vec{l}}^{\ 2} \, Y_{l}^{m_l} =  \hbar^2 \, l(l+1)\, Y_{l}^{m_l}$. 
Using the spherical harmonic addition theorem, the plane wave in Eq.~\eqref{asymexpansion} can be expanded in spherical harmonics of quantum numbers $l,m_l$: 
\begin{eqnarray} \label{psiincdev1}
 \psi_{inc}  &=& {\cal A} \, e^{i \vec{k}_{\alpha}.\vec{r}} \, \phi_{\alpha}(\vec{\rho}_1, \vec{\rho}_2) \nonumber \\
 &=& {\cal A} \ 4\pi \, \sum_{l=0}^{\infty} \sum_{m_l=-l}^{l} 
i^l \, j_l(k_{\alpha}r) \, [Y_l^{m_l}(\hat{k}_{\alpha})]^* \,  Y_l^{m_l}(\hat{r}) \, \phi_{\alpha}(\vec{\rho}_1, \vec{\rho}_2) .
\end{eqnarray}
$j_l$ is a regular spherical Bessel function which behaves at large distances as:
\begin{eqnarray}
j_l(k_{\alpha}r) &\underset{r \to \infty}{\to}& \frac{\sin(k_\alpha r - l\pi/2)}{k_\alpha r} \nonumber \\
& \to & \frac{i}{2 k_\alpha r} \, \bigg[ e^{-i(k_\alpha r - l\pi/2)} - e^{i(k_\alpha r - l\pi/2)}\bigg] .
\end{eqnarray}
The asymptotic behavior of $\psi_{inc}$ is then:
\begin{eqnarray} \label{psiincdev2}
 \psi_{inc}  &\underset{r \to \infty}{\to}& \sum_{l=0}^{\infty} \sum_{m_l=-l}^{l} \, 
 N^{inc}_{\alpha \, l \, m_l}(\vec{k}_{\alpha}) \, 
\psi^{inc}_{\alpha \, l \, m_l}(\vec{\rho}_1, \vec{\rho}_2, \vec{r}) ,
\end{eqnarray}
where $N^{inc}_{\alpha \, l \, m_l}(\vec{k}_{\alpha}) \equiv {\cal A} \, (2 \pi i)/k_\alpha^{1/2} \, i^l \, \, [Y_l^{m_l}(\hat{k}_{\alpha})]^*$ is a normalization factor
independent of $r$. The functions:
\begin{eqnarray} \label{psiincdev3}
\psi^{inc}_{\alpha \, l \, m_l}(\vec{\rho}_1, \vec{\rho}_2, \vec{r})
 \equiv  \frac{f^{inc}(r)}{r} \, Y_l^{m_l}(\hat{r}) \, \phi_{\alpha}(\vec{\rho}_1, \vec{\rho}_2) 
\end{eqnarray}
are called the {\bf partial waves}, where 
$f^{inc}(r) = (e^{-i(k_\alpha r - l\pi/2)} - e^{i(k_\alpha r - l\pi/2)})/k_\alpha^{1/2}$ is 
the incident radial function (not to be mistaken with the scattering amplitude $f^+$) .
The expansion over the quantum numbers $l,m_l$ is called the partial wave expansion which represents the description of the colliding system in terms of components of the orbital angular momentum of the translational (collisional) motion.
We then extend to all $r$ 
the partial wave expansion Eq. \eqref{psiincdev2} for the total wavefunction 
$\psi^{E_{tot}}_{\alpha, \vec{k}_{\alpha}}$, for a given total energy $E_{tot}$, an initial internal quantum state of the molecules $\alpha$ and an initial wavevector $\vec{k}_{\alpha}$:
\begin{eqnarray} \label{totalexpansion}
\psi^{E_{tot}}_{\alpha,\vec{k}_{\alpha}}(\vec{\rho}_1, \vec{\rho}_2, \vec{r}) = \sum_{l=0}^{\infty} \sum_{m_l=-l}^{l}  \, N_{\alpha \, l \, m_l}(\vec{k}_{\alpha}) \, \psi^{E_{tot}}_{\alpha \, l \, m_l}(\vec{\rho}_1, \vec{\rho}_2, \vec{r}),
\end{eqnarray}
where $N_{\alpha \, l \, m_l}(\vec{k}_{\alpha})$ is the normalization factor for the wavefunction $\psi^{E_{tot}}_{\alpha \, l \, m_l}$. It will be defined, in 
Eq. \eqref{normalization}, by matching $\psi^{E_{tot}}$ 
to the asymptotic form of the wavefunction.
At finite $r$, the partial waves in Eq. \eqref{psiincdev3} are now expressed by:
\begin{eqnarray} \label{partialexpansion}
\psi^{E_{tot}}_{\alpha \, l \, m_l}(\vec{\rho}_1, \vec{\rho}_2, \vec{r}) =  \sum_{\alpha'} \sum_{l'=0}^{\infty} \sum_{m_l'=-l'}^{l'} \, \frac{f^{E_{tot}}_{\alpha' \, l' \, m_l' , \alpha \, l \, m_l}(r)}{r} \, Y_{l'}^{m_l'}(\hat{r}) \, \phi_{\alpha'}(\vec{\rho}_1, \vec{\rho}_2) . 
\end{eqnarray}
In contrast with Eq. \eqref{psiincdev3} for the incident wavefunction, we now allow 
in Eq. \eqref{partialexpansion} the components $\psi^{E_{tot}}_{\alpha \, l \, m_l}(\vec{\rho}_1, \vec{\rho}_2, \vec{r})$ to be a general, linear combination of the other final internal states $\alpha'$ and final orbital quantum numbers $l', m_l'$. This is due to the presence at finite $r$ of the potential energy term which can couple the initial state to all the other final states. 
The quantum numbers ($\alpha \, l \, m_l$) and ($\alpha' \, l' \, m_l'$) define the {\bf collisional channels} for the initial and final states, respectively.
The functions $f_{\alpha' \, l' \, m_l' , \alpha \, l \, m_l}$ will be responsible for the  transitions $\alpha \, l \, m_l \to \alpha' \, l' \, m_l' $. The transition is inelastic if 
$\alpha \ne \alpha'$ or elastic if $\alpha = \alpha'$ as illustrated in Fig. \ref{FIG2}.
In Eq. \eqref{partialexpansion}, one can define the {\bf basis set functions}:
\begin{eqnarray} \label{uncoupledbasis}
\Phi_{\alpha' \, l' \, m_l'}(\vec{\rho}_1, \vec{\rho}_2, \hat{r}) \equiv Y_{l'}^{m_l'}(\hat{r})  \, \phi_{\alpha'}(\vec{\rho}_1, \vec{\rho}_2)
\end{eqnarray}
which include all but the radial colliding motion degrees of freedom.
If we take the example mentioned above of two diatomic molecules with no spin, this basis set is:
\begin{eqnarray} \label{uncoupledbasisdiatomic}
\Phi_{\alpha' \, l' \, m_l'}(\vec{\rho}_1, \vec{\rho}_2, \hat{r}) 
 =  \frac{\chi_{v_1',n_1'}(\rho_1)}{\rho_1} \, 
\frac{\chi_{v_2',n_2'}(\rho_2)}{\rho_2} \, Y_{n'_1}^{m_{n_1}'}(\hat{\rho}_1) \, Y_{n'_2}^{m_{n_2}'}(\hat{\rho}_2) \, Y_{l'}^{m_l'}(\hat{r}) .
\end{eqnarray}
The basis set in Eq. \eqref{uncoupledbasis} is independent of the particles separation $r$, referred to as a {\bf diabatic representation}. There are other possible representations, such as the the {\bf diabatic-by-sector representation} or the {\bf adiabatic representation} which involves additional coupling terms. Such cases are beyond the scope of this paper.
Note that the formalism chosen here uses coordinate axes that 
point in the space-fixed frame directions, in particular the
quantization axis is oriented along the space-fixed unit vector $\vec{e}_Z$. 
For that reason, this is referred to a {\bf space-fixed frame formalism}
\cite{Takayanagi_PTP_11_557_1954,Arthurs_PRS_256_540_1960, Pack_JCP_60_633_1974,Green_JCP_62_2271_1975,Alexander_JCP_66_2166_1977}. 
In contrast, there is also a {\bf body-fixed frame formalism} 
\cite{Curtiss_JCP_21_2045_1953,Pack_JCP_60_633_1974,Launay_JPBAMOP_9_1823_1976, Alexander_JCP_66_2166_1977,Heil_JCP_68_2562_1978}
where the coordinate axes follow the body-fixed axes (see Fig.~\ref{FIG1}).
The two formulations are equivalent. The appropriate formalism depends on the treated problem. Generally, it is more efficient to use the space-fixed frame approach for large $J$, long range, weak coupling collisions, and the body-fixed frame approach for small $J$, short range, strong coupling collisions~\cite{Pack_JCP_60_633_1974}. 
Finally, because the basis set in Eq.~\eqref{uncoupledbasis} uses uncoupled functions of angular momentum, this is called the {\bf uncoupled representation} of the wavefunction \cite{Curtiss_JCP_21_2045_1953,Takayanagi_PTP_11_557_1954}.
In constrast, one can use a {\bf coupled representation} 
\cite{Curtiss_JCP_21_2045_1953,Takayanagi_PTP_11_557_1954,Arthurs_PRS_256_540_1960,Takayanagi_AAMP_1_149_1965,Zarur_JCP_60_2057_1974, Green_JCP_62_2271_1975,Alexander_JCP_66_2166_1977}
where all composite functions of angular momentum 
are coupled together to form a total angular momentum.
Taking as example the basis set of a diatomic molecule in Eq. \eqref{uncoupledbasisdiatomic},
one can couple their rotational angular momenta operators $\widehat{\vec{n}}_1$ and 
$\widehat{\vec{n}}_2$ and projections $\widehat{n}_{Z_1}$ and $\widehat{n}_{Z_2}$, 
into a coupled rotational angular momentum operator $\widehat{\vec{n}}_{12}$
and projection $\widehat{n}_{Z_{12}}$ with characteristic quantum numbers 
$n_{12}, m_{n_{12}}$. The coupled rotational wavefunction becomes:
\begin{eqnarray}
Y_{n_{12}}^{m_{n_{12}}} = \sum_{m_{n_1},  m_{n_2}} \, \langle n_1, m_{n_1}, n_2, m_{n_2} | n_{12}, m_{n_{12}} \rangle \, Y_{n_1}^{m_{n_1}}(\hat{\rho}_1) \, Y_{n_2}^{m_{n_2}}(\hat{\rho}_2) . 
\end{eqnarray}
The coupled rotational angular momentum operator $\widehat{\vec{n}}_{12}$ 
and orbital angular momentum $\widehat{\vec{l}}$, and projections 
$\widehat{n}_{Z_{12}}$ and $\widehat{l}_{Z}$, 
can be further coupled to form the total 
angular momentum operator $\widehat{\vec{J}}$ and projection $\widehat{J}_{Z}$
with characteristic quantum numbers $J, M_J$.
The fully coupled wavefunction becomes:
\begin{eqnarray} \label{coupledbasisdiatomic}
Y_{J}^{M_J} = \sum_{m_{n_{12}}, m_{l}} \, \langle n_{12}, m_{n_{12}}, l, m_{l} | J, M_J \rangle \, Y_{n_{12}}^{m_{n_{12}}} \, Y_{l}^{m_{l}}(\hat{r}) 
\end{eqnarray}
with $M_J = m_{n_{12}} + m_{l} = m_{n_1} + m_{n_2} + m_{l}$. 
The basis set is completed by combining this angular basis set with the internal radial functions $ \frac{\chi_{v_1,n_1}(\rho_1)}{\rho_1} \, \frac{\chi_{v_2,n_2}(\rho_2)}{\rho_2}$. 
The quantum numbers describing the wavefunction in the coupled representation are now $v_1$, $v_2$, $n_1$, $n_2$, $n_{12}$, $l$, $J$, $M_J$ in contrast with $v_1$, $v_2$, $n_1$, $m_{n_1}$, $n_2$, $m_{n_2}$, $l$, $m_l$ for the uncoupled representation. 
If the total potential energy satisfies 
$V(-\vec{\rho}_1, -\vec{\rho}_2, -\vec{r}) =  V(\vec{\rho}_1, \vec{\rho}_2, \vec{r})$, 
then $J$ and $M_J$ are good quantum numbers and are conserved during the collision. 
This is the case when no external potentials $V_{ext}$ are applied, as the potential energy surface $V_{int}$ does not depend on the global orientation of the two particles.
Then when no external field is applied it is useful to use the coupled representation since $J$ and $M_J$ are good quantum numbers, which leads to efficient and fast numerical calculations.
When external potentials $V_{ext}$ are applied, such as with arbitrary external electric or magnetic fields, different states with different values of $J$ become coupled so that $J$ is no more a good quantum number. Therefore the coupled representation loses its advantage. 
In contrast $M_J$ is still conserved if only one of the electric or the magnetic field is present at a time or if both fields share the same quantization axis. Once both fields are not aligned~\cite{Quemener_PRA_88_012706_2013}, $M_J$ is not a good quantum number anymore.
To treat ultracold collisions in any arbitrary external fields, including both weak and strong regimes, the uncoupled representation is generally preferred. The weak (strong) regime corresponds respectively to an interaction between the particle and the field much smaller (bigger) than the typical
zero-field particle energy. For example for a diatomic molecule with a permanent electric dipole moment $d$, one has to compare the magnitude $d \, E$ of the interaction of the 
molecule with an electric field $E$, with its typical energy without field, that is the rotational constant of the molecule $B_{rot}$. The weak (strong) regime is reached when typically $d \, E \ll B_{rot}$ ($d \, E \gg B_{rot}$). 
Note though that in case of strongly dominated anisotropic collisions 
\cite{Tscherbul_JCP_133_184104_2010,Tscherbul_PRA_85_052710_2012} the 
use of a body-fixed coupled representation can still be beneficial for efficient calculations, 
provided an appropriate treatment of unphysical states.
In the following, we will use the uncoupled representation in the collisional formalism 
since it is more intuitive to think in term of the individual quantum numbers of the separated particles. Besides, the last section of the paper will illustrate the case of dipolar molecules collisions in an electric field where the uncoupled representation is preferred and where the following formalism applies.

\subsection{Coupled equations}

We look now for the equations satisfied by the radial functions in Eq.~\eqref{partialexpansion}. The Schr\"odinger equation for a given partial wave $l,m_l$ is:
\begin{eqnarray} \label{Hamiltonianpartial}
\widehat{H} \, \psi^{E_{tot}}_{\alpha \, l \, m_l} = E_{tot} \, \psi^{E_{tot}}_{\alpha \, l \, m_l}
\end{eqnarray}
with $E_{tot} = \varepsilon_{\alpha} + \frac{\hbar^2 k_{\alpha}^2}{2 m_{red}} = \varepsilon_{\alpha} + E_{c}$ and $\widehat{H}$ given in Eq.~\eqref{Hamiltonianrel}. Inserting 
Eq.~\eqref{partialexpansion} into Eq.~\eqref{Hamiltonianpartial} and 
using Eq. \eqref{kineticoperatorspherical}, we obtain:
\begin{multline} \label{SEwithpartial}
\sum_{\alpha'} \sum_{l'=0}^{\infty} \sum_{m_l'=-l'}^{l'} \, 
\bigg[ - \frac{\hbar^2}{2 m_{red} } \, \frac{d^2}{d r^2} + \frac{\hbar^2 \, l'(l'+1)}{2 m_{red}  r^2}+ U_{int}(\vec{\rho}_1, \vec{\rho}_2, \vec{r}) + \varepsilon_{\alpha'} - E_{tot} \bigg] \\
\times f^{E_{tot}}_{\alpha' \, l' \, m_l' , \alpha \, l \, m_l}(r)
\, Y_{l'}^{m_l'}(\hat{r}) \, \phi_{\alpha'}(\vec{\rho}_1, \vec{\rho}_2)   = 0.
\end{multline}
The first derivatives and the $1/r$ term have disappeared in Eq.~\eqref{SEwithpartial} due to the choice of the form $f(r)/r$ in Eq.~\eqref{partialexpansion}. 
If we multiply the left-hand side of Eq. \eqref{SEwithpartial} by $[Y_{l''}^{m_l''}(\hat{r})]^* \, \phi^*_{\alpha''}(\vec{\rho}_1, \vec{\rho}_2)$ and integrate over all but the radial coordinate $r$, we are led to a system of coupled equations:
\begin{multline} \label{coupledequations}
\sum_{\alpha'} \sum_{l'=0}^{\infty} \sum_{m_l'=-l'}^{l'} \, 
\bigg[ \bigg\{ - \frac{\hbar^2}{2 m_{red} } \, \frac{d^2}{d r^2} + \frac{\hbar^2 \, l'(l'+1)}{2 m_{red}  r^2} + \varepsilon_{\alpha'} - E_{tot} \bigg\} \, \delta_{\alpha',\alpha''} \, \delta_{l',l''} \, \delta_{m_l',m_l''} \\
+ {\cal U}^{int}_{\alpha'' \, l'' \, m_l'' , \alpha' \, l' \, m_l'}(r) \bigg] \, f^{E_{tot}}_{\alpha' \, l' \, m_l' , \alpha \, l \, m_l}(r)  = 0,
\end{multline}
where:
\begin{multline} \label{couplingmatrix}
{\cal U}^{int}_{\alpha'' \, l'' \, m_l'' , \alpha' \, l' \, m_l'}(r) = \\
\int \, d\vec{\rho}_1 \, d\vec{\rho}_2 \, d\hat{r} \ [Y_{l''}^{m_l''}(\hat{r})]^* \, \phi^*_{\alpha''}(\vec{\rho}_1, \vec{\rho}_2) \, U_{int}(\vec{\rho}_1, \vec{\rho}_2, \vec{r}) 
\, Y_{l'}^{m_l'}(\hat{r}) \, \phi_{\alpha'}(\vec{\rho}_1, \vec{\rho}_2) \\
= \int \, d\vec{\rho}_1 \, d\vec{\rho}_2 \, d\hat{r} \ \Phi^*_{\alpha'' \, l'' \, m_l''}(\vec{\rho}_1, \vec{\rho}_2, \hat{r}) \, U_{int}(\vec{\rho}_1, \vec{\rho}_2, \vec{r}) 
\, \Phi_{\alpha' \, l' \, m_l'}(\vec{\rho}_1, \vec{\rho}_2, \hat{r})
 \end{multline}
is a matrix element of the {\bf coupling matrix} $\mathbf{{\cal U}}^{int}$. This matrix is real, symmetric and in general 
non-diagonal. It provides the couplings between the collisional channel $\alpha'' \, l'' \, m_l''$ to $\alpha' \, l' \, m_l'$ and is responsible for the inelastic transition in the collision. 
There are as many line equations of Eq.~\eqref{coupledequations} as there are $\alpha'' \, l'' \, m_l''$ numbers. All but the translational radial motion $r$, including the vibration and rotation of the molecules and the orbital angular momentum of the collision,
has been integrated out in Eq.~\eqref{couplingmatrix}. 
This provides a set of second-order coupled differential equations 
for the radial functions $f^{E_{tot}}_{\alpha' \, l' \, m_l' , \alpha \, l \, m_l}(r)$ for a given $\alpha \, l \, m_l$ and $E_{tot}$. 
A {\bf centrifugal term} has appeared in Eq.~\eqref{SEwithpartial} and Eq.~\eqref{coupledequations} coming from the development of the kinetic energy operator into an angular term proportional to the operator $\widehat{\vec{l}}^{\ 2}$.
One can define the corresponding (diagonal) matrix $\mathbf{{\cal U}}^{cent}$ with (diagonal) matrix elements:
\begin{eqnarray} \label{centrifugalterm}
{\cal U}^{cent}_{\alpha'' \, l'' \, m_l'' , \alpha' \, l' \, m_l'}(r) = 
\frac{\hbar^2 \, l'(l'+1)}{2 m_{red}  r^2} \ 
\delta_{\alpha',\alpha''} \, \delta_{l',l''} \, \delta_{m_l',m_l''}.
\end{eqnarray}
At ultralow collision energy, only a few, low quantum numbers of $l$ are required to describe the collision in Eq. \eqref{coupledequations} 
since higher values of $l$ implies higher values of the centrifugal barrier elements. 
Then this will imply lower values of the tunneling probability, which prevents the particles to come close to each other as $E_{c} \to 0$. 
As more values of $l$ are required for higher $E_{c}$, 
the time-independent partial wave method is more adapted to study collision
at ultralow energies than at high energies. \\

\noindent It is often useful to plot some elements of the set of equations
to get a knowledge of how strong the system is coupled. 
Defining the indexes $i'' \equiv \alpha'' \, l'' \, m_l''$ and $i' \equiv \alpha' \, l' \, m_l'$, one can extract an effective potential matrix 
$\mathbf{{\cal U}}^{eff}$ in Eq. \eqref{coupledequations}
with the following matrix elements:
\begin{eqnarray} \label{poteff}
{\cal U}^{eff}_{i'',i'}(r) = 
{\cal U}^{cent}_{i'',i'}(r)  
+ {\cal U}^{int}_{i'',i'}(r) 
+ \varepsilon_{\alpha'} \, \delta_{i',i''} , 
\end{eqnarray}
which includes the diagonal centrifugal term elements, 
the coupling matrix elements and
the energy thresholds of the two particles.
One can plot each diagonal element of this matrix as a function of $r$.
The corresponding curves are called {\bf the diabatic energy curves} and each of them 
tend at large $r$ to one of the threshold energies of the two particles. 
These curves provide a set of all possible effective potentials for the radial motion of 
the two colliding particles, 
when the non-diagonal terms of the coupling matrix are not present.
One can also include the effects of 
the non-diagonal terms of the coupling matrix in Eq.~\eqref{poteff} by
diagonalizing first the matrix $\mathbf{{\cal U}}^{eff}$ and then plot the eigenvalues 
as a function of $r$. The resulting curves are called {\bf the adiabatic energy curves},
each of them tend as well to the threshold energies of the two particles at large $r$.
These curves also provide a set of effective potentials for the radial motion, 
but now when the effect of the couplings is present.
When comparing both types of curves, one can see directly how and where 
the non-diagonal couplings elements affect the diagonal elements. 
If the adiabatic curves are quite comparable to the diabatic curves, then the system is weakly coupled. However, it is strongly coupled if both types of curves differ significantly.
This is illustrated later in the Section \ref{sec:krb} in Fig.~\ref{FIG7}. \\

\noindent Finally, the system of coupled equations can be expressed in a very compact form 
\cite{Johnson_JCP_13_445_1973,Johnson_JCP_69_4678_1978,Manolopoulos_JCP_85_6425_1986},
using a matrix notation:
\begin{eqnarray} \label{matrixnotation}
\bigg\{\mathbf{D}^2  + \mathbf{W} \bigg\} \, \mathbf{F}  = \mathbf{0}.
\end{eqnarray}
The matrix:
\begin{eqnarray} 
\mathbf{D}^2 =   \mathbf{I} \, \frac{d^2}{d r^2}
\end{eqnarray}
is a diagonal matrix, $\mathbf{I}$ being the identity matrix. 
$\mathbf{W}$ and $\mathbf{F}$ are real and symmetric matrices.
The matrix elements of $\mathbf{W}$ are:
\begin{eqnarray}
W_{i'',i'} = - \frac{2 m_{red}}{\hbar^2} \bigg[ 
  {\cal U}^{cent}_{i'',i'}(r)
+ {\cal U}^{int}_{i'',i'}(r)
+ ( \varepsilon_{\alpha'}  - E_{tot} ) \, \delta_{i',i''}   \bigg] .
\end{eqnarray}
The square matrix $\mathbf{F}$ involves the radial functions, and its elements are given by $F_{i',i} = f^{E_{tot}}_{i',i}(r)$ for which the line $i'\equiv \alpha' \, l' \, m_l'$ refers to the final state and the column $i \equiv \alpha \, l \, m_l$ to the initial state.
There are as many initial states as there are open channels for a given total energy $E_{tot}$. 
For example in Figure~\ref{FIG2}, there are 4 open channels at the given total energy, say $i=1,2,3,4$ with increasing energies $\varepsilon_1,\varepsilon_2,\varepsilon_3,\varepsilon_4$. The initial state is two molecules in $i=2$ with internal energy $\varepsilon_2$ with a collision energy $E_c$ and total energy $E_{tot} = \varepsilon_2 + E_c$. The solution of the wavefunction is the column $i=2$ of $\mathbf{F}$. The other columns of $\mathbf{F}$ represent the other independent solutions of the wavefunction corresponding to a total energy $E_{tot}$ but different initial conditions: $i=1$ corresponds to an initial state where the molecules start with internal energy $\varepsilon_1$ and collision energy $E_c = E_{tot} - \varepsilon_1$, $i=3$ to $\varepsilon_3$ and $E_c = E_{tot} - \varepsilon_3$ and finally $i=4$ to $\varepsilon_4$ and $E_c = E_{tot} - \varepsilon_4$.
Each column of $\mathbf{F}$ then represents a linearly independent solution of the problem. 

\subsection{Case of long-range interactions described by an electrostatic multipole-multipole expansion}

In practical, one has to compute all the elements ${\cal U}^{int}$ of the coupling matrix in Eq.~\eqref{couplingmatrix}. This requires the knowledge of the full potential energy surface $U_{int}(\vec{\rho}_1, \vec{\rho}_2, \vec{r})$. 
At long-range, the potential energy surface can be described in terms of an electrostatic multipole-multipole expansion~\cite{Stone_Book_1996}:
\begin{multline}
 U_{mult} = \frac{1}{4 \pi \varepsilon_0} \sum_{\lambda_1 \, \lambda_2 \, \lambda} \, \sum_{\omega_{\lambda_1} \, \omega_{\lambda_2}} \, (-1)^{\lambda_1} \ \left( \frac{(2 \lambda_1 + 2 \lambda_2 + 1)!}{(2 \lambda_1)!(2 \lambda_2)!}  \right)^{1/2}
 \,  \frac{Q_{\lambda_1  \omega_{\lambda_1}} \, Q_{\lambda_2  \omega_{\lambda_2}}}{ r^{\lambda+1}}   \\
 \times \delta_{\lambda,\lambda_1+\lambda_2} \,
\sum_{m_{\lambda_1} \, m_{\lambda_2} \, m_\lambda}  {\cal A}(\hat{\rho}_1,\hat{\rho}_2,\hat{r})
\end{multline}
with $\lambda  = \lambda_1 + \lambda_2$. The angular part is given by:
\begin{multline} \label{multipole}
{\cal A}(\hat{\rho}_1,\hat{\rho}_2,\hat{r})  = 
\, \left( \begin{array}{ccc}  \lambda_1 & \lambda_2 & \lambda \\ m_{\lambda_1} & m_{\lambda_2} & -m_\lambda \end{array}  \right) \\
 \times  [D^{\lambda_1}_{m_{\lambda_1} \omega_{\lambda_1}} (\hat{\rho}_1)]^* \ [D^{\lambda_2}_{m_{\lambda_2} \omega_{\lambda_2}}(\hat{\rho}_2)]^*  \  [D^{\lambda}_{-m_{\lambda} 0}(\hat{r})]^*.
\end{multline}
The symbol $(:::)$ is a Wigner 3-j symbol related to a Clebsch--Gordan coefficient and it is non-zero only if $m_\lambda = m_{\lambda_1} + m_{\lambda_2} $ and if $\lambda_1, \lambda_2, \lambda$ satisfy the triangle relation.
$Q_{\lambda_i \omega_{\lambda_i}}$ is a generalized multipole in the body-fixed frame 
of the molecule where we choose the unit vector $\vec{\rho}_i/|\vec{\rho}_i|$ for
molecule $i=1,2$ in Fig.~\ref{FIG1} to characterize the quantization axis.  
$\lambda_i$ is an angular momentum quantum number corresponding to the electronic charge distribution in the molecules $i=1,2$. $\lambda_i=0,1,2,3, ...$ correspond respectively to the {\bf charge, dipole, quadrupole, octopole moments}, and so on. $m_{\lambda_i}=[-\lambda_i,+\lambda_i]$
are the projection of these angular momenta onto the space-fixed frame quantization axis 
$\vec{e}_Z$. 
$\omega_{\lambda_i}=[-\lambda_i,+\lambda_i]$ is the projection onto the body-fixed frame quantization axis $\vec{\rho}_i/|\vec{\rho}_i|$ of molecule $i$.
In the case of $\Sigma$ electronic diatomic molecules, $\omega_{\lambda_1}=\omega_{\lambda_2}=0$ and one can write Eq.~\eqref{multipole} using the rotational eigenfunctions $|{n}_1, m_{n_1} \rangle$ and $|{n}_2, m_{n_2} \rangle$ of the molecules 1 and 2 for the internal wavefunction $\phi_{\alpha}$:
\begin{multline} \label{multipoleinbasis}
\langle {n}_1, m_{n_1}, {n}_2, m_{n_2}, l, m_l  | U_{mult} | {n}_1', m_{n_1}', {n}_2', m_{n_2}', l', m_l' \rangle  =   \\
 \frac{1}{4 \pi \varepsilon_0} \sum_{\lambda_1 \, \lambda_2 }
\, (-1)^{\lambda_1} \ \left( \frac{(2 \lambda_1 + 2 \lambda_2 + 1)!}{(2 \lambda_1)!(2 \lambda_2)!}  \right)^{1/2}
\, \frac{Q_{\lambda_1  0} \, Q_{\lambda_2  0}}{ r^{\lambda_1+\lambda_2+1}}   \\
  \sum_{m_{\lambda_1} \, m_{\lambda_2} }
 (-1)^{m_{n_1}+m_{n_2}+m_l}
 \, \left( \begin{array}{ccc}  \lambda_1 & \lambda_2 & \lambda_1 + \lambda_2 \\ m_{\lambda_1} & m_{\lambda_2} & - (m_{\lambda_1} + m_{\lambda_2} ) \end{array}  \right)  \\
\times  \sqrt{(2n_1+1) \, (2n_1'+1)}
\, \left( \begin{array}{ccc} n_1 & \lambda_1 & n_1' \\ 0 & 0 & 0 \end{array} \right)
\, \left( \begin{array}{ccc} n_1 & \lambda_1 & n_1' \\ -m_{n_1} & m_{\lambda_1} & m_{n_1}' \end{array}  \right) \\
\times  \sqrt{(2n_2+1) \, (2n_2'+1)}
\, \left( \begin{array}{ccc} n_2 & \lambda_2 & n_2' \\ 0 & 0 & 0 \end{array} \right)
\, \left( \begin{array}{ccc} n_2 & \lambda_2 & n_2' \\ -m_{n_2} & m_{\lambda_2} & m_{n_2}' \end{array}  \right) \\
 \times  \sqrt{(2l+1) \, (2l'+1)}
\, \left( \begin{array}{ccc} l & \lambda_1 + \lambda_2 & l' \\ 0 & 0 & 0 \end{array} \right)
\, \left( \begin{array}{ccc} l & \lambda_1 + \lambda_2 & l' \\ -m_{l} & - (m_{\lambda_1} + m_{\lambda_2}) & m_{l}' \end{array}  \right) .
\end{multline}
This provides the elements of the coupling matrix in Eq.~\eqref{couplingmatrix} at long-range.
From the properties of the 3-j symbols in Eq.~\eqref{multipoleinbasis}, we find the following selection rules (in addition to the triangle relation selection rule): 

(i)  $-m_{n_1} + m_{\lambda_1} + m_{n_1}'=0$,

(ii) $-m_{n_2} + m_{\lambda_2} + m_{n_2}'=0$,

(iii) $-m_{l} - (m_{\lambda_1} + m_{\lambda_2}) + m_{l}'=0$, 

\noindent which imply $m_{n_1} + m_{n_2} + m_{l} = m_{n_1}' + m_{n_2}' + m_{l}'$
or $M_J = M_J'$. There are no couplings if $M_J' \ne M_J$ showing that $M_J$ is conserved during the collision.

\subsection{Propagation. Log-derivative $\mathbf{Z}$ matrix}

To get all the radial functions $f(r)$, we need to solve the system of coupled equations Eq.~\eqref{coupledequations}. In reality, for practical and numerical reasons, the log-derivative of the radial functions is computed, rather than the functions themselves. This avoids numerical instabilities of the radial functions when a classically forbidden region is reached, and it avoids the necessity to compute the normalization of the functions at each $r$. We define the {\bf log-derivative matrix} of the matrix $\mathbf{F}(r)$ 
in Eq. \eqref{matrixnotation} by:
\begin{eqnarray}
\mathbf{Z}(r) = \mathbf{F}' \, \mathbf{F}^{-1}  = \bigg[ \frac{d}{dr} \, \mathbf{F}(r) \bigg] \, \bigg[ \mathbf{F}(r) \bigg]^{-1} .
\end{eqnarray}
The log-derivative matrix is a real and symmetric matrix 
so that $\mathbf{Z}^* = \mathbf{Z}$ and $\mathbf{Z}^t = \mathbf{Z}$.
When $r \to r_{min} \simeq 0$, the potential energy surface becomes very repulsive due to the impenetrability of the particles. Then the radial functions become zero with no couplings, so that $\mathbf{F}$ and $\mathbf{Z}$ are diagonal. We then impose the initial log-derivative at $r=r_{min}$ to be:
\begin{eqnarray} \label{initlogder}
\mathbf{Z}(r_{min}) = \infty \times \mathbf{I} .
\end{eqnarray}
If we divide the range of the radial coordinate $r$ from $r_{min}$ to $r_{max}$ into small segments of width $\Delta r$ (called sectors), one can propagate the log-derivative from sectors to sectors. Knowing what the log-derivative is in the previous sector, one can know what it is in the current sector. Because we know the log-derivative at $r_{min} \simeq 0$ we can propagate it to $r_{max} \simeq \infty$. We solve this way the system of coupled equations Eq.~\eqref{coupledequations} for each $r$, called a close-coupling calculation.
There are several efficient numerical methods to solve this set of equations, for example Refs.~\cite{Johnson_JCP_13_445_1973,Johnson_JCP_69_4678_1978,Manolopoulos_JCP_85_6425_1986},  which present no specific problems and can be routinely implemented. Those methods can compute not only the scattering properties of the coupled system with positive collision energies above the energy threshold of two initial separated particles, but also the presence of bound states with negative energies below the same threshold~\cite{Hutson_CPC_84_1_1994}.

\subsection{Symmetry considerations}

We discuss in this section the role of the inversion and permutation 
symmetries and how they are handled in the quantum formalism.
Inversion symmetry is considered when the potential energy does not include
potentials that depend on the absolute position of the particles,
while permutation symmetry is required when dealing with collisions of identical particles.
Including those symmetries will reduce the 
number of equations that are coupled in Eq.~\eqref{coupledequations}.
Finally, we briefly discuss how the quantum formalism is modified when dealing with external potentials when for example electric or magnetic fields are applied.

\subsubsection*{Inversion symmetry}

We consider here a potential energy $V_{int}$ that do not 
depend on the absolute position of the particles,
such as a potential energy surface of a system.
The basis function in Eq.~\eqref{uncoupledbasis} turns out to be also an eigenfunction of the inversion parity operator $\widehat{I}$. This operator corresponds to the transformation $(\vec{\rho}_1,\vec{\rho}_2,\vec{r}) \to (-\vec{\rho}_1,-\vec{\rho}_2,-\vec{r})$. This gives $\widehat{I} \, \Phi_{\alpha \, l \, m_l}(\vec{\rho}_1,\vec{\rho}_2,\hat{r}) = \epsilon_I \, \Phi_{\alpha \, l \, m_l}(\vec{\rho}_1,\vec{\rho}_2,\hat{r})$ with the inversion parity quantum number $\epsilon_I \equiv (-1)^{n_1+n_2+l} = \pm 1 $. 
This comes from the fact that the inversion symmetry of a vector $\vec{x} \to -\vec{x}$ is equivalent to $(x,\theta_x,\varphi_x) \to (x,\pi-\theta_x,\varphi_x+\pi)$ and thus implies $Y_{j}^{m_j}(-\hat{x}) = (-1)^{j} \, Y_{j}^{m_j}(\hat{x})$, while the radial wavefunction remains unchanged. 
If so, applying the inversion operator to a function depending on the radial coordinate
like the coupling elements ${\cal U}^{int}(r)$ in Eq.~\eqref{couplingmatrix} will let the function unchanged, $\widehat{I} \, {\cal U}^{int}(r) = {\cal U}^{int}(r)$. On the other hand:
\begin{eqnarray}
\widehat{I} \, {\cal U}^{int}(r) &=& \int \, d\vec{\rho}_1 \, d\vec{\rho}_2 \, d\hat{r} \ \Phi^*_{\alpha'' \, l'' \, m_l''}(-\vec{\rho}_1, -\vec{\rho}_2, -\hat{r}) \,  U_{int}(-\vec{\rho}_1, -\vec{\rho}_2, -\vec{r}) \, \Phi_{\alpha' \, l' \, m_l'}(-\vec{\rho}_1, -\vec{\rho}_2, -\hat{r}) \nonumber \\
&=& \epsilon_I \, \epsilon_I' \, \int \, d\vec{\rho}_1 \, d\vec{\rho}_2 \, d\hat{r} \ \Phi^*_{\alpha'' \, l'' \, m_l''}(\vec{\rho}_1, \vec{\rho}_2, \hat{r})  \, U_{int}(-\vec{\rho}_1, -\vec{\rho}_2, -\vec{r}) \, \Phi_{\alpha' \, l' \, m_l'}(\vec{\rho}_1, \vec{\rho}_2, \hat{r})  .
 \end{eqnarray}
As the potential energy surface satisfies 
$U_{int}(-\vec{\rho}_1, -\vec{\rho}_2, -\vec{r}) = U_{int}(\vec{\rho}_1, \vec{\rho}_2, \vec{r})$, then $\epsilon_I \, \epsilon_I' = 1$ or $\epsilon_I = \epsilon_I'$. Inversion parity is then conserved in a collision involving a potential energy surface. This can be checked directly in Eq.~\eqref{multipoleinbasis}. From the property of the 3-j symbols which contains the zero elements, these three symbols are non-zero if $(-1)^{n_1+\lambda_1+n_1'}=1$, $(-1)^{n_2+\lambda_2+n_2'}=1$, and $(-1)^{l+\lambda_1+\lambda_2+l'}=1$. By arranging the $(-1)^{\lambda_1+\lambda_2}$ term, this implies $(-1)^{n_1+n_2+l}=(-1)^{n_1'+n_2'+l'}$ and then $\epsilon_I=\epsilon_I'$. This applies to collision of either identical or different molecules.
Note that inversion is not always conserved in a collision
if external potentials $V_{ext}$ are included. 

\subsubsection*{Permutation symmetry} 

If the two particles are identical, one also has to symmetrize the internal wavefunction $\phi_{\alpha}(\vec{\rho}_1, \vec{\rho}_2) = \phi_{\alpha_1}(\vec{\rho}_1) \, \phi_{\alpha_2}(\vec{\rho}_2)$ with respect to the permutation of the two particles operator $\widehat{P}$.
The permutation of the two particles is equivalent to the transformation $(\vec{\rho}_1, \vec{\rho}_2,\vec{r}) \to (\vec{\rho}_2, \vec{\rho}_1,-\vec{r})$.
The properly symmetrized internal wavefunction is given by:
\begin{eqnarray} \label{internalbasissym}
\phi_{\alpha \, \eta}(\vec{\rho}_1, \vec{\rho}_2) = 
\frac{1}{\sqrt{2(1+\delta_{\alpha_1,\alpha_2})}} \, 
\bigg\{
\phi_{\alpha_1}(\vec{\rho}_1) \, \phi_{\alpha_2}(\vec{\rho}_2)
+ \eta \,
\phi_{\alpha_2}(\vec{\rho}_1) \, \phi_{\alpha_1}(\vec{\rho}_2)
\bigg\} .
\end{eqnarray}
$\eta = \pm 1$ describes, respectively, a symmetric and anti-symmetric internal wavefunction with respect to the permutation, so that $\widehat{P} \, \phi_{\alpha \, \eta} = \eta \, \phi_{\alpha,\eta}$.
Now the basis set functions in Eq.~\eqref{uncoupledbasis} become:
\begin{eqnarray} \label{uncoupledbasissym}
\Phi_{\alpha \, l \, m_l \, \eta}(\vec{\rho}_1, \vec{\rho}_2, \hat{r})  \equiv  \phi_{\alpha \, \eta}(\vec{\rho}_1, \vec{\rho}_2) \, Y_{l}^{m_l}(\hat{r}) ,
\end{eqnarray}
and $\widehat{P} \, \Phi_{\alpha \, l \, m_l \, \eta} = \eta \, (-1)^{l} \, \Phi_{\alpha \, l \, m_l \, \eta}$ since $\vec{r} \to -\vec{r}$ is equivalent to $(r, \theta_r,\varphi_r) \to (r, \pi-\theta_r,\varphi_r+\pi)$ and $Y_{l}^{m_l}(-\hat{r}) = (-1)^{l} \, Y_{l}^{m_l}(\hat{r})$. One can show, using the above properly symmetrized basis set, that the coupled equations are diagonal in $\eta$. \\

\noindent Additionally, under permutation $\widehat{P}$ of two identical particles, the total wavefunction has to obey the symmetrization principle:
\begin{eqnarray} 
\widehat{P} \, \psi^{E_{tot}} = \epsilon_P \, \psi^{E_{tot}}
\end{eqnarray}
with $\epsilon_P=+1$ if the (composite) particles are identical bosons and $\epsilon_P=-1$ if the (composite) particles are identical fermions. On the basis set functions, it gives $\widehat{P} \, \Phi =  \epsilon_P \, \Phi$.
Then this implies specific selection rules for $\eta$ and 
$l$ following the fact that $\eta \, (-1)^{l} = \epsilon_P$. In the case of identical bosons $\epsilon_P=+1$, internal wavefunctions of $\eta = +1$ (resp. $\eta = -1$) symmetry imply even partial waves $l=0,2,4...$ (resp. odd partial waves $l=1,3,5...$). In the case of identical fermions $\epsilon_P=-1$, internal wavefunctions of $\eta = +1$ (resp. $\eta = -1$) symmetry imply odd partial waves $l=1,3,5...$ (resp. even partial waves $l=0,2,4...$). 
Note that all values of $l$ are included in the dynamics since both symmetries of $\eta$ are generally allowed. \\

\noindent In the special case of indistinguishable particles, meaning particles in the same quantum state so that $\alpha_1 = \alpha_2$, Eq.~\eqref{internalbasissym} implies that the wavefunction for the $\eta=-1$ symmetry does not exist. In this special case, the number of partial waves describing the dynamics is reduced following the rules just mentioned above since only the $\eta=+1$ symmetry survives. 
This implies even partial waves $l=0,2,4...$ for indistinguishable bosons and odd partial waves $l=1,3,5...$ for indistinguishable fermions.
At ultralow energy $E_c \to 0$, only the first and lowest partial wave is important for the dynamics. It is usually common to say that identical bosons in indistinguishable states collide in the $s$-wave (to refer to $l=0$) and identical fermions in indistinguishable states collide in the $p$-wave (to refer to $l=1$). 

\subsubsection*{Collisions in external fields}

Often in ultracold physics, additional external 
fields, such as electric or magnetic fields, are present to control the properties of the individual particles~\cite{Krems_IRPC_24_99_2005,Krems_PCCP_10_4079_2008} and their interactions~\cite{Quemener_CR_112_4949_2012}. Then, additional external potentials $V_{ext}$ 
appear in the Hamiltonian of the system. 
The previous formalism remains unchanged except that the individual particles $i=1,2$ are now perturbated by the external field. As a consequence, the (bare) internal state of the particle $\phi_{\alpha_i}$ in the absence of an external field is replaced with the corresponding (dressed) internal state $\tilde{\phi}_{\alpha_i}$ in the presence of the field.
The dressed states are a linear combination of the bare states with given coefficients 
due to the interaction of the particle with the field. 
In the collision formalism, we 
just replace the individual bare states $\phi_{\alpha_i}$ of the particles $i=1,2$ with their dressed states $\tilde{\phi}_{\alpha_i}$. To compute the elements of the coupling matrix in Eq.~\eqref{couplingmatrix} between the dressed states
$\tilde{\Phi}_{\alpha \, l \, m_l} = $ $\tilde{\phi}_{\alpha_1} \, \tilde{\phi}_{\alpha_2} \, Y_l^{m_l}$, there is now just an additional step. We replace the dressed states by the expression of their linear combination of bare states and we compute the corresponding sum of all the bare elements. This presents no difficulties and is routinely done numerically.
The other consequence is that $J$ is not a good quantum number anymore, as mentioned above, and an uncoupled representation basis set is generally preferred.
The last section of this paper will illustrate such an example, where ultracold collisions of electric dipolar molecules of KRb occur in an external electric field. 

\section{Matching the two regions}
\label{sec:matching}

To relate the observables far from the collision region to the potential energy and radial functions in the collision region, we will equate Eq.~\eqref{asymexpansion} and Eq.~\eqref{totalexpansion}.

\subsection{Reactance matrix $\mathbf{K}$. Relation with $\mathbf{Z}$}

For practical and numerical reasons, the matching is not done at $r_{max} \simeq \infty$ but rather at $r_{max}$ for which $|{\cal U}^{int}| \ll |{\cal U}^{cent}|$. This the distance
for which the interaction terms (diagonal and non-diagonal) 
can be safely neglected compared to the centrifugal ones.
The set of coupled equations Eq.~\eqref{matrixnotation} becomes diagonal,
each diagonal elements taking the form:
\begin{eqnarray} \label{uncoupledequation}
\bigg\{ - \frac{\hbar^2}{2 m_{red}} \, \frac{d^2}{d r^2} + \frac{\hbar^2 \, l(l+1)}{2 m_{red} r^2} + \varepsilon - E_{tot} \bigg\} \, f(r) = 0 ,
\end{eqnarray}
each equations only differing by the values $\varepsilon$ of the thresholds. This can also be written: 
\begin{eqnarray}
r^2 \, f''(r) + [k^2 r^2 - l(l+1)] \, f(r) = 0 
\end{eqnarray}
with the wavevector $k = \sqrt{2 m_{red}  (E_{tot}-\varepsilon)/\hbar^2}$.
Two independent solutions are given by $\tilde{j}$ and $\tilde{n}$, the Ricatti-Bessel functions and Ricatti-Neumann functions \cite{Abramowitz_Stegun_Book_1964}. They are related to the spherical Bessel and spherical Neumann functions by $\tilde{j}_l=kr \, j_l(kr)$ and $\tilde{n}_l= kr \, n_l(kr)$ and to the Bessel and Neumann functions by $j_l(kr) = \sqrt{\pi/2kr} \, J_{l+1/2}(kr)$ and $n_l(kr)= \sqrt{\pi/2kr} \, N_{l+1/2}(kr)$. 
If we set $\rho=kr$, the solutions for the first $l$'s are:
\begin{align} \label{jntable}
\tilde{j}_0(\rho) &= \sin(\rho) & \tilde{j}_1 &= \frac{\sin(\rho)}{\rho} - \cos(\rho) \\
\tilde{n}_0(\rho) &= - \cos(\rho) & \tilde{n}_1 &= - \frac{\cos(\rho)}{\rho} - \sin(\rho)  .
\end{align}
The behaviour for $\rho \to 0$ is:
%
\begin{align} \label{jnrho0}
\tilde{j}_l(\rho) &\underset{\rho \to 0}{\propto} \frac{\rho^{l+1}}{(2l+1)!!} & 
\tilde{n}_l(\rho) &\underset{\rho \to 0}{\propto} -(2l-1)!! \, \rho^{-l}
\end{align}
with $x!!=x (x-2) (x-4) ...$ 
$\tilde{j}$ are often called regular functions since $\tilde{j} \to 0$ as $\rho \to 0$ and $\tilde{n}$ are often called irregular functions since $\tilde{n} \to \pm \infty$ as $\rho \to 0$. 
For $\rho \to \infty$: 
\begin{align} \label{jnrhoinfty}
\tilde{j}_l(\rho) &\underset{\rho \to \infty}{\propto}  \sin(\rho - l\pi/2) & 
\tilde{n}_l(\rho) &\underset{\rho \to \infty}{\propto} -\cos(\rho - l\pi/2).
\end{align}
A general solution of Eq.~\eqref{uncoupledequation} for the radial functions
at $r=r_{max}$ is given by:
\begin{eqnarray} \label{generalasymformK}
\mathbf{F}(r) = \mathbf{F}^{(1)} \, \mathbf{A} + \mathbf{F}^{(2)} \,  \mathbf{B} \ \bigg|_{r=r_{max}}
\end{eqnarray}
where:
\begin{align}
&
F^{(1)}_{i',i} = \delta_{i',i} \frac{1}{k_{\alpha'}^{1/2}} \, \tilde{j}_{l'}(k_{\alpha'} r) 
&
F^{(2)}_{i',i} = \delta_{i',i} \frac{1}{k_{\alpha'}^{1/2}} \, \tilde{n}_{l'}(k_{\alpha'} r) .
\end{align}
$\mathbf{A}, \mathbf{B}$ are real constant matrices, independent of $r$.
In the special case without coupling terms, that is no off-diagonal terms in Eq.~\eqref{matrixnotation} $\forall r$,
the system is uncoupled and $\mathbf{A}$, $\mathbf{B}$ will be diagonal
at $r=r_{max}$. 
More generally when coupling terms are present 
for $r < r_{max}$ in Eq.~\eqref{matrixnotation},
the system is coupled and $\mathbf{A}$, $\mathbf{B}$ will be full matrices in general
at $r=r_{max}$.
We can also write Eq.~\eqref{generalasymformK} as:
\begin{eqnarray} \label{Kasymform1}
\mathbf{F}(r) = \mathbf{F}^{K}(r) \, \mathbf{N}^{K} \  \bigg|_{r=r_{max}}
\end{eqnarray}
with:
\begin{eqnarray} \label{Kasymform2}
\mathbf{F}^{K}(r) = \{ \mathbf{F}^{(1)} - \mathbf{F}^{(2)} \,  \mathbf{K} \} .
\end{eqnarray}
$\mathbf{K}$ is called the {\bf reactance matrix}. $\mathbf{N}^{K}$ is a real normalisation matrix. 
From Eq.~\eqref{generalasymformK}, Eq.~\eqref{Kasymform1} and Eq.~\eqref{Kasymform2},
$\mathbf{K} \equiv - \mathbf{B} \, \mathbf{A}^{-1}$ and $\mathbf{N}^{K} \equiv \mathbf{A}$. 
The superscript $K$ indicates that the radial functions obey boundary conditions of the $\mathbf{K}$ matrix.
The $\mathbf{K}$ matrix is real as the matrices $\mathbf{A}, \mathbf{B}$ are real.
The off-diagonal matrix elements of $\mathbf{K}$ provide an indication
of the character of the other final channels due to the couplings from the interaction potential energy of the system in the wavefunction, for a given incident initial colliding channel.
We chose the factors $k_{\alpha'}^{-1/2}$ in the two linearly independent functions $\mathbf{F}^{(1)}, \mathbf{F}^{(2)}$ so that the {\bf Wronskian matrix} $ \mathbf{W} = \mathbf{F}^{(1)} \mathbf{F}^{'(2)} - \mathbf{F}^{'(1)} \mathbf{F}^{(2)} $ is the identity matrix $\mathbf{I}$. If so, $\mathbf{K}$ is also a symmetric matrix. 
This is shown in Proof 1 of the appendix of this paper.
$\mathbf{K}$ is related to the $\mathbf{Z}$ matrix by (the order of the matrix multiplication is important to get a symmetric matrix):
\begin{eqnarray} \label{ZtoK} 
\mathbf{K} = \bigg\{ \mathbf{Z} \, \mathbf{F}^{(2)} - \mathbf{F}^{'(2)} \bigg\}^{-1} \, \bigg\{ \mathbf{Z} \, \mathbf{F}^{(1)} - \mathbf{F}^{'(1)} \bigg\} \,  \bigg|_{r=r_{max}} .
\end{eqnarray}
This is often referred to as the matching procedure, performed at $r=r_{max}$.
This is shown in Proof 2 of the appendix. From the proof, one can see that the reactance matrix is independent of the choice of the normalisation matrix $\mathbf{N}^{K}$ of the radial functions. It depends only on its log-derivative matrix $\mathbf{Z}$ at $r_{max}$: if $\mathbf{Z}$ is diagonal (non-diagonal) due to the uncoupled (coupled) Schr{\"o}dinger equations, $\mathbf{K}$ is diagonal (non-diagonal). 

\subsection{Scattering matrix $\mathbf{S}$. Relation with $\mathbf{K}$}

The problem with Eq.~\eqref{Kasymform1} is that the functions are not written in terms of incoming and outgoing radial functions, as the ones appearing in the asymptotic wavefunction in Eq.~\eqref{asymexpansion}.
When $r \to \infty$ in Eq.~\eqref{Kasymform1}, Eq.~\eqref{jnrhoinfty} shows that the Ricatti-Bessel and Ricatti-Neumann functions behave as sine and cosine functions which can also be written in terms of incoming/outgoing spherical wave. 
Another general solution of Eq.~\eqref{uncoupledequation}
for the radial functions is then given by:
\begin{eqnarray} \label{generalasymformS}
\mathbf{F}(r) &\underset{r \to \infty}{=}& \mathbf{F}^{(-)} \, \mathbf{A'} + \mathbf{F}^{(+)} \,  \mathbf{B'},
\end{eqnarray}
where
\begin{eqnarray} \label{radialasymformS}
F^{\pm}_{i',i} = \delta_{i',i} \frac{1}{k_{\alpha'}^{1/2}} \, e^{\pm i(k_{\alpha'} r - l'\pi/2)}
\end{eqnarray}
are incoming (-) or outgoing (+) spherical waves and $\mathbf{A'}, \mathbf{B'}$ are complex constant matrices, independent of $r$.
Again, in the special case without coupling terms
$\forall r$ in Eq.~\eqref{matrixnotation}, 
$\mathbf{A'}$ and $\mathbf{B'}$ will be diagonal
while they will be full matrices 
if coupling terms are present.
We can also write Eq.~\eqref{generalasymformS} as:
\begin{eqnarray} \label{Sasymform1}
\mathbf{F}(r) &\underset{r \to \infty}{=}& \mathbf{F}^{S}(r)  \,  \mathbf{N}^S
\end{eqnarray}
with:
\begin{eqnarray} \label{Sasymform2}
\mathbf{F}^{S}(r) = \{ \mathbf{F}^- - \mathbf{F}^+ \,  \mathbf{S} \}.
\end{eqnarray}
$\mathbf{S}$ is the {\bf scattering matrix}. $\mathbf{N}^{S}$ is a complex normalisation matrix. The superscript $S$ indicates now that the radial functions obey boundary conditions of 
the $\mathbf{S}$ matrix.
Eq.~\eqref{Sasymform1} is the useful form to match with the asymptotic one in 
Eq.~\eqref{asymexpansion} because it uses incoming and outgoing radial functions as well. 
From Eq.~\eqref{generalasymformS}, Eq.~\eqref{Sasymform1} and Eq.~\eqref{Sasymform2},
$\mathbf{S} \equiv - \mathbf{B'} \, \mathbf{A'}^{-1}$ and $\mathbf{N}^{S} \equiv \mathbf{A'}$. %
$\mathbf{S}$ is related to the $\mathbf{K}$ matrix by:
\begin{eqnarray} \label{KtoS}
\mathbf{S} = \frac{\mathbf{I}+i\mathbf{K}}{\mathbf{I}-i\mathbf{K}} .
\end{eqnarray}
This is shown in Proof 3 of the appendix. Again from the proof, one can see 
that the scattering matrix is independent of the normalization matrix $\mathbf{N}^{S}$ of the radial functions. It depends only on the reactance matrix, and hence the log-derivative matrix. 
$\mathbf{S}$ is a symmetric matrix: $\mathbf{S}^t = \mathbf{S}$, and a unitary matrix: $\mathbf{S} \,  \mathbf{S}^\dagger = \mathbf{S}^\dagger \, \mathbf{S} = \mathbf{I}$,
as shown in Proof 4 of the appendix.
$\mathbf{S}$ in general is a complex matrix including a real and imaginary part.
The coefficient of the outgoing waves in the channel $i'$ coming from an incoming wave in the channel $i$ is given by the element $S_{i',i}$. The elements $|S_{i',i}|^2$ correspond to the ratio of the outgoing flux $4 \pi \hbar |S_{i',i}|^2 / m_{{red}}$ over the incoming one $4 \pi \hbar / m_{{red}}$ in absolute value (one can compute the flux using Eq.~\eqref{flux} and Eq.~\eqref{current}, using the radial functions in Eq.~\eqref{radialasymformS} and integrating over the whole solid angle $d\hat{r}$). Then the probability to collide from a state $i$ to a state $i'$ is simply given by
\begin{eqnarray}
P_{i \to i'} = |S_{i',i}|^2 \qquad \qquad \text{with:} \qquad \sum_{i'} \, P_{i \to i'} = 1.
\end{eqnarray}
Finally, we impose a diagonal normalization matrix $\mathbf{N}^{S}$ in Eq.~\eqref{Sasymform1}. This enables that an independent solution of the Schr{\"o}dinger equation, corresponding to a given column of the $\mathbf{F}$ matrix given by Eq.~\eqref{Sasymform1}, has the same overall normalization in a multiplicative factor, as suggested by Eq.~\eqref{totalexpansion}. In that way, the diagonal elements of this matrix identify directly with the normalization factor we have already defined in Eq.~\eqref{totalexpansion} so that ${N}^{S}_{\alpha \, l \, m_l, \alpha \, l \, m_l} \equiv N_{\alpha \, l \, m_l}$.

\subsection{Transition matrix $\mathbf{T}$. Relation with observables}

At $r_{max} \simeq \infty$, ${\cal U}^{int}, {\cal U}^{cent} \to 0$ 
in Eq.~\eqref{SEwithpartial}, the wavefunction tends to Eq.~\eqref{asymexpansion} 
far from the collision region:
\begin{eqnarray}
\psi^{E_{tot}}_{\alpha, \vec{k}_{\alpha}} \underset{r \to \infty}{=} {\cal A} \, \bigg[ e^{i \vec{k}_{\alpha} . \vec{r} } \, \phi_{\alpha} + \sum_{\alpha'} \, f^+_{\alpha \to \alpha'} \, \frac{ e^{i k_{\alpha'} r} }{r} \, \phi_{\alpha'} \bigg]  = \psi_{inc} + \psi_{scat},
\end{eqnarray}
where $\psi_{inc}$ has the form of Eq.~\eqref{psiincdev1}.
When no interaction potential energy is present, no scattering is present ($\psi_{scat} =0$), we see that $\psi_{\alpha, \vec{k}_{\alpha}} = \psi_{inc}$ contains only the initial internal state $\phi_{\alpha}$ and for which the radial function is a superposition of an incoming and outgoing spherical wave $ e^{\pm i(k_\alpha r - l\pi/2)}/r $, of same amplitudes. 
In the presence of the interaction potential energy term $U_{int}$, the scattering wave $\psi_{scat}$ will additionally produce outgoing spherical waves $e^{i(k_{\alpha'} r - l'\pi/2)}/r $  in final states $\phi_{\alpha'}$, responsible for inelastic transitions. 
Both the asymptotic expansion Eq.~\eqref{asymexpansion} and the partial wave expansion Eq.~\eqref{totalexpansion} and Eq.~\eqref{partialexpansion}, using Eq.~\eqref{Sasymform1}, contain now an incoming and outgoing spherical wave term. One can then identify their expressions. This leads to the expression of the normalization factor of each partial waves in Eq.~\eqref{totalexpansion}:
\begin{eqnarray}\label{normalization}
N_{\alpha \, l \, m_l}(\vec{k}_\alpha) = {\cal A} \, \frac{2 \pi i}{k_\alpha^{1/2}} \, i^l \, [Y_l^{m_l}(\hat{k}_\alpha)]^*  .
\end{eqnarray}
Similarly, the scattering amplitude in Eq.~\eqref{asymexpansion} writes:
\begin{multline} \label{scatamplitude}
f^+_{\alpha \to \alpha'}(\vec{k}_\alpha,\hat{r}) = \frac{2 \pi}{i \, k_\alpha^{1/2} \, k_{\alpha'}^{1/2}} \\ 
\sum_{l=0}^{\infty} \sum_{m_l=-l}^{l} \sum_{l'=0}^{\infty} \sum_{m_l'=-l'}^{l'}  i^{l-l'} \, [Y_l^{m_l}(\hat{k}_\alpha)]^* \, Y_{l'}^{m_l'}(\hat{r}) \, T_{\alpha' \, l' \, m_l',\alpha \, l \, m_l}(k_\alpha)
\end{multline}
in terms of the {\bf transition matrix}:
\begin{eqnarray}
\mathbf{T} = \mathbf{S} - \mathbf{I} .
\end{eqnarray}
Note that some references use a definition $\mathbf{T} = \mathbf{I} - \mathbf{S}$ but the scattering amplitude is then defined with a factor of $2 \pi i$ instead of $2 \pi / i $ in Eq.~\eqref{scatamplitude}, which provides in any case the same scattering amplitude.
One can then get the observables in terms of the $\mathbf{T}$ matrix. The differential cross section is given by Eq.~\eqref{diffXS2}:
\begin{multline} \label{diffXS3}
\frac{\partial \sigma_{\alpha \to \alpha'}(\vec{k}_{\alpha},\hat{r})}{\partial \hat{r} \, \partial \vec{k}_{\alpha}} = \frac{k_{\alpha'}}{k_\alpha} \, |f^+_{\alpha \to \alpha'}(\vec{k}_\alpha,\hat{r})|^2 \\
= \frac{4 \pi^2}{k_\alpha^2} \sum_{l_a}^{} \sum_{m_{l_a}}^{} \sum_{l_b}^{} \sum_{m_{l_b}}^{} \sum_{l_c}^{} \sum_{m_{l_c}}^{} \sum_{l_d}^{} \sum_{m_{l_d}}^{}
i^{-l_a+l_b+l_c-l_d} \\
\times Y_{l_a}^{m_{l_a}}(\hat{k}_\alpha) \, [Y_{l_b}^{m_{l_b}}(\hat{r})]^* \, [Y_{l_c}^{m_{l_c}}(\hat{k}_\alpha)]^* \, Y_{l_d}^{m_{l_d}}(\hat{r}) \\
\times  T^*_{\alpha' l_a m_{l_a},\alpha l_b m_{l_b}}(k_\alpha) \, T_{\alpha' \, l_c \, m_{l_c},\alpha \, l_d \, m_{l_d}}(k_\alpha),
\end{multline}
where running indexes $l_a, m_{l_a}, ..., l_d, m_{l_d}$ have been used in the expression of 
the modulus squared of the scattering amplitude.
The averaged integral cross section is given by Eq.~\eqref{intXS2}:
\begin{eqnarray} \label{avintXS}
\sigma_{\alpha \to \alpha'}(E_{c}) &=& \Delta \times
\frac{\pi}{k_\alpha^2} \, \sum_{l}^{} \sum_{m_{l}}^{} \sum_{l'}^{} \sum_{m_{l}'}^{}  \,
|T_{\alpha' \, l' \, m_l',\alpha \, l \, m_l}(k_\alpha)|^2 \nonumber \\
&=& \sum_{l}^{} \sum_{m_{l}}^{} \sigma_{\alpha \to \alpha', l \, m_l}(E_{c}), 
\end{eqnarray}
where we can define a partial wave cross section $\sigma_{\alpha \to \alpha', l \, m_l} $.
From Eq.~\eqref{diffXS3} to Eq.~\eqref{avintXS}, we used the fact that the integration over 
$\hat{k}_{\alpha}$ gives $\delta_{l_a,l_c} \, \delta_{m_{l_a},m_{l_c}}$ and the integration over $\hat{r}$ gives $\delta_{l_b,l_d} \, \delta_{m_{l_b},m_{l_d}}$ .
In the case of identical particles starting in indistinguishable states ($\phi_\alpha = \phi_{\alpha_1} \phi_{\alpha_2}$ with $\alpha_1 = \alpha_2$), 
one has to multiply the cross sections by a factor $\Delta = 2$ for symmetry reasons as the differential cross sections have to be integrated 
over half space only~\cite{Burke_PhDThesis_1999,Tscherbul_NJP_11_055021_2009}. Note that in this case the number of partial waves is halved compared to the case of identical but distinguishable or different particles, due to the specific rules mentioned above for the partial waves. 
In the case of identical particles starting in distinguishable states ($\alpha_1 \ne \alpha_2$),
or in the case of different particles, $\Delta = 1$.
Eq.~\eqref{rate} is used to obtain the corresponding rate coefficient.
In a numerical calculation, one usually computes the $\mathbf{Z},\mathbf{K},\mathbf{S},\mathbf{T}$ matrices in this order to get the observables.

\subsection{Link to scattering of structureless particles. The central potential problem}

It is interesting to see how to recover 
the central potential problem for elastic scattering of structureless particles 
(that can be found in many textbooks \cite{Brandsen_Joachain_Book_2003,Cohen-Tannoudji_Book_1997,Friedrich_Book_2005,Landau_Book_1958,Child_Book_1996,
Atkins_Friedman_Book_2005}),
from the more general elastic and inelastic scattering formalism 
of particles with internal structure presented in this paper.
First, in the central potential problem the interaction is assumed to be
isotropic, $U_{int}(\vec{r})=U_{int}(r)$, so that it 
does not depend on the angles $\hat{r}$.
Then the operators ${\widehat{H},\widehat{L}^2,\widehat{L}_z}$ commute and $l,m_l$ are good quantum numbers which are conserved during the collision, in addition with the total energy. So $l'=l$ and $m_l'=m_l$.
Secondly, for an elastic collision, $\alpha'=\alpha$. 
Finally, the collision does not depend on the direction of the incident particles since the potential is isotropic. One can choose for example the direction $\hat{k}_{\alpha} \equiv \hat{z} = (0,0)$. Then $[Y_l^{m_l}(0,0)]^* \equiv \sqrt{2l+1/4\pi} \, \delta_{m_l,0}$, this implies $m_l=0$.
The asymptotic expansion writes:
\begin{eqnarray}
\psi^{E_{tot}}_{k} &\underset{r \to \infty}{=}& A \bigg[ e^{i {k}  \, z } + f^+(k,\hat{r}) \, \frac{ e^{i k r} }{r}  \bigg] .
\end{eqnarray}
Then the scattering amplitude reduces to:
\begin{eqnarray} \label{scatamplcentral}
f^+_{\alpha \to \alpha}(\vec{k}_\alpha,\hat{r}) &=& f^+(k_\alpha,\hat{r}) \nonumber \\
&=& \frac{2 \pi}{i \, k_\alpha^{1/2} \, k_{\alpha'}^{1/2}} \
\sum_{l=0}^{\infty} \sum_{m_l=-l}^{l} \sum_{l'=0}^{\infty} \sum_{m_l'=-l'}^{l'}  i^{l-l'}  [Y_l^{m_l}(\hat{k}_\alpha)]^*  \, Y_{l'}^{m_l'}(\hat{r}) \, T_{\alpha' \, l' \, m_l',\alpha \, l \, m_l}(k_\alpha) \nonumber \\
&=& \frac{2 \pi}{i \, k_\alpha} \,
\sum_{l=0}^{\infty} i^{0} \, \sqrt{\frac{2l+1}{4\pi}} \, \delta_{m_l,0} \, Y_{l}^{0}(\hat{r}) \, T_{\alpha \, l \, 0,\alpha \, l \, 0}(k_\alpha) \nonumber \\
&=& \frac{2 \pi}{i \, k} \,
\sum_{l=0}^{\infty} i^{0} \, \sqrt{\frac{2l+1}{4\pi}} \,  \sqrt{\frac{2l+1}{4\pi}} \, P_l^0(\cos \theta) \,  T_{l} \nonumber \\
&=& \frac{1}{2 \, i \, k} \,
\sum_{l=0}^{\infty} (2l+1) P_l^0(\cos \theta) \,  T_{l} 
\end{eqnarray}
and the cross section reduces to:
\begin{eqnarray} \label{XScentral}
\sigma(k) &=& \int \, d\hat{r} \, |f^+|^2 \nonumber \\
&=& \frac{1}{4 k^2} \,
\sum_{l=0}^{\infty} \sum_{l'=0}^{\infty} \, (2l+1) \, (2l'+1) 
\, \bigg[ \int \, d\hat{r} \, P_l^0(\cos \theta) \,  P_{l'}^0(\cos \theta) \bigg] \, T^*_{l} \, T_{l'} \nonumber \\
&=& \frac{\pi}{k^2} \,
\sum_{l=0}^{\infty} \, (2l+1) \, |T_{l}|^2 .
\end{eqnarray}
From Eq.~\eqref{scatamplcentral} to Eq.~\eqref{XScentral}, 
we used $\int_0^{\pi} \, P_l^0 \,  P_{l'}^0 \, \sin{\theta} \, d\theta = 2/(2l+1) \ \delta_{l,l'}$ and $\int_0^{2\pi} \, d\varphi = 2 \pi$.
Because for elastic collisions, the $\mathbf{S}$ matrix reduces to an element for a given $l$, it can be written $S_l=e^{2i \delta_l(k)}$. $\delta_l(k)$ is called the {\bf scattering phase shift} in the partial wave $l$. Since there are no inelastic channels then $|S_l|^2=1$. The role of the central potential is then to shift the phase of the outgoing wave by $\delta_l(k)$. 
By noting that $1 -  e^{2i \delta_l(k)} = e^{i \delta_l(k)} ( e^{-i \delta_l(k)} - e^{i \delta_l(k)} ) = - 2i \, e^{i \delta_l(k)} \sin\delta_l(k)$,
one can also find: 
\begin{eqnarray} \label{XSinBook}
\sigma(k) =  \frac{4 \pi}{k^2} \, \sum_{l=0}^{\infty} \, (2l+1) \, \sin^2\delta_l(k)
\end{eqnarray}
which is a formula often quoted in textbooks.  
The phase shift is related to the K matrix by $K_l=\tan \delta_l$. Note that we recover Eq.~\eqref{KtoS} because:
\begin{eqnarray}
S_l = e^{2i \delta_l} = \frac{1+i\tan \delta_l}{1-i\tan \delta_l}=\frac{1+i K_l}{1-iK_l}.
\end{eqnarray}
We used the fact that $1 \pm i \tan{\delta} = 1 \pm \frac{e^{i\delta}-e^{-i\delta}}{e^{i\delta}+e^{-i\delta}} =  \frac{2 e^{\pm i \delta}}{e^{i\delta}+e^{-i\delta}}$.

\section{Behaviour at ultralow energy. Scattering length and threshold laws}
\label{sec:ultracold}

We now present how the dynamics of two colliding particles behaves at ultralow energy when $E_c \to 0$. To simplify the discussion, we will take the case of an elastic collision of structureless particles interacting with a central potential $U_{int}(r)$, as described in the previous section. The Schr\"odinger equation writes:
\begin{eqnarray}
\bigg\{ - \frac{\hbar^2}{2 m_{red} } \, \frac{d^2}{d r^2} + \frac{\hbar^2 \, l(l+1)}{ 2 m_{red} r^2} + {\cal U}^{int}(r) - E_c \bigg\} \, f(r) = 0  
\end{eqnarray}
where $E_c = \hbar^2 k^2/2m_{red}$ (we take the energy of the two separated particles as the reference energy). 
The matching procedure Eq.~\eqref{ZtoK} is performed at $r_{max}=r_0$ where
$r_0$ denotes the typical distance for which $|{\cal U}^{int}(r_0)| \ll |{\cal U}^{cent}(r_0)|$. 
On one hand, there is always a typical collision energy $E_c^*$ for and below which $E_c \ll |{\cal U}^{int}(r_0)|, |{\cal U}^{cent}(r_0)|$ so that the Schr{\"o}dinger equation is in this limit independent of $E_c$ at $r_0$. Then, the function and its derivative at $r = r_0$ are also independent of $E_c$. Its log-derivative is then a given constant $Z = C$ at $r = r_0$. 
On the other hand, from Eq.~\eqref{Kasymform2}, 
we know the general form of $f(r) = f^{(1)}(\rho) - f^{(2)}(\rho) \  K_l = \tilde{j}(\rho)/\sqrt{k} - \tilde{n}(\rho)/\sqrt{k} \  K_l$ (using $\rho = kr$) and its derivative $f'(r) = f^{'(1)}(kr) - f^{'(2)}(kr) \  K_l $, the prime being a derivative with respect to $r$. If we use $d/dr = k \, d/d\rho$, we have $f'(r) = k \, (df^{(1)}(\rho)/d\rho) \, - \, k \, (df^{(2)}(\rho)/d\rho) \  K_l = \sqrt{k} \, (d\tilde{j}(\rho)/d\rho) - \sqrt{k} \, (d\tilde{n}(\rho)/d\rho) \ K_l$. 
We perform the matching procedure at $r_{max}=r_0$, using Eq.~\eqref{jnrho0} for the functions and their derivatives as $E_c, k \to 0$, using a constant energy-independent value of the log-derivative $Z = C$, and using the fact that $K_l = \tan(\delta_l)$ where $\delta_l$ is the scattering phase shift (see the central potential problem above). 
Eq.~\eqref{ZtoK} gives \cite{Brandsen_Joachain_Book_2003}:
\begin{eqnarray} \label{tandeltal1}
\tan(\delta_l) &=& \frac{Z_l \, f^{(1)} - f^{'(1)}}{Z_l \, f^{(2)} - f^{'(2)}} \nonumber \\
&=& \frac{C \, \tilde{j}(\rho)/\sqrt{k} - \sqrt{k} \, (d\tilde{j}(\rho)/d\rho)}{C \, \tilde{n}(\rho)/\sqrt{k} - \sqrt{k} \, (d\tilde{n}(\rho)/d\rho)}  \nonumber \\
&=& \frac{C \, \tilde{j}(\rho) - k \, (d\tilde{j}(\rho)/d\rho)}{C \, \tilde{n}(\rho) - k \, (d\tilde{n}(\rho)/d\rho)}  \nonumber \\
&\underset{\rho \to 0}{=}& -\frac{1}{(2l+1)!!(2l-1)!!} \ \frac{C \, D \, \rho^{l+1} - E \, k \rho^{l}}{C \, F \, \rho^{-l} - G \, k \rho^{-l-1}}  \nonumber \\
&\underset{k \to 0}{=}& -\frac{(2l+1)}{[(2l+1)!!]^2} \ \frac{C \, D \, k^{l+1} \, r_0^{l+1} - E \, k^{l+1} \, r_0^{l}}{C \, F \, k^{-l} \, r_0^{-l} - G \, k^{-l} \, r_0^{-l-1}}  \nonumber \\
&\underset{k\to 0}{=}& -\frac{(2l+1)}{[(2l+1)!!]^2} \bigg( \frac{C \, D \, r_0^{l+1} - E \, r_0^{l}}{C \, F \, r_0^{-l} - G \, r_0^{-l-1}}\bigg) \, k^{2l+1} \nonumber \\
&\underset{k \to 0}{=}& - {\cal L} \, k^{2l+1}
\end{eqnarray}
where $D,E,F,G$ are dimensionless proportionality factors in Eq.~\eqref{jnrho0}. 
Since $C$ and $k$ have the dimension of an inverse length and $\tan(\delta_l)$ has no units, 
the constant ${\cal L}$ has the dimension of a length to the power $2l+1$. 
The most important partial wave to describe the collision at ultralow energies
corresponds to the first lowest partial wave. 
For identical and indistinguishable bosonic particles or for different particles, the first partial wave is $l=0$ as mentioned earlier, then ${\cal L}$ has the dimension of a length. We define the {\bf $\bf{s}$-wave scattering length} by:
\begin{eqnarray} \label{as}
a_s = \underset{k\to 0}{\lim} - \frac{\tan \delta_{l=0}(k)}{k}.
\end{eqnarray}
The cross section can be linked to the scattering length by:
\begin{eqnarray} \label{XStoscat}
\sigma_{l=0}(k) &=& \frac{4\pi}{k^2} \sin^2\delta_0(k) = \frac{4\pi}{k^2} \frac{1}{\frac{\sin^2\delta_0(k)+\cos^2\delta_0(k)}{\sin^2\delta_0(k)}} \nonumber \\
&=& \frac{4\pi}{k^2} \frac{1}{1 + \frac{1}{\tan^2\delta_0(k)}} 
= \frac{4\pi}{k^2} \frac{1}{1 + \frac{1}{(a_s k)^2}} \nonumber \\
&\underset{k\to 0}{\to}& 4 \pi a_s^2.
\end{eqnarray}
This cross section is the same than the one provided by a hard sphere potential of 
radius $a_s$, that is ${\cal U}^{int}(r) = \infty$ if $r \le a_s$, 0 otherwise.
Then at ultralow energy, one can safely replace a complicated interaction potential energy
by a simple hard sphere model potential, since the cross sections will be the same.
The model potential represents a simple, effective potential for the collision of the system,
for which the scattering length plays the essential parameter.
In ultracold physics in many-body interacting systems, 
the scattering length plays a crucial role 
in terms of which the many-body physics is described. It appears, for example, in the Gross--Pitaevskii equations to describe the physics of ultracold gases of particles in interaction \cite{Pethick_Smith_Book_2001,Pitaevskii_Stringari_Book_2003}.
For identical and indistinguishable fermionic particles, the first partial wave is $l=1$, then ${\cal L}$ is a volume. We define the {\bf $\bf{p}$-wave scattering length} (the volume ${\cal L}$ is the cube of this length) by:
\begin{eqnarray} \label{ap}
a^3_p = \underset{k\to 0}{\lim} - \frac{\tan \delta_{l=1}(k)}{k^3}.
\end{eqnarray}
The result in Eq.~\eqref{tandeltal1}, Eq.~\eqref{as} and Eq.~\eqref{ap}
are not generally valid for potentials falling off asymptotically as an inverse power of the distance $r$.
Also, for interaction potential ${\cal U}^{int}(r) = \pm C_s / r^s$, with $s>2$,
the threshold behaviour in Eq.~\eqref{tandeltal1} is dominant for partial waves
$l < (s-3)/2$ \cite{Friedrich_Book_2005,Child_Book_1996,Sadeghpour_JPBAMOP_33_93_2000,
Hutson_BookChapter_2009}.
For partial waves $l > (s-3)/2$, the dominant threshold behaviour becomes:
\begin{eqnarray} \label{tandeltal2}
\tan(\delta_l) \underset{k\to 0}{\propto} k^{s-2}.
\end{eqnarray}
For partial waves $l = (s-3)/2$, both contributions Eq.~\eqref{tandeltal1}
and Eq.~\eqref{tandeltal2} are taken to describe the threshold behaviour.
Using Eq.~\eqref{XSinBook} and Eq.~\eqref{XStoscat}, the behaviour of the elastic cross sections and rate coefficients at a vanishing collision energy becomes 
when using the threshold behaviour Eq.~\eqref{tandeltal1}:
\begin{align} \label{Wignerlaws1}
&
\sigma^{el}_l \underset{k, E_c\to 0}{\propto} k^{4l} \propto E_c^{2l} 
&
\beta^{el}_l \underset{k, E_c\to 0}{\propto} k^{4l+1} \propto E_c^{2l+1/2}.
\end{align}
When using the threshold behaviour Eq.~\eqref{tandeltal2}, it becomes:
\begin{align} \label{Wignerlaws2}
&
\sigma^{el}_l \underset{k, E_c\to 0}{\propto} k^{2s-6} \propto E_c^{s-3}
&
\beta^{el}_l \underset{k, E_c\to 0}{\propto} k^{2s-5} \propto E_c^{s-5/2}.
\end{align}
Inelastic/reactive cross sections and rate coefficients behaviours 
are given without proof \cite{Wigner_PR_73_1002_1948}:
\begin{align} \label{Wignerlaws3}
&
\sigma^{in/re}_l \underset{k, E_c\to 0}{\propto} k^{2l-1} \propto E_c^{l-1/2}
&
\beta^{in/re}_l \underset{k, E_c\to 0}{\propto} k^{2l} \propto E_c^{l} .
\end{align}
These expressions are called the {\bf threshold laws} or {\bf Wigner laws}
\cite{Wigner_PR_73_1002_1948}. 

\section{Application to ultracold collisions of dipolar molecules in electric fields}
\label{sec:krb}

In 2008, a major breakthrough has been made in the field of ultracold molecular physics
with the production of a dense and coherent gas of 
ultracold dipolar fermionic $^{40}$K$^{87}$Rb 
molecules~\cite{Ni_S_322_231_2008}.
In contrast with the previous experiments of that time \cite{Quemener_CR_112_4949_2012},
these molecules were produced 
in their ground electronic state $^1\Sigma^+$, their ground vibrational state $v=0$,
and their ground rotational state $n=0$,
with additional control over the hyperfine 
states~\cite{Ospelkaus_PRL_104_030402_2010}.
Therefore, the experimentalists were able 
to address the internal state of all the molecules of a dense gas 
to the absolute ground state.
The molecule of KRb possesses in its own frame 
a permanent electric dipole moment of $d=0.57$~D\cite{Ni_S_322_231_2008}. 
Therefore, the energy of the molecules and their interactions can be manipulated with an external electric field.
KRb molecules are also chemically reactive even 
in their absolute ground state~\cite{Zuchowski_PRA_81_060703_2010,
Byrd_PRA_82_010502_2010,Meyer_PRA_82_042707_2010}
so that KRb + KRb $\to$ K$_2$ + Rb$_2$ is an exoergic process.
On the one hand, this is a drawback for creating long-lived gases of strong dipolar ultracold molecules in experiments since this chemical reaction will lead to large molecular losses. But on the other hand if an electric field is applied,
the molecular losses, which can be quite easily measured 
in a experiment as a function of time, will directly provide a signature
of the dipolar interaction of the colliding molecules.
It is therefore important to understand the collisional properties of the dipolar gas, in terms of its stability and lifetime. Collisions are also driving the thermal equilibrium of the gas and are very important to perform evaporative cooling to further decrease the temperature and reach eventually quantum degeneracy, as it was performed for ultracold gases of atoms~\cite{Cornell_RMP_74_875_2002,Ketterle_RMP_74_1131_2002}.
As an illustration of the formalism studied in this paper, we will present in this section the collisional properties of KRb + KRb $\to$ K$_2$ + Rb$_2$ as a function of an electric field, for $^1\Sigma^+$, $v=0$ molecules initially in the ground rotational state $n=0$ and in the first excited rotational state $n=1$, for both fermionic $^{40}$K$^{87}$Rb molecules and bosonic $^{41}$K$^{87}$Rb molecules. The spin structure of the molecules will not be taken into account in the following.

\subsection{A simplified problem}

The full time-independent quantum mechanical formalism developed previously still represents a numerical challenge for diatom-diatom or polyatomic molecular collisions at the present time: 

(i) Firstly, full potential energy surfaces of polyatomic systems (involving all degrees of freedom) are generally challenging to compute, especially in the region of the complex where the atoms are close to each other. This is still feasible for tri-atomic systems but becomes in general difficult for tetra-atomic ones.

(ii) Secondly, when systems are chemically reactive, the Jacobi coordinates used in the present formalism are not appropriate anymore. Instead, one has to use hyperspherical coordinates~\cite{Whitten_JMP_9_1103_1968,Johnson_JCP_79_1916_1983} as already mentioned, which treat, in a symmetric way, the polyatomic system formed by the atoms. The hyperspherical formalism~\cite{Pack_JCP_87_3888_1987,Launay_CPL_163_178_1989, Rittenhouse_JPBAMOP_44_172001_2011} is well adapted for proper symmetrization of the overall wavefunction with respect to identical atom exchange as well as treating the  products of a chemical reaction. 
However, the formalism becomes difficult to handle numerically, especially using a full potential energy surface.
Consequently, chemically reactive collision of diatomic molecules have to be tackled in another way at the present time. To overcome those problems, we will use two assumptions to treat the collisions of two diatomic reactive molecules.

\subsubsection*{Long-range interaction}

First we will consider only the long-range interaction of the potential energy
so that $U_{int} = U_{mult}$. 
At ultralow collision energies, the dynamics becomes more and more sensitive
to the term that is the most longer-ranged in the potential energy.
In the case of neutral diatomic molecules which possess an electric dipole moment, like KRb, the most longer-ranged term in the multipole-multipole interaction
is the dipole-dipole interaction ($\lambda_1 = \lambda_2 = 1$, $\lambda=2$ in Eq.~\eqref{multipoleinbasis}), so that $U_{mult} = U_{dd}$. 
The matrix elements in the uncoupled basis presented above 
for a diatomic molecule are given by: 
\begin{multline}
\langle {n}_1, m_{n_1}, {n}_2, m_{n_2}, l, m_l  | U_{dd} | {n}_1', m_{n_1}', {n}_2', m_{n_2}', l', m_l' \rangle  =   \\
 -\sqrt{30} \, \frac{d^2}{4 \pi \varepsilon_0 \, r^3} 
  \sum_{m_{\lambda_1} \, m_{\lambda_2}}
 (-1)^{m_{n_1}+m_{n_2}+m_l}
 \, \left( \begin{array}{ccc}  1 & 1 & 2 \\ m_{\lambda_1} & m_{\lambda_2} & -(m_{\lambda_1} + m_{\lambda_2}) \end{array}  \right)  \\
\times  \sqrt{(2n_1+1) \, (2n_1'+1)}
\, \left( \begin{array}{ccc} n_1 & 1 & n_1' \\ 0 & 0 & 0 \end{array} \right)
\, \left( \begin{array}{ccc} n_1 & 1 & n_1' \\ -m_{n_1} & m_{\lambda_1} & m_{n_1}' \end{array}  \right) \\
\times  \sqrt{(2n_2+1) \, (2n_2'+1)}
\, \left( \begin{array}{ccc} n_2 & 1 & n_2' \\ 0 & 0 & 0 \end{array} \right)
\, \left( \begin{array}{ccc} n_2 & 1 & n_2' \\ -m_{n_2} & m_{\lambda_2} & m_{n_2}' \end{array}  \right) \\
 \times  \sqrt{(2l+1) \, (2l'+1)}
\, \left( \begin{array}{ccc} l & 2 & l' \\ 0 & 0 & 0 \end{array} \right)
\, \left( \begin{array}{ccc} l & 2 & l' \\ -m_{l} & -(m_{\lambda_1} + m_{\lambda_2}) & m_{l}' \end{array}  \right) 
\label{dipoledipole}
\end{multline}
where $d \equiv Q_{1 0}$ is the electric dipole moment. 
Higher multipole terms such as the quadrupole and octopole terms
\cite{Byrd_PRA_86_032711_2012} can become important at higher collision energies
\cite{Wang_NJP_17_035015_2015}.
In addition to the dipole-dipole term, we include 
a diagonal electronic $-C_6/r^6$ van der Waals interaction \cite{Kotochigova_NJP_12_073041_2010,Lepers_PRA_88_032709_2013,Zuchowski_PRA_87_022706_2013}.

\subsubsection*{A short-range tunable condition}

Secondly, we will use a phenomenological approach to treat the molecular collisions at short-range. The initial condition for the propagation of the radial wavefunction was given by a diagonal matrix in Eq.~\eqref{initlogder} corresponding to an infinite wall at $r=r_{min}$. We will now slightly modify this condition. We still keep the matrix diagonal, meaning no couplings between channels at short-range, 
but we now allow some additional effective scattering phase-shift 
and some effective loss for each channels due to the result of the (unknown) potential energy surface at short-range. We then construct a flexible and tunable log-derivative
matrix, where the diagonal elements for a channel $i$ are given 
by~\cite{Wang_NJP_17_035015_2015}:
\begin{eqnarray}
Z_{i,i}(r=r_{{min}}) &=&  
\frac{4 \, k_{{min}} \, s \, c \, \sqrt{1-p_{SR}}}{c^2 \, (\sqrt{1-p_{SR}}-1)^2 + s^2 \, (\sqrt{1-p_{SR}}+1)^2} \nonumber \\
&-& i \ \frac{k_{{min}} \, p_{SR}}
{c^2 \, (\sqrt{1-p_{SR}}-1)^2 + s^2 \, (\sqrt{1-p_{SR}}+1)^2},
\label{logder}
\end{eqnarray}
where:
\begin{eqnarray}
k_{{min}} = \sqrt{\frac{2 \, m_{{red}} \, [E_{{tot}} - {\cal U}^{{eff}}_{i,i}(r=r_{{min}})]} {\hbar^2}} 
\end{eqnarray}
and:
\begin{eqnarray}
c = \cos(k_{{min}} \, r_{{min}} + \delta_{SR})  \qquad \qquad 
s = \sin(k_{{min}} \, r_{{min}} + \delta_{SR}). 
\end{eqnarray}
The log-derivative at $r_{{min}}$ can be continuously tuned by two 
parameters $0 \le p_{SR} \le 1$ and $0 \le \delta_{SR} \le \pi$.
$p_{SR}$ represents a loss probability for the flux coming from the long-range region
$r > r_{{min}}$ describing phenomenologically a loss at short-range,
while $\delta_{SR}$ represents a phase shift accumulated from the short-range region
$r < r_{{min}}$, describing phenomenologically the result of 
the (unknown) potential energy surface there.
The above log-derivative condition has been constructed at $r=r_{min}$
so that it describes a square well of constant depth ${\cal U}^{eff}_{i,i}(r_{min})$ given by Eq.~\eqref{poteff} from $r=0$ to $r=r_{min}$,
with a tunable complex phase shift 
$\delta  = \delta_r + i \, \delta_i$ and a corresponding amplitude    
$ e^{ 2i\delta}  = e^{-2 \delta_i} \, e^{ 2i \delta_r }$
appearing in front of the outgoing solution of the radial wavefunction of the square well potential 
$ e^{- i k_{{min}} r} - e^{ 2i\delta}  \, e^{+ i k_{{min}} r}$
\cite{Wang_NJP_17_035015_2015}. 
$e^{ 2i\delta}$ corresponds to a $S_{SR}$ matrix element at short-range which probability
$|S_{SR}|^2=e^{-4\delta_i}$ is a number between 0 and 1 depending on $\delta_i$
and represents the probability for the flux going to the long-range region $r > r_{{min}}$.
The loss probability $p_{SR}$ is then defined as $p_{SR}=1-e^{-4\delta_i}$
so that $e^{-2 \delta_i} \equiv \sqrt{1 - p_{SR}}$.
We also note $\delta_r \equiv \delta_{SR}$. \\

\noindent The condition for full loss of the flux at short range is given by $p_{SR}=1$ 
and gives $Z_{i,i}(r_{{min}}) = - i \ k_{{min}} $ for all diagonal elements. This is often called the universal regime since 
no resonances appear in the cross sections or rate 
coefficients \cite{Idziaszek_PRA_82_020703_2010,Wang_NJP_17_035015_2015}
and they are independent of the phase shifts $\delta_{SR}$
\cite{Idziaszek_PRL_104_113202_2010}. 
The results then become independent of 
the short-range interaction of the systems.
The opposite condition for full reflection of the flux is given by $p_{SR}=0$ and gives $Z(r_{{min}}) = k_{{min}} c/s $ which is the usual case for a square potential and depends on the tunable phase-shift $\delta_{SR}$. With an adequate choice of $\delta_{SR} = - k_{{min}} \, r_{{min}}$ (modulo $\pi$), we can recover the infinite wall condition from Eq.~\eqref{initlogder}.
A number of $0 < p_{SR} < 1$ in between with $0 < \delta_{SR} < \pi$ 
describes an intermediate case where we can have 
both loss and reflection~\cite{Idziaszek_PRA_82_020703_2010}. Actually, this can be a way to fit the theoretical results with experimental data~\cite{Bishof_PRA_84_052716_2011,Ludlow_PRA_84_052724_2011,Jachymski_PRL_110_213202_2013} since the short-range potentials are not known generally. \\

\noindent The form of this initial tunable log-derivative is then  flexible and can treat  
the possibility of loss at short-range in a phenomenological way. The complex log-derivative matrix provides a complex $\mathbf{K}$ matrix and a $\mathbf{S}$ matrix which is not a unitary matrix anymore. The difference of the sum of the $|S|^2$ matrix element for one channel with unity provides the overall loss probability of this channel which translates into a loss cross section and a loss rate coefficient. This is an overall loss as we cannot determine each final state-to-state loss probabilities. \\

\noindent When describing ultracold collisions of reactive molecules, 
the universal regime condition at short-range $p_{SR}=1$ is often chosen as we know nothing about the full potential energy surface. It is convenient since this condition is
independent of the short-range interaction of the systems as mentioned above.
It means that when the two molecules meet at short-range, the probability of reaction is one. 
Comparison with experimental data will eventually tell if one deviates from this regime or not.
For the case of non-reactive molecules with a high density of Fano-Feshbach resonances 
around the collisional threshold \cite{Mayle_PRA_85_062712_2012,Mayle_PRA_87_012709_2013}, this condition is also often chosen. 
In this case, it has been supposed that the molecules might form a molecule-molecule complex
for a certain time. 
The higher the density of Fano-Feshbach resonances, the longer the lifetime of this forming complex. As a consequence, in this high density regime, it has been shown that 
the rate of two molecules being formed in the tetra-atomic complex is exactly the same as the rate of two molecules being destroyed at short-range with a full loss probability $p_{SR}=1$~\cite{Mayle_PRA_87_012709_2013}. 
Subsequently, the complex can be destroyed by a collision with a third molecule,
resulting in losses of the molecules.
Recent experiments observed losses of non-reactive molecules in their absolute ground 
state for RbCs \cite{Takekoshi_PRL_113_205301_2014},
NaK \cite{Park_PRL_114_205302_2015},
and NaRb \cite{Guo_PRL_116_205303_2016} molecules.
Even though a direct observation of the forming complexes was not obtained,
the hypothesis formulated in \cite{Mayle_PRA_85_062712_2012,Mayle_PRA_87_012709_2013}
could be a possible explanation of the experimental molecular losses. \\

\noindent We end up with: (i) a long-range interaction from $r=r_{min}$ to $r=r_{max}$ and (ii) a short-range tunable boundary condition at $r=r_{min}$ which describes phenomenologically
scattering phase-shifts and additional losses from short-range. To study the collision KRb + KRb $\to$ K$_2$ + Rb$_2$, we will use the full loss (universal) condition at short-range $p_{SR}=1$ so that $Z_{i,i}(r_{min}) = - i \ k_{min} $ for each diagonal elements.

\subsection{Molecules in an electric field}

\noindent We consider KRb molecules in their ground electronic state $^1\Sigma^+$ and their ground vibrational state $v=0$. We do not take into account any spin structure as mentioned earlier. Then, only their rotational structure can change in a collision. The bare internal rotational states of a molecule are usual described by spherical harmonics $Y_{n_i}^{m_{n_i}}$ noted by the ket $|n_i \, m_{n_i} \rangle$ for molecule $i=1,2$.
In this basis set, the rotational Hamiltonian is given by $ \langle n_i \, m_{n_i} | h_{{rot}} | n_i' \, m_{n_i}' \rangle$ = $B_{{rot}} \, n_i(n_i+1) \, \delta_{n_i,n_i'} \, \delta_{m_{n_i},m_{n_i}'}$ where $B_{{rot}}$ is the rotational constant of the molecule.
We take $B_{{rot}} = 1.113950$~GHz~\cite{Ospelkaus_PRL_104_030402_2010} for the fermionic $^{40}$K$^{87}$Rb molecule and $B_{{rot}} = 1.095362$~GHz~\cite{Aikawa_PRL_105_203001_2010} for the bosonic $^{41}$K$^{87}$Rb molecule.
In an electric field, we add the Stark term given by the interaction $h_S = - \vec{d} . \vec{E} $ between the permanent electric dipole moment $\vec{d}$ of the molecule and an electric field $\vec{E} = E \, \vec{e}_Z$ taken along the $OZ$ direction.
In the basis set $|n_i \, m_{n_i} \rangle $, the Stark term is written  
\cite{Bohn_BookChapter_2009}:
\begin{eqnarray} \label{VS}
 \langle n_i \, m_{n_i} | h_{S} | n_i' \, m_{n_i}' \rangle   & = &
 - d \, E  \ \delta_{m_{n_i},m_{n_i}'} \, (-1)^{m_{n_i}} \, \sqrt{2n_i+1} \, \sqrt{2n_i'+1} \nonumber \\
& \times &  \left( \begin{array}{ccc} n_i & 1 & n_i' \\ 0 & 0 & 0 \end{array} \right)
\, \left( \begin{array}{ccc} n_i & 1 & n_i' \\ -m_{n_i} & 0 & m_{n_i}' \end{array}  \right) .
\end{eqnarray}
A permanent electric dipole moment $\vec{d}$ is defined  
in the frame of the individual molecule, 
where the inter-atomic axis is chosen as quantization axis.
We choose the convention that the orientation of the permanent dipole moment 
$\vec{d}$ points from the negative to the positive distribution of charge
\cite{Stone_Book_1996}.
The sign of the vector $\vec{d}$ depends on the 
inter-atomic axis orientation in the frame of the individual molecule. 
This is an arbitrary choice but needs to be specified to avoid confusion.
Here we assume that the inter-atomic axis is oriented
from the lightest atom to the heaviest one \cite{Aymar_JCP_122_204302_2005} 
(for identical atoms of same mass, there is no electric dipole moment),
as shown in Fig.~\ref{FIG1} where 
the unit vector $\vec{\rho}_i/|\vec{\rho}_i|$ 
for molecule $i=1,2$ points from the lightest atom (represented by a small blue circle) 
to the heaviest one (represented by a big red circle).
Using the above convention and the orientation of the inter-atomic axis, 
a positive vector $\vec{d}$ would then mean that the negative distribution of 
charge is on the lightest atom while the positive distribution is on the heaviest one.
A negative vector would mean the opposite.
As an example, the permanent dipole moment $\vec{d}$ for KRb is a positive vector with 
a magnitude of $d=0.57$~D. It means that the negative distribution of charge 
is on the K atom while the positive one is on Rb, in the individual molecular frame.
If we diagonalize the internal Hamiltonian matrix $h_i = h_{{rot}} + h_{S}$ for molecule $i=1,2$ in the basis set $|n_i \, m_{n_i} \rangle$, we get the corresponding eigenvectors
(often called dressed states) $|\tilde{n}_i \, m_{n_i} \rangle $ for a given electric field, which are a linear combination of the bare state $|n_i \, m_{n_i} \rangle$. The quantum number $m_{n_i}$ is conserved. The tilde corresponds to a certain admixture of different rotational quantum numbers due to the electric field but when $E \to 0$, the dressed states $|\tilde{n}_i \, m_{n_i} \rangle$ tend to the bare states $|n_i \, m_{n_i} \rangle $. The number of significantly admixed bare states increases with the magnitude of the electric field.
The eigenenergies $\varepsilon_{\alpha_i}$ for molecule $i=1,2$ are shown in Fig.~\ref{FIG5}-a for the fermionic $^{40}$K$^{87}$Rb molecule, for $E=[0-50]$~kV/cm where we used $n=[0-5]$ to insure convergence of the results. 

\begin{figure}[h]
\begin{center}
\includegraphics*[width=5cm,keepaspectratio=true,angle=-90,trim=0 0 0 0]{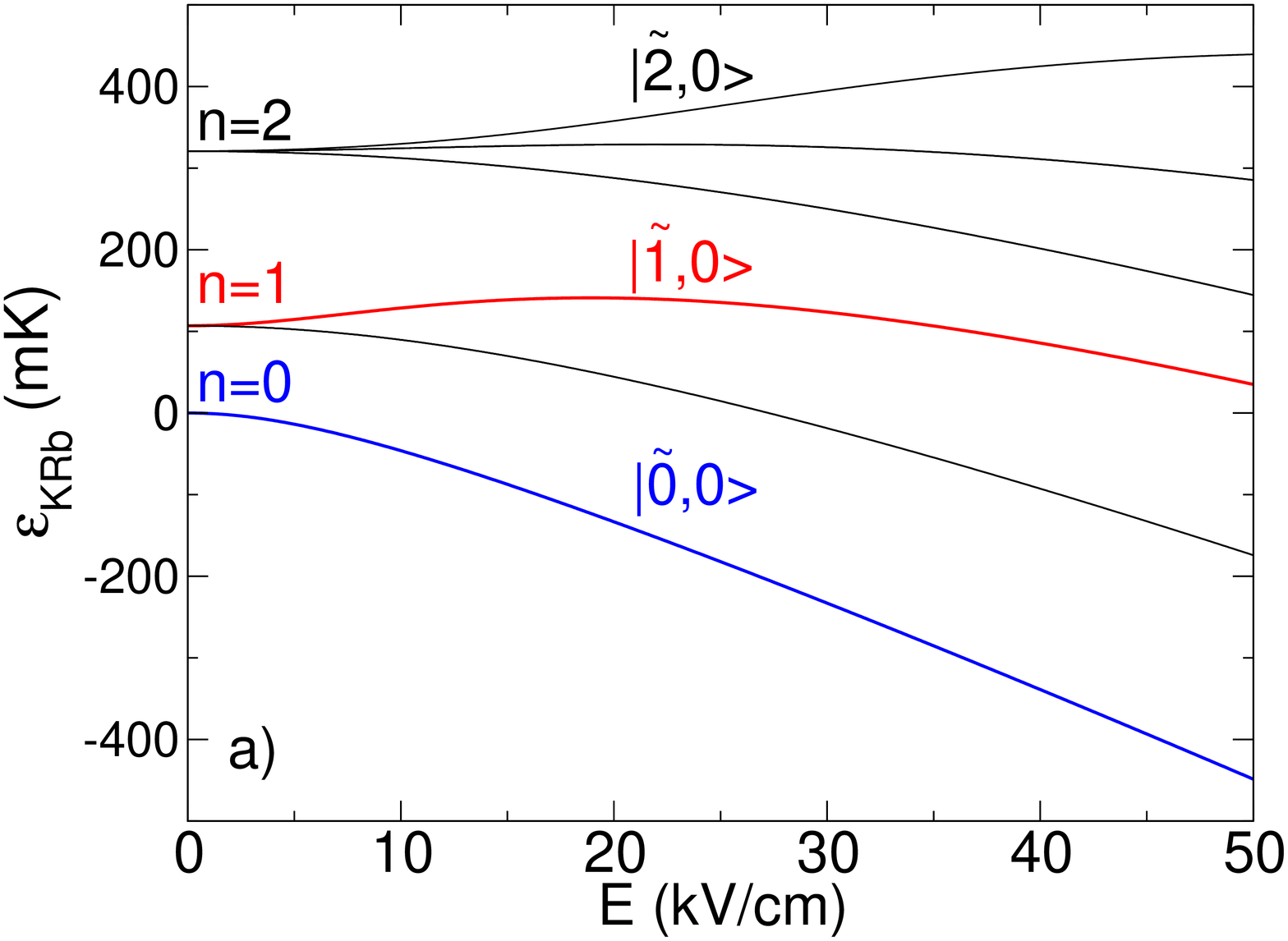} 
\includegraphics*[width=5cm,keepaspectratio=true,angle=-90,trim=0 0 0 0]{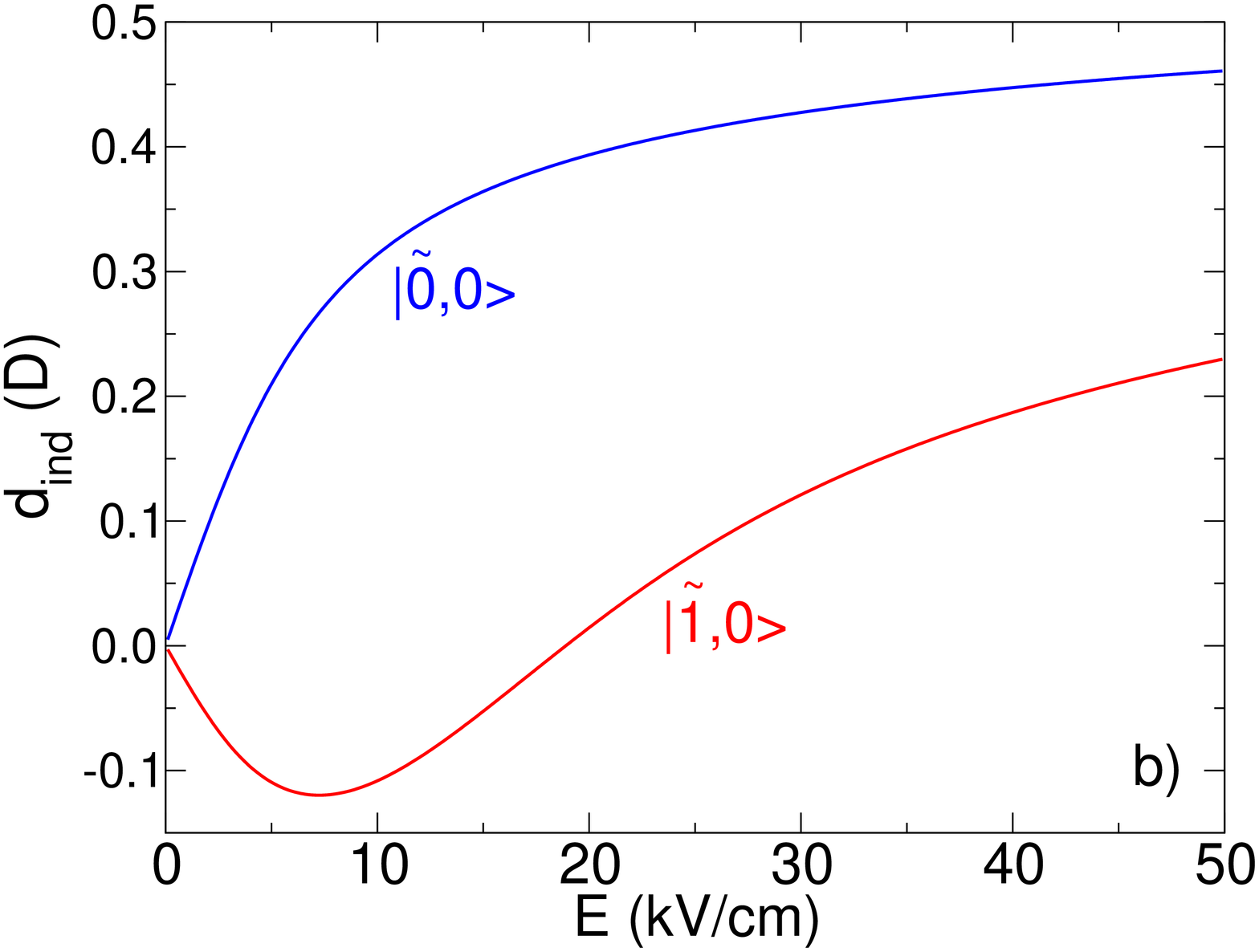}
\caption{a) Energy of a fermionic $^{40}$K$^{87}$Rb molecule as a function of an electric field for different internal states. b) Corresponding induced dipole moment $d_{{ind}}$ 
in the direction of the electric field 
in the space-fixed frame for the $|\tilde{0}, 0 \rangle$ and $|\tilde{1}, 0 \rangle$ state.}
\label{FIG5}
\end{center}
\end{figure}

\noindent It is also useful to plot the induced dipole moment 
in the electric field direction in the space-fixed frame. 
The induced dipole moment is the mean value 
of the permanent dipole moment over the dressed state $|\tilde{n}_i \, m_{n_i} \rangle$
at a given electric field $E=E_0$:
\begin{eqnarray}
d_{{ind}}(E_0) = \langle \tilde{n}_i  \, m_{n_i} \, | \, \vec{d} \cdot \vec{e}_Z \, | \, \tilde{n}_i \, m_{n_i}\rangle \, \bigg|_{E_0} = - \frac{d \varepsilon_{\alpha_i}}{dE} \, \bigg|_{E_0} .
\end{eqnarray}
The sign of the induced dipole moment represents now 
the sign of the mean value of the permanent dipole moment for a given state 
in the direction of the electric field in the space-fixed frame. 
A positive sign represents a mean value
pointing along the field while a negative sign 
represents a mean value pointing against the field. 
The induced dipole moments for different rotational states are shown in Fig.~\ref{FIG5}-b for the fermionic $^{40}$K$^{87}$Rb molecule. The ground rotational state $|\tilde{0}, 0 \rangle $ has a positive induced dipole moment growing in a monotonic way from 0 to $d=0.57$~D. 
For the first excited state $|\tilde{1}, 0 \rangle $ this is different.
The induced dipole moment of $|\tilde{1}, 0 \rangle $ is first negative from $E=0$ to $E = 19$~kV/cm, with an increase in magnitude up to $E = 7.25$~kV/cm and a decrease after. Then it becomes positive at $E \ge 19$~kV/cm. \\

\noindent Finally,  
the energy of the combined initial dressed 
states $\varepsilon_\alpha = \varepsilon_{\alpha_1}  + \varepsilon_{\alpha_2}$ 
for two fermionic $^{40}$K$^{87}$Rb molecules $i=1,2$ in an electric field,
is shown in Fig.~\ref{FIG6} as a function of the electric field. 
This gives an indication of the 
energy thresholds of the possible collisional states. 

\begin{figure}[h]
\begin{center}
\includegraphics*[width=5cm,keepaspectratio=true,angle=-90,trim=0 0 0 0]{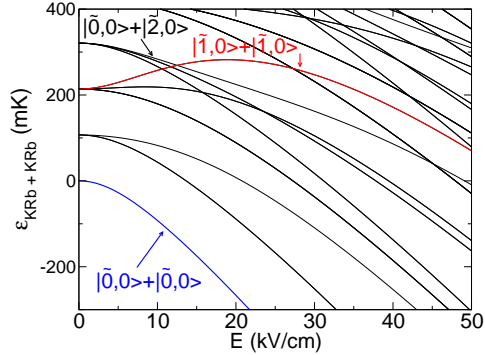}
\caption{Energy of two separated fermionic $^{40}$K$^{87}$Rb molecules as a function of an electric field. The blue (red) curve corresponds to the energy 
of two separated molecules in the ground (first excited) rotational state.}
\label{FIG6}
\end{center}
\end{figure}

\subsection{Collisions of molecules in an electric field}

\noindent We use now a fixed collision energy $E_c = 500$~nK since this is a typical 
value reached in experiments of ultracold molecules. 
From the previous section, we know as well the energy of the individual 
molecules as a function of an applied electric field.
We use the dipole-dipole interaction in Eq.~\eqref{dipoledipole}. The interaction varies as
$-C_3/r^3$ and depends on the applied electric field. 
For the van der Waals interaction we use
a value of $C_6=12636$~a.u.~\cite{Lepers_PRA_88_032709_2013} for KRb.
The molecules are identical and start in the same internal state so that they are indistinguishable. The partial waves used are $l=1,3,5$ for the fermionic molecules and $l=0,2,4$ for the bosonic ones.
The initial quantum numbers for the individual molecules $i=1,2$ are $m_{n_i}=0$. Those numbers are still good quantum numbers even in an electric field.
The total $M_J = m_{n_1} + m_{n_2} + m_l = m_{n_1}' + m_{n_2}' + m_l'$ is conserved during the collision. As we start with $m_{n_1},m_{n_2}=0$, then $M_J=m_l$. At such an ultralow energy$E_c = 500$~nK, the most important partial wave is the first and lowest one. For fermions, the lowest partial wave quantum number is $l=1$ (p-wave), so that $m_l=0,\pm1$. Then we restrict the calculation to $M_J=0,\pm1$. For bosons, the lowest partial wave quantum number is $l=0$ (s-wave), so that $m_l=0$, and then we restrict to $M_J=0$.
The corresponding diabatic and adiabatic energies for 
two fermionic $^{40}$K$^{87}$Rb molecules 
in the ground rotational state $|\tilde{0},0\rangle$ at an electric field 
of $E$ = 5 kV/cm are plotted
in Fig.~\ref{FIG7} as a function of $r$ 
(see definition in Section \ref{sec:short}).
We selected the component $M_J=0$ for this figure so that $m_l=0$. 
At large distances, 
the energies tend to the energy of two separated molecules recovering the results
in Fig.~\ref{FIG6}.
At short distances, one can see the onset of the centrifugal terms 
characterized by the partial wave numbers $l=1,3,5$
and the corresponding barriers.
The diabatic curves are shown in black while the adiabatic ones are shown in red.
The effect of the dipole-dipole coupling elements in Eq.~\eqref{dipoledipole}
can be seen in this figure where the adiabatic energies differ from the diabatic ones. 

\begin{figure}[h]
\begin{center}
\includegraphics*[width=5cm,keepaspectratio=true,angle=-90,trim=0 0 0 0]{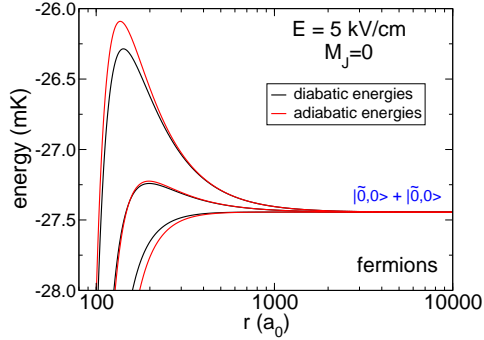}
\caption{Diabatic (black) and adiabatic (red) energies 
for two fermionic $^{40}$K$^{87}$Rb molecules in the ground rotational state
at an electric field of $E$ = 5 kV/cm.}
\label{FIG7}
\end{center}
\end{figure}

\noindent We apply the quantum formalism that we have presented in this paper.
This is what it is referred to as the close-coupling 
quantum calculation in the following.
Starting with a boundary condition 
at $r_{min} = 10 \ a_0$, where $a_0$ is the Bohr radius,
corresponding to a full loss condition at short-range,
we propagate the log-derivative matrix $\mathbf{Z}$
up to $r_{max} = 10000 \ a_0$. At this distance,
we obtain the reactance, scattering and transition matrices
$\mathbf{K},\mathbf{S},\mathbf{T}$,
and finally the cross sections and rate coefficients.
As we use a boundary condition with full loss at short-range, there are three collisional processes possible: elastic, inelastic and loss processes. 
The loss processes mimic chemical reaction processes for reactive molecules. 
For non-reactive molecules, they would mimic
the losses of two free molecules into a
molecule-molecule complex, subsequently destroyed by a collision with a third molecule.
In the following, we will call quenching processes the sum of inelastic and loss processes, that is everything that leads to molecular losses and compare with elastic processes. 
We present two cases: collisions of molecules (i) in the ground rotational state 
and (ii) in the first rotational excited state. For the former case, we also introduce
an insightful model, a quantum threshold model, that semi-quantitatively explains the collisional results.

\subsubsection*{Molecules in the ground rotational state: enhancement of the loss rates}

\begin{figure}[h]
\begin{center}
\includegraphics*[width=5cm,keepaspectratio=true,angle=-90,trim=0 0 0 0]{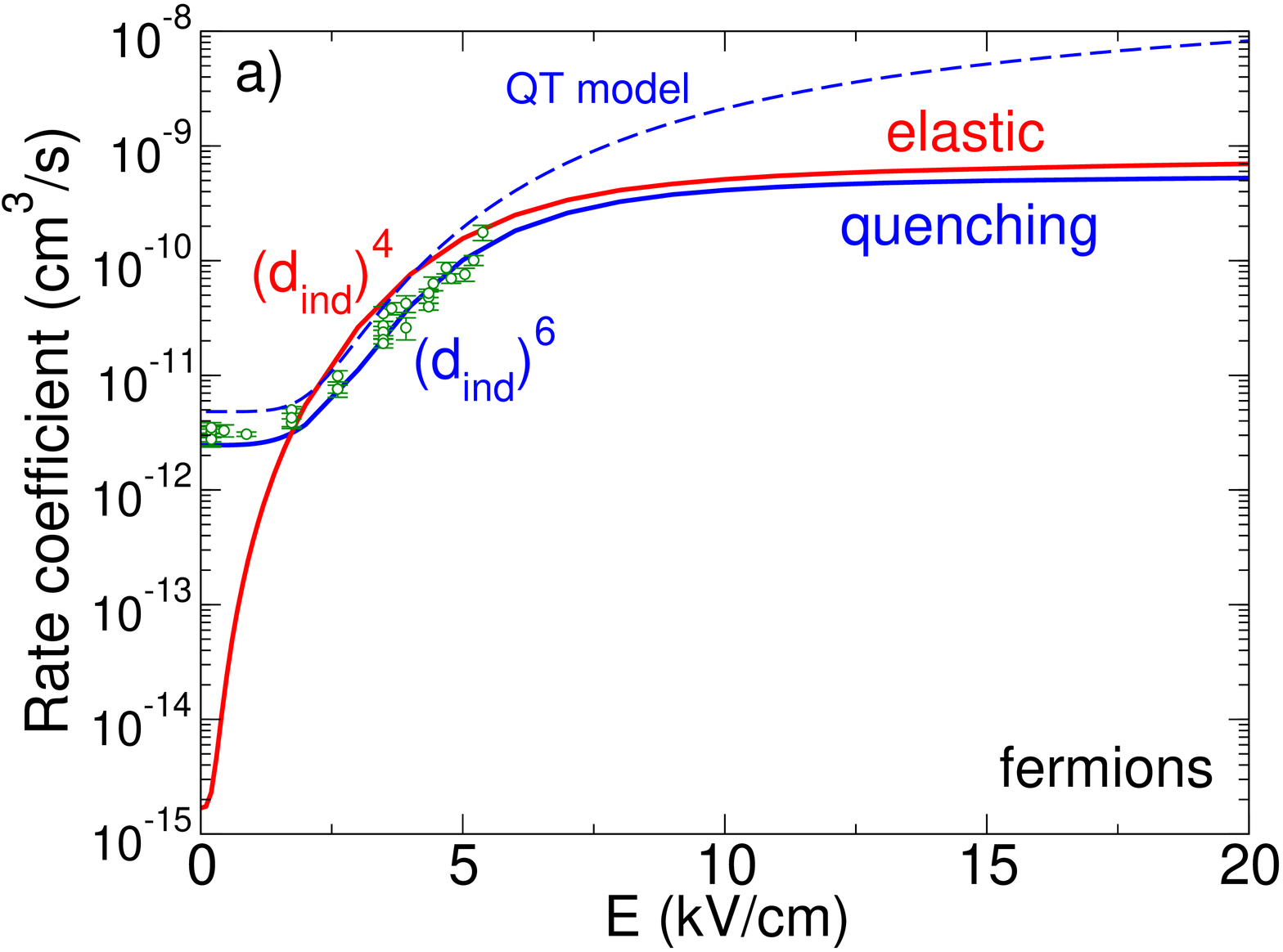}
\includegraphics*[width=5cm,keepaspectratio=true,angle=-90,trim=0 0 0 0]{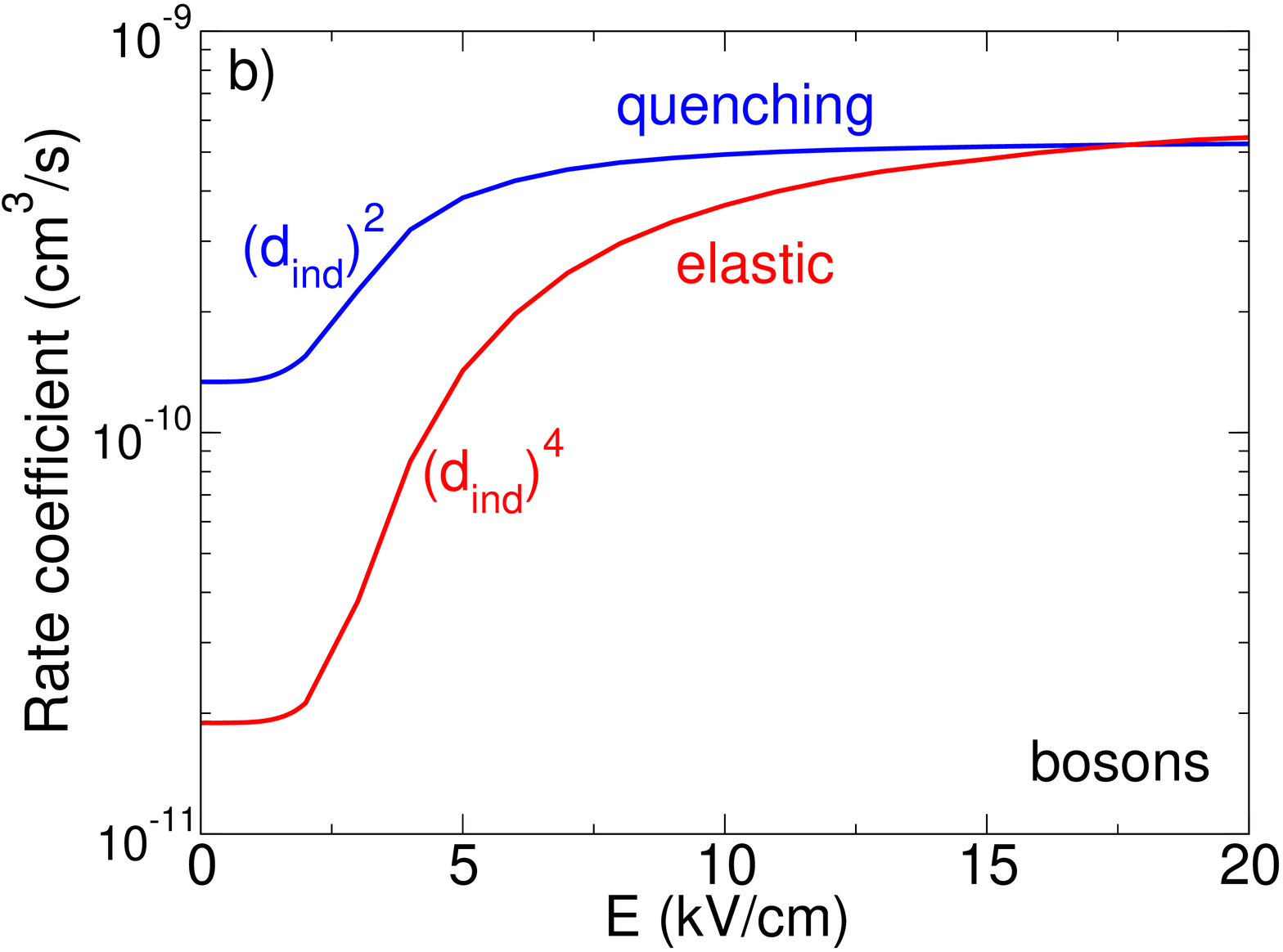}
\caption{Rate coefficient as a function of an electric field for molecules in 
the ground rotational state $|\tilde{0},0\rangle$. Elastic (quenching) processes
correspond to the red (blue) curves. 
The solid lines are the results from the close-coupling 
quantum formalism presented in this paper
(see text for details).
a): Fermionic $^{40}$K$^{87}$Rb + $^{40}$K$^{87}$Rb collisions. 
The data points (green circles) are the 
experimental results of Ref.~\cite{Ni_N_464_1324_2010}.
The dashed line comes from a Quantum Threshold model 
\cite{Quemener_PRA_81_022702_2010}, 
see Eq.~\eqref{ratelossQT}.
b): Bosonic $^{41}$K$^{87}$Rb + $^{41}$K$^{87}$Rb collisions.}
\label{FIG8}
\end{center}
\end{figure}

\noindent We present in Fig.~\ref{FIG8} the elastic (red) and quenching (blue) 
rate coefficients for fermions (Fig.~\ref{FIG8}-a) and bosons (Fig.~\ref{FIG8}-b) for two molecules in the ground rotational state $|\tilde{0},0\rangle$. 
The results were obtained using the close-coupling quantum calculation aforementioned.
The energy threshold for the two molecules  $|\tilde{0},0\rangle+|\tilde{0},0\rangle$ is shown in blue in Fig.~\ref{FIG6}.
For fermions, experimental data of Ref.~\cite{Ni_N_464_1324_2010} are also included.
Globally for both cases, the quenching rate dominates over the elastic rate or they have the same order of magnitude. This is a bad outcome
for example for evaporative cooling purpose where elastic collisions have to be important while quenching collisions have to be negligible.
Comparing fermions to bosons, similar behaviour is seen except that the bosonic rates are globally higher than the fermionic ones. This is expected from the parity of 
the $l$ quantum numbers. For bosons, the $l$ numbers are even and include the s-wave $l=0$
curve, for which there is no centrifugal barrier (barrierless case). For fermions, the $l$
numbers are odd and include the p-wave $l=1$ curve, for which there is a centrifugal barrier. 
In the former case, the particles approach each other easily without any barier
so that the rate is high while in the latter case, 
the particles approach less easily due to the presence of the p-wave centrifugal barrier.  \\

\noindent Both rates increase with increasing electric field. They display the same behaviour as their induced dipole moment. When the electric field increases, the induced dipole moment
increases monotonically (see Fig.\ref{FIG5}-b), so does the magnitude of the dipole-dipole interaction and then the rate coefficient.
This can be explained by the fact that for fermions or bosons at ultralow energies, the main contribution to the rates comes from an attractive dipole-dipole interaction from the $m_l=0$ component~\cite{Quemener_PRA_81_022702_2010,Quemener_PRA_84_062703_2011}. When the electric field increases the dipole-dipole interaction becomes more and more attractive, favouring the meeting of molecules at short-range and then molecular losses. \\

\noindent Fermionic and bosonic elastic rates behave as 
$d_{{ind}}^4$ as predicted in Ref.~\cite{Bohn_NJP_11_055039_2009}. 
Quenching rate coefficients have a strong dependence with increasing electric field and induced dipole moment. The fermionic quenching rates display a $d_{{ind}}^6$ behaviour as found in Ref.~\cite{Quemener_PRA_81_022702_2010} while the bosonic quenching rates display a $d_{{ind}}^2$ one as found in Ref.~\cite{Quemener_PRA_84_062703_2011}. 

\subsubsection*{A Quantum Threshold model}

\begin{figure}[h]
\begin{center}
\includegraphics*[width=6cm,keepaspectratio=true,angle=-90,trim=150 100 150 150]{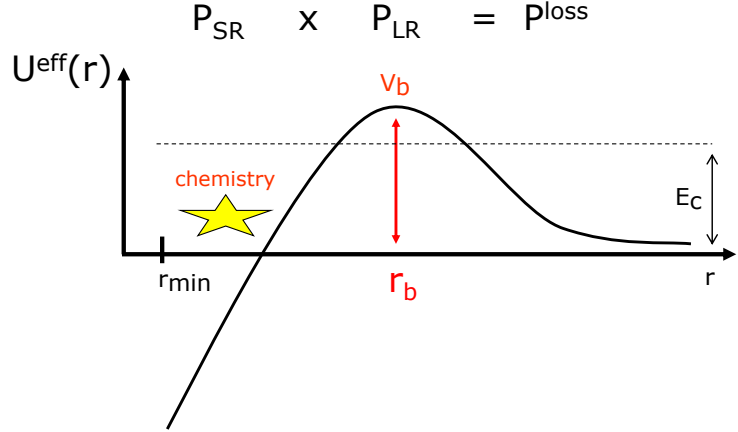}
\caption{Basis of a simple Quantum Threshold model \cite{Quemener_PRA_81_022702_2010}.
The incident particles have to overcome a barrier of height $V_b$ at a position $r_b$
at long-range. The probability to reach the short-range region 
is governed by the tunneling probability through the barrier and depends 
on the collision energy $E_c$ and the height of the barrier, see Eq.~\eqref{QT}.}
\label{FIG9}
\end{center}
\end{figure}

\noindent The behaviour of the quenching rate coefficients 
can be found semi-quantitatively using a Quantum Threshold (QT) model~\cite{Quemener_PRA_81_022702_2010}. 
The method consists in taking into account only the lowest channel curve
of the initial colliding state corresponding to the lowest partial wave quantum number.
The energy of the two initial free particles is taken as reference.
This curve is described by the interaction potential ${\cal U}^{eff}$:
\begin{eqnarray}
{\cal U}^{eff}(r)= {\cal U}^{cent}(r) + {\cal U}^{int}(r) 
= \frac{\hbar^2 \, l (l+1)}{2 \, m_{red} r^2} - \frac{C_s}{r^s},
\end{eqnarray}
for an attractive interaction $- C_s / r^s$ with $s>2$ and $C_s > 0$.
The competition between the repulsive centrifugal potential and the attractive interaction 
creates a potential energy barrier for the incident colliding motion (or incident barrier), 
of height $V_b$ at position $r_b$ (see Fig.~\ref{FIG9}).
The position and the height of the barrier are given by:
\begin{align}
r_b &= \bigg[\frac{m_{red} \, s \, C_s}{\hbar^2 \, l(l+1)} \bigg]^{1/(s-2)} &
V_b &= \frac{\hbar^2 \, l(l+1)}{2 m_{red} r_b^2} - \frac{C_s}{r_b^s}.
\end{align}
In the case of a barrierless collisions ($l=0$), one cannot define a position $r_b$ and 
height $V_b$ of a barrier. Instead the characteristic length and energy of the 
$- C_s / r^s$ interaction are taken in the model\cite{Quemener_PRA_84_062703_2011,Gao_PRA_78_012702_2008}: 
\begin{align}
a_s &= \bigg[ \frac{2 m_{red} C_s}{\hbar^2} \bigg]^{1/(s-2)} &
E_s &= \frac{\hbar^2}{2 m_{red} a_s^2}.
\end{align}
The model simply uses two probabilities of collision: 
one at long-range and one at short-range.
At long range, the two molecules see the incident barrier and tunnel through it.
This is described by a long-range (tunneling) probability $P_{LR}$.
The molecules enter then the short-range region,
where they can chemically react or form a complex and be lost from the trap, with a probability $P_{SR}$. We will assume full probability of loss at short range so that $P_{SR}=1$. 
Then the probability of loss is $P^{loss} = P_{SR} \times P_{LR} = P_{LR}$. To estimate the tunneling probability $P_{LR}$, we use:

(i) a classical Langevin model~\cite{Langevin_ACP_5_245_1905}: 
when $E_c \ge V_b$:
\begin{eqnarray} \label{pcond1}
P_{LR}(E_c=V_b)=1
\end{eqnarray}
that is if the molecules have enough energy to overcome the barrier, the probability of passing above is one, 

(ii) the form of the threshold laws for the loss probability (Eq.~\eqref{Wignerlaws3}): when $E_c \to 0$, the probability which is proportional to the cross section multiplied by $k^2 \sim E_c$ (see Eq.~\eqref{avintXS}) should obey: 
\begin{eqnarray} \label{pcond2}
P_{LR}(E_c)=\gamma E_c^{l+1/2} ,
\end{eqnarray}

(iii) Eq.~\eqref{pcond1} to determine the constant $\gamma$ in Eq.~\eqref{pcond2}, so that $P_{LR}(E_c=V_b)=1=\gamma V_b^{l+1/2}$. Then we get $\gamma = 1 / V_b^{l+1/2}$. Within the QT model, the total loss probability is: 
\begin{eqnarray} \label{QT}
P^{loss}(E_c)=  \bigg( \frac{E_c}{V_b} \bigg)^{l+1/2}.
\end{eqnarray}
Replacing Eq.~\eqref{QT} into Eq.~\eqref{avintXS} leads to the quenching rate coefficient within the QT model for a given $l,m_l$:
\begin{eqnarray} \label{ratelossQT}
\beta^{{qu}}_{l,m_l}(E_c) = \frac{\hbar^2 \pi}{\sqrt{2 m_{red}^3}} \frac{{E_c}^{l}}{{V_b}^{l+1/2}} \, \Delta ,
\end{eqnarray}
with $\Delta = 2$ if the particles are identical and indistinguishable and $\Delta = 1$ otherwise. This is a simple way to estimate the characteristics of 
loss collisions. Once we know the height of the barrier $V_b$ we know how the rate coefficient scales. 
At zero electric field, the dominant interaction is the attractive van der Waals interaction with $s=6$. For $l=1$, $V_b = [ 8 \hbar^2 / 54 m_{red}^3 C_6 ]^{1/2}$. For $l=0$, $E_6 = \hbar^3 / [8 m_{red}^3 C_6 ]^{1/2}$.
In the electric field regime, $V_b = (25 \hbar^6 / 108 m_{red}^3) \times (d_{{ind}}^2/ 4 \pi \varepsilon_0)^{-2}$ for $l=1$ and $E_4 = (15 \hbar^6 / 16 m_{red}^3) \times (d_{{ind}}^2/ 4 \pi \varepsilon_0)^{-2}$ for $l=0$, where the characteristic interaction is $s=4$, see Ref.~\cite{Quemener_PRA_84_062703_2011} for more details.
Inserting these expressions into Eq.~\eqref{ratelossQT}, we see that the quenching rate behaves then as $d_{{ind}}^6$ and $d_{{ind}}^2$ for indistinguishable fermions ($l=1$) and bosons ($l=0$) respectively, and in general as $d_{{ind}}^{4(l+1/2)}$.
For $l=0$, the quenching rate coefficients are independent of the collision energy and hence of the temperature. 
For $l=1$ and to get the rate coefficients as a function of the temperature $T$, 
one can replace $E_c$ by $\langle E_c \rangle = 3 k_B T/2$, the mean collision energy of a Maxwell-Boltzmann distribution. \\

\noindent The QT rate coefficient is shown in Fig.~\ref{FIG9} as a dashed line for the fermionic case. It gives the proper scaling law and transition zone 
between the Van der Waals regime (where we took $l=1,m_l=0,\pm1$) and the electric field regime (where we only took $l=1,m_l=0$). 
However it overestimates both quantum results and experimental data 
by about a factor of 2. This can be traced back in the classical Langevin criterion where we chose a unit probability when $E_c = V_b$. This is true in classical mechanics but in quantum mechanics, the colliding particles are described by a wave. Close to and at the top 
a barrier, a wave has a transmission probability but also a reflection probability, the sum of both being one. It implies that the transmission probability of the wave function is not equal to unity, in contrast with what is assumed by the classical Langevin model. This explains why the QT model gives an upper value of the quenching rates for fermions. \\

\noindent Comparing the QT quenching rates with the ones using the quantum formalism for different molecular systems of dipolar alkali molecules 
\cite{Quemener_PRA_84_062703_2011}
provides the corrections to make for the model. The correction is a factor $p$ of order of unity in front of Eq.~\eqref{ratelossQT}. 
For $l=1$, the corrections are $p=0.53$ for the van der Waals regime and $p=0.54$ for the electric field regime, while for $l=0$ they are $p=1.92$ and $p=3.74$ respectively.
Some of those values can also be found using a Quantum-Defect Theory (QDT) formalism~\cite{Idziaszek_PRL_104_113202_2010,Gao_PRL_105_263203_2010}.
The QT model in Eq.~\eqref{ratelossQT} provides then an underestimating rate for the barrierless case $l=0$ since the correction factor $p > 1$, 
while it gives an overestimating rate for the barrier case $l=1$
since the correction factor $p < 1$.
With the corrections of about 0.5 for $l=1$ on this figure, one can see that 
the QT model will then agree with the numerical 
close-coupling quantum calculation and the experimental data. 

\subsubsection*{Molecules in the first rotational excited state: suppression of the loss rates}

\begin{figure}[h]
\begin{center}
\includegraphics*[width=5cm,keepaspectratio=true,angle=-90,trim=0 0 0 0]{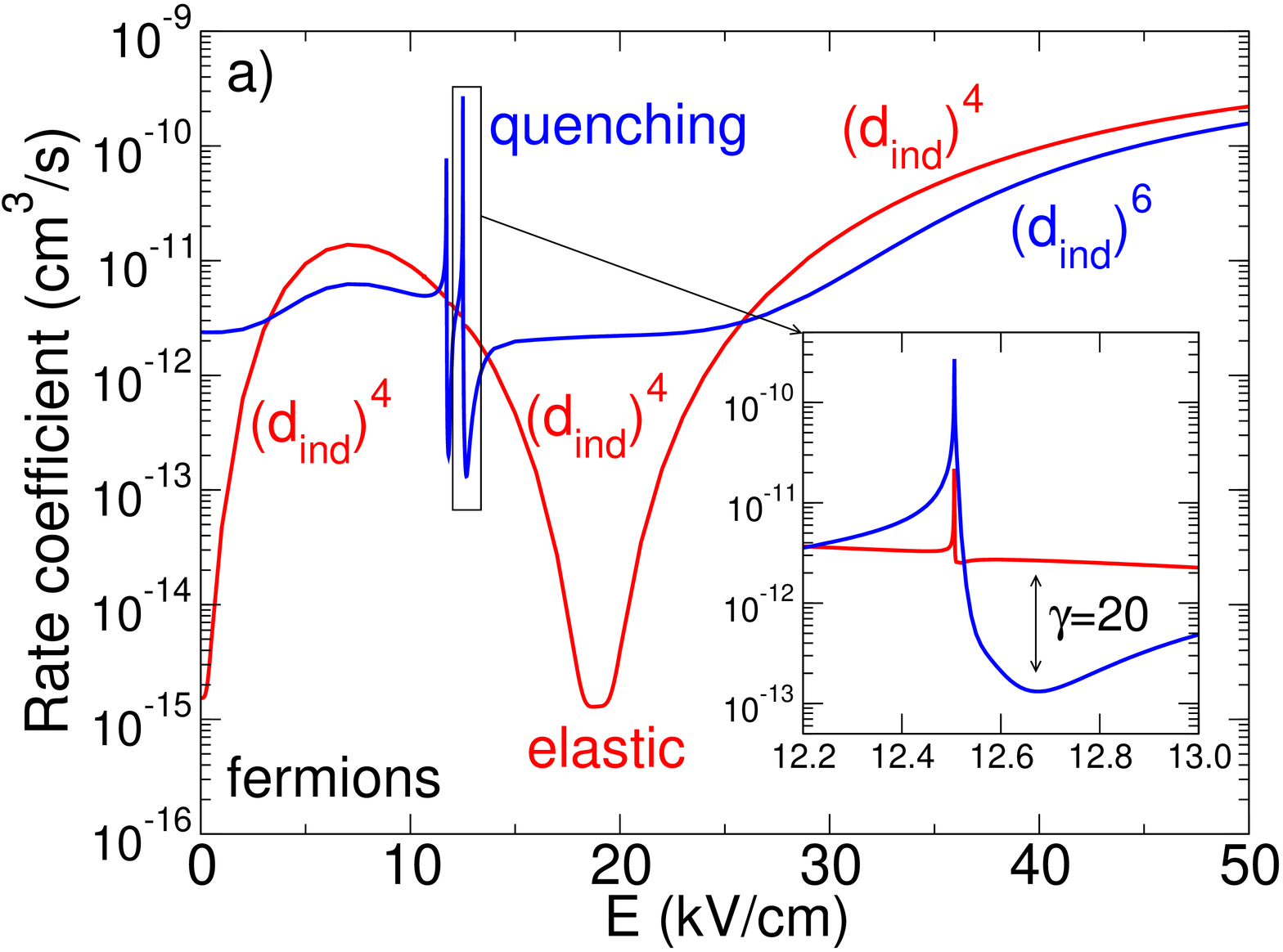}
\includegraphics*[width=5cm,keepaspectratio=true,angle=-90,trim=0 0 0 0]{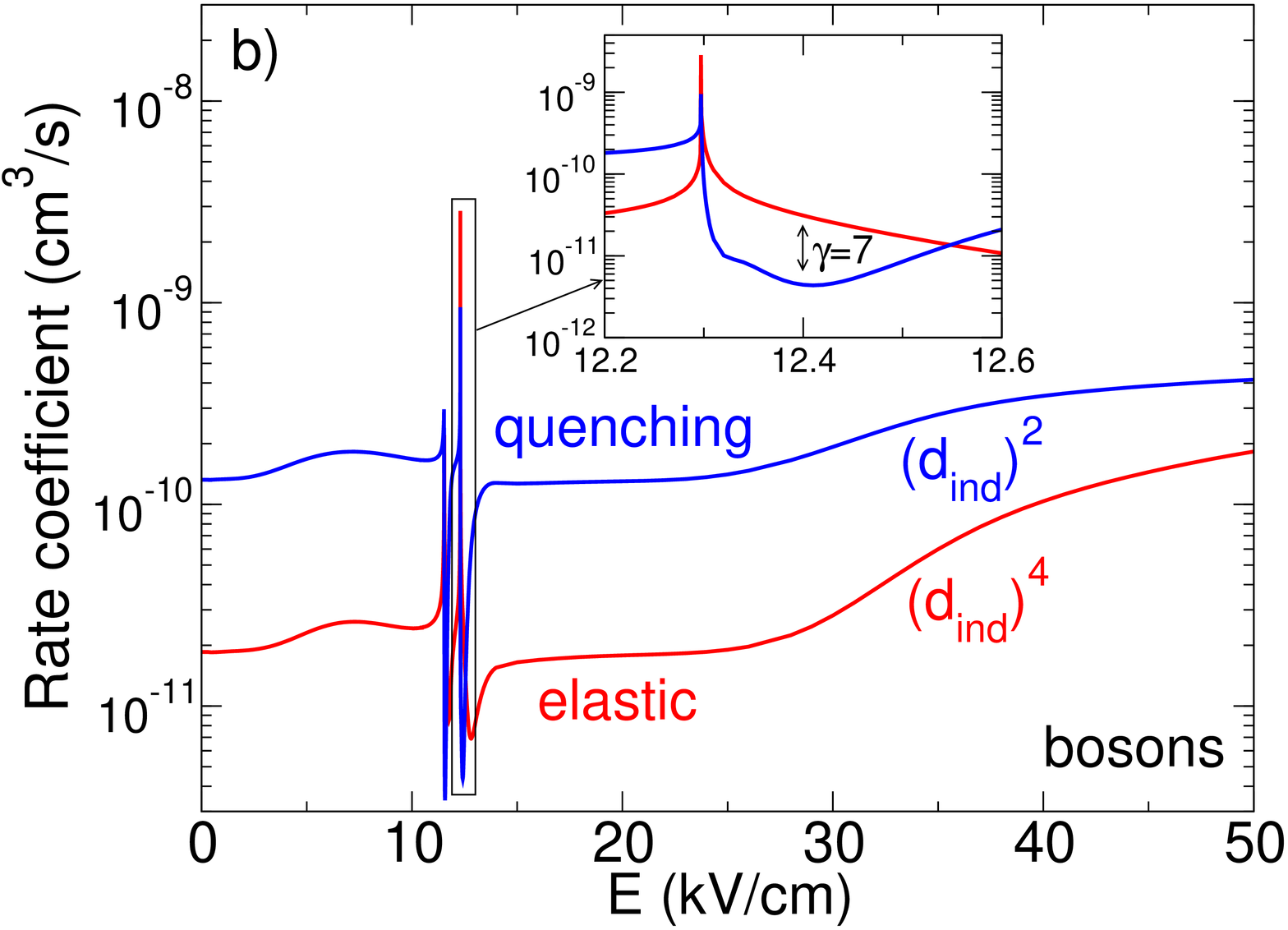}
\caption{Rate coefficient as a function of an electric field for molecules in 
the first excited rotational state $|\tilde{1},0\rangle$. 
The solid lines are the results from the close-coupling 
quantum formalism presented in this paper
(see text for details).
a): Fermionic $^{40}$K$^{87}$Rb + $^{40}$K$^{87}$Rb collisions. 
b): Bosonic $^{41}$K$^{87}$Rb + $^{41}$K$^{87}$Rb collisions.}
\label{FIG10}
\end{center}
\end{figure}

\noindent What happens now if the molecules are prepared in the first excited rotational state $|\tilde{1},0\rangle$? 
The corresponding rate coefficients 
\cite{Wang_NJP_17_035015_2015} are presented in Fig.~\ref{FIG10}
using the close-coupling quantum calculation.
Globally, we found the same overall trend as for two molecules in the ground rotational state.
The rates follow again the behaviour of the induced dipole moment as a function of the electric field: when $|d_{{ind}}|$ increases from $E=0$ to $E = 7.25$ kV/cm and from $E=19$ kV/cm, the rate increases, and inversely when it decreases from $E = 7.25$ kV/cm to $E=19$ kV/cm, the rate decreases. 
We found again that the quenching rate behaves as $d_{{ind}}^6$ and $d_{{ind}}^2$ for fermions and bosons and that the elastic rate behaves as $d_{{ind}}^4$. \\

\noindent The main interesting feature of Fig.~\ref{FIG10} comes from the presence of sharply varying structures for the rates near $E^* \sim 12.5$ kV/cm and $E^* \sim 11.5$ kV/cm 
(two smoother ones appear near $E^* \sim 10.5$ kV/cm and $E^* \sim 27$ kV/cm but cannot be 
seen in the figure).
This is in strike contrast with collisions of ground rotational states molecules.
These features appear at the specific electric fields $E^*$ where the energy threshold of other combined molecular states crosses the initial one $|\tilde{1},0\rangle+|\tilde{1},0\rangle$, shown in red in Fig.~\ref{FIG6}. Those states are for example the $|\tilde{0},0\rangle+|\tilde{2},0\rangle$ and the $|\tilde{0},0\rangle+|\tilde{2},\pm1\rangle$ for the two most prominent features respectively.
Slightly below $E^*$, the quenching rate first increases when the electric field is increased, then above $E^*$, it suddenly drops. Eventually it gets back to a steady value far from $E^*$. In the region where the quenching rate is suppressed, the elastic rate remains quite high so that elastic processes are bigger than the loss processes, by a factor of $\gamma=20$ for fermions and $\gamma=7$ for bosons. 
The principle of this mechanism was originally explored in 
Ref.~\cite{Avdeenkov_PRA_73_022707_2006} for molecules without 
losses at short-range ($P_{SR}=0$). 
We consider the initial colliding state of interest, here $|\tilde{1},0\rangle+|\tilde{1},0\rangle$ and we take the second prominent structure in the rates (insets of Fig.~\ref{FIG10}) as an example. 
When the electric field is increased starting from below $E^*$, 
the energy of the coupling state, $|\tilde{0},0\rangle+|\tilde{2},0\rangle$ in this example, approaches the one of the initial state $|\tilde{1},0\rangle+|\tilde{1},0\rangle$ from above. 
The effective potential curve (${\cal U}^{{eff}}(r)$ in Eq.~\eqref{poteff}) 
of the coupling state pushes the one of the initial state downward, due to the dipole-dipole coupling between the two channels. This results in lowering the curve of the initial state,
making it more attractive, hence favouring the molecules to come close to each other and react/be lost at short-range. The quenching rate is thus enhanced.
When the electric field is further increased but from above $E^*$ now,
the energy of the coupling state lies below the one of the initial state. Its effective potential curve pushes upwards the one of the incident state. This results now 
in increasing the curve of the initial state, making it more repulsive, 
hence preventing the molecules to come close to each other. The quenching rate is suppressed. \\

\noindent Even though this mechanism has to be confirmed by experimental results, this is a promising way of suppressing molecular losses due to any reasons (inelastic collisions, chemical reactions, complex-forming losses). This is also promising to perform evaporative 
cooling of a dipolar gas since elastic processes are more efficient 
than quenching ones. In order to perform efficient evaporative cooling, 
a ratio of $\gamma \simeq 100$ has to be reached 
\cite{Cornell_RMP_74_875_2002,Ketterle_RMP_74_1131_2002},
with perhaps a safer estimation using $\gamma \simeq 1000$. 
As described above, this is not the case for the KRb system where $\gamma \simeq 10$
so that evaporative cooling might not be an efficient method to further cool down the gas. 
However, the suppression of the quenching processes becomes more effective as the permanent electric dipole moment of the molecules increases~\cite{Quemener_PRA_93_012704_2016}.
For those molecules, the ratio $\gamma$ can reach values of 1000 or more, so that the conditions for efficient evaporative cooling are fullfilled 
to further cool down dipolar gases and hopefully reach quantum degeneracy.

\section{Conclusion and perspectives}
\label{sec:conclusion}

In this paper we presented a time-independent quantum formalism to describe ultracold collisions of particles with internal structure, also accounting for the presence of an external field. It was shown, taking the dipolar KRb molecule as an example, how collisional properties can be tuned with an electric field, from enhancing the quenching 
rates to suppressing them. 

Of course many other configurations could be engineered to control the molecules dynamics and could be implemented within the present formalism.
For example, collisions of ultracold molecules in a confined geometry is possible by adding 
in the quantum formalism an external harmonic oscillator trap that can mimic the presence 
in an experiment of a one-dimensional optical lattice
\cite{Quemener_PRA_83_012705_2011,DeMiranda_NP_7_502_2011}. 
For sufficiently high induced dipole moments and strong confinements, 
fermionic and bosonic collisional losses in two dimensions can be suppressed 
due to the side-by-side repulsive dipole-dipole interaction. 
In the particular case where the confinement of the lattice is not strong enough, 
only fermionic collisional losses can be suppressed due to appropriate selection rules
related to the fermionic character of the system
\cite{Quemener_PRA_83_012705_2011,DeMiranda_NP_7_502_2011}.
The long-range ultracold dipolar physics is also
quite general since experiments with ultracold
magnetic dipolar molecules \cite{Frisch_PRL_115_203201_2015} 
lead to the same conclusions than for electric dipolar ones.
In addition, any arbitrary electric or magnetic field 
with an arbitrary direction could also be added into the quantum formalism
\cite{Quemener_PRA_88_012706_2013,Quemener_PRA_92_042706_2015},
which can be interesting 
to control ultracold molecules that both possess 
electric and magnetic dipole moments.
Another interesting tool of control is to employ electromagnetic waves and especially microwaves to control the rotational degree of freedom of the molecules \cite{Gorshkov_PRL_101_073201_2008,Alyabyshev_PRA_80_033419_2009, Avdeenkov_PRA_86_022707_2012}.
Finally, in addition to two-body collisions, three-body collisions
\cite{Ticknor_PRL_105_013201_2010,Wang_PRL_106_233201_2011,Wang_PRL_107_233201_2011} 
and more \cite{Rittenhouse_JPBAMOP_44_172001_2011} can start to play a role for high density 
of the ultracold molecular cloud. The few- and many-body characters of the dipolar 
interactions can also start to reveal the increasing anisotropic complexity of the systems 
\cite{Lepers_JPBAMOP_49_014004_2016}. 

Treating all those additional possibilities goes beyond the scope of this paper. We introduced here only a small and simple part of the ultracold collision formalism. In the future, one could increase at will the versatility and the flexibility of the formalism to cover all possible configurations accessible in an experiment, certainly enabling the exploration of all new kinds of ultracold, ultra-controlled dynamics of molecules! 

\section*{Appendix}
\label{sec:appendix}

\subsection*{Proof 1}

Let's start with Eq.~\eqref{matrixnotation} (first equation) and its transpose (second equation) using the fact that $\mathbf{U}$ is real and symmetric. Let's multiply by $\mathbf{F}^t$ on the left for the first equation and by $\mathbf{F}$ on the right for the second equation:
\begin{eqnarray}
\mathbf{F}^t &\times& \{ \mathbf{D}^2 \, \mathbf{F} + \mathbf{U} \, \mathbf{F} \} = 0 \nonumber \\
& & \{ \mathbf{D}^2 \, \mathbf{F}^t + \mathbf{F}^t \mathbf{U} \} = 0 \quad \times \quad \mathbf{F} \nonumber .
\end{eqnarray}
By retrieving both equations one gets:
\begin{eqnarray}
\mathbf{F}^t \, (\mathbf{D}^2 \, \mathbf{F}) - (\mathbf{D}^2 \, \mathbf{F}^t )\, \mathbf{F} = 0 \nonumber 
\end{eqnarray}
which implies:
\begin{eqnarray}
\mathbf{D} [ \mathbf{F}^t \, (\mathbf{D} \, \mathbf{F}) - (\mathbf{D} \, \mathbf{F}^t) \, \mathbf{F} ] = 0 \nonumber 
\end{eqnarray}
where $\mathbf{D} \equiv \mathbf{I} \, \frac{d}{dr}$. This means that the matrix $\mathbf{F}^t \, (\mathbf{D} \, \mathbf{F}) - (\mathbf{D} \, \mathbf{F}^t) \, \mathbf{F}$ is independent of $r$. Moreover at $r = r_{min}$, we took $\mathbf{F}=0$ as mentioned by Eq.~\eqref{initlogder}, so then:
\begin{eqnarray}
\mathbf{F}^t \, (\mathbf{D} \, \mathbf{F}) - (\mathbf{D} \, \mathbf{F}^t) \, \mathbf{F} = 0
 \qquad \forall \, r \nonumber .
\end{eqnarray}
By inserting Eq.~\eqref{Kasymform1} and its transpose into this expression, one gets:
\begin{eqnarray}
(\mathbf{F}^{(1)}  - \mathbf{K}^t \, \mathbf{F}^{(2)} ) \, (\mathbf{F}^{'(1)}  - \mathbf{F}^{'(2)} \, \mathbf{K}) - (\mathbf{F}^{'(1)} - \mathbf{K}^t \, \mathbf{F}^{'(2)}) \, (\mathbf{F}^{(1)}  - \mathbf{F}^{(2)} \, \mathbf{K}) = 0 
\nonumber  
\end{eqnarray}
by factorizing the matrices $(\mathbf{N}^K)^t$ and $\mathbf{N}^K$,
and by developing:
\begin{multline}
\mathbf{F}^{(1)} \, \mathbf{F}^{'(1)} 
- \mathbf{F}^{(1)} \, \mathbf{F}^{'(2)} \, \mathbf{K} 
- \mathbf{K}^t \, \mathbf{F}^{(2)} \, \mathbf{F}^{'(1)} 
+ \mathbf{K}^t \, \mathbf{F}^{(2)} \, \mathbf{F}^{'(2)} \, \mathbf{K}
\nonumber \\
- \mathbf{F}^{'(1)} \, \mathbf{F}^{(1)} 
+ \mathbf{F}^{'(1)} \, \mathbf{F}^{(2)} \, \mathbf{K} 
+ \mathbf{K}^t \, \mathbf{F}^{'(2)} \, \mathbf{F}^{(1)}  
- \mathbf{K}^t \, \mathbf{F}^{'(2)} \, \mathbf{F}^{(2)} \, \mathbf{K} = 0
\nonumber .
\end{multline}
Using the fact that diagonal matrices commute, we finally get:
\begin{eqnarray}
\mathbf{K}^t \, (\mathbf{F}^{'(2)} \, \mathbf{F}^{(1)} - \mathbf{F}^{(2)} \, \mathbf{F}^{'(1)})
=
(\mathbf{F}^{(1)} \, \mathbf{F}^{'(2)} - \mathbf{F}^{'(1)} \, \mathbf{F}^{(2)}) \, \mathbf{K} 
\nonumber  
\end{eqnarray}
or in term of the Wronskian matrix:
\begin{eqnarray}
\mathbf{K}^t \, \mathbf{W} = \mathbf{W} \, \mathbf{K} .
\nonumber  
\end{eqnarray}
Since $\mathbf{W} = \mathbf{I}$ due to the $k_{\alpha'}^{-1/2}$ factors, this implies that $\mathbf{K}^t = \mathbf{K}$ so that $\mathbf{K}$ is symmetric.

\subsection*{Proof 2}

\begin{eqnarray}
\mathbf{Z} &=& \mathbf{F}' \, \mathbf{F}^{-1}  \nonumber \\
&=& \{ \mathbf{F}^{'(1)}-\mathbf{F}^{'(2)} \, \mathbf{K} \} \, \mathbf{N}^{K} \, [ \{ \mathbf{F}^{(1)}-\mathbf{F}^{(2)} \, \mathbf{K} \} \, \mathbf{N}^{K} ]^{-1} \nonumber \\
&=&  \{ \mathbf{F}^{'(1)}-\mathbf{F}^{'(2)} \, \mathbf{K} \} \, \mathbf{N}^{K}  \,  [\mathbf{N}^{K}]^{-1} \, \{ \mathbf{F}^{(1)}-\mathbf{F}^{(2)} \, \mathbf{K} \}^{-1} \nonumber \\
&=& \{ \mathbf{F}^{'(1)}-\mathbf{F}^{'(2)} \, \mathbf{K} \} \, \{ \mathbf{F}^{(1)}-\mathbf{F}^{(2)} \, \mathbf{K} \}^{-1} . \nonumber
\end{eqnarray}
Then:
\begin{eqnarray}
\mathbf{Z} \, \{ \mathbf{F}^{(1)}-\mathbf{F}^{(2)} \, \mathbf{K} \} = \{ \mathbf{F}^{'(1)}-\mathbf{F}^{'(2)} \, \mathbf{K} \} \nonumber \\
\mathbf{Z} \, \mathbf{F}^{(1)}-\mathbf{F}^{'(1)}  = \{ \mathbf{Z} \, \mathbf{F}^{(2)}-\mathbf{F}^{'(2)} \} \, \mathbf{K} \nonumber \\
\mathbf{K} = \{ \mathbf{Z} \, \mathbf{F}^{(2)}-\mathbf{F}^{'(2)} \}^{-1} \, \{ \mathbf{Z} \, \mathbf{F}^{(1)}-\mathbf{F}^{'(1)} \} . \nonumber
\end{eqnarray}

\subsection*{Proof 3}

First as $r \to \infty$: $\mathbf{F}^{\pm} = - \mathbf{F}^{(2)} \pm i \, \mathbf{F}^{(1)}$. This implies $\mathbf{F}^{(1)} = (\mathbf{F}^{+} - \mathbf{F}^{-})/2i$ and $\mathbf{F}^{(2)} = -(\mathbf{F}^{+} + \mathbf{F}^{-})/2$. \\

\noindent Then:
\begin{eqnarray}
\mathbf{F} &=& \mathbf{F}^{(1)} \, \mathbf{A}  + \mathbf{F}^{(2)} \, \mathbf{B} \nonumber \\
&=& \{(\mathbf{F}^{+} - \mathbf{F}^{-})/2i \} \, \mathbf{A} - \{ (\mathbf{F}^{+} + \mathbf{F}^{-})/2 \} \, \mathbf{B}  \nonumber \\
&=& (\mathbf{F}^{+} \, \mathbf{A})/2i  - (\mathbf{F}^{-} \, \mathbf{A})/2i
- (\mathbf{F}^{+} \, \mathbf{B})/2  - (\mathbf{F}^{-} \, \mathbf{B})/2 \nonumber  \\
&=& \mathbf{F}^{-} \, [- (\mathbf{B} - i \, \mathbf{A})/2] 
+ \mathbf{F}^{+} \, [- (\mathbf{B} + i \, \mathbf{A})/2] \nonumber \\
& \equiv & \mathbf{F}^{-} \, \mathbf{A'}  + \mathbf{F}^{+} \, \mathbf{B'} \nonumber.
\end{eqnarray} 
From that we identify:
\begin{align}
\mathbf{A'} &=  i/2 \ (\mathbf{A} + i \, \mathbf{B}) 
& \mathbf{B'} &= - i/2 \ (\mathbf{A} - i \, \mathbf{B}) \nonumber \\
\mathbf{A} &= -i \, (\mathbf{A'} - \mathbf{B'}) & \mathbf{B} &= - (\mathbf{A'} + \mathbf{B'}) \nonumber .
\end{align} 
We know from Eq.~\eqref{Sasymform1} and Eq.~\eqref{Sasymform2} that:
\begin{eqnarray}
\mathbf{S} & \equiv & - \mathbf{B'} \, \mathbf{A'}^{-1} \nonumber \\
&=& [\mathbf{A} - i \, \mathbf{B}] \, [\mathbf{A} + i \, \mathbf{B}]^{-1} \nonumber \\
&=&  [ (\mathbf{I} - i \, \mathbf{B} \, \mathbf{A}^{-1}) \, \mathbf{A}] \, [(\mathbf{I} + i \, \mathbf{B} \, \mathbf{A}^{-1}) \mathbf{A} ]^{-1} \nonumber \\
&=&  (\mathbf{I} - i \, \mathbf{B} \, \mathbf{A}^{-1}) \, \mathbf{A} \, \mathbf{A}^{-1} \, (\mathbf{I} + i \, \mathbf{B} \, \mathbf{A}^{-1})^{-1} \nonumber \\
&=&  (\mathbf{I} + i \, \mathbf{K}) \, (\mathbf{I} - i \, \mathbf{K})^{-1} \nonumber 
\end{eqnarray} 
where we used $\mathbf{K}  \equiv  - \mathbf{B} \, \mathbf{A}^{-1}$ from Eq.~\eqref{Kasymform1} and Eq.~\eqref{Kasymform2}.
The matrix $\mathbf{M} = \mathbf{I}+i\mathbf{K}$ is what is called 
a normal matrix since $\mathbf{M}$ and $\mathbf{M}^\dagger$ 
commute ($\mathbf{M} \, \mathbf{M}^\dagger = \mathbf{M}^\dagger \, \mathbf{M}$, 
easy to show using the fact that $\mathbf{K}$ is real and symmetric). 
Then $\mathbf{M}$ and $\mathbf{M}^\dagger$ can be expressed by $\mathbf{P} \, \mathbf{D}_M \, \mathbf{P}^{-1}$ and $\mathbf{P} \, \mathbf{D}_{M^\dagger} \, \mathbf{P}^{-1}$ 
with the same invertible matrix $\mathbf{P}$. $\mathbf{D}_{M,M^\dagger}$ are 
diagonal matrices with different complex eigenvalues since $\mathbf{M}$ is 
not a hermitian matrix. Then:
\begin{eqnarray}
\mathbf{M} \, [\mathbf{M}^\dagger]^{-1} &=&  \mathbf{P} \, \mathbf{D}_M \, \mathbf{P}^{-1} \, 
[\mathbf{P} \, \mathbf{D}_{M^\dagger} \, \mathbf{P}^{-1} ]^{-1} \nonumber \\
&=&  \mathbf{P} \, \mathbf{D}_M \, \mathbf{P}^{-1} \, \mathbf{P} \, \mathbf{D}_{M^\dagger}^{-1} \, \mathbf{P}^{-1} 
= \mathbf{P} \, \mathbf{D}_M \, \mathbf{D}_{M^\dagger}^{-1} \, \mathbf{P}^{-1} \nonumber \\
&=& \mathbf{P} \, \mathbf{D}_{M^\dagger}^{-1} \, \mathbf{D}_M \, \mathbf{P}^{-1} 
=  \mathbf{P} \, \mathbf{D}_{M^\dagger}^{-1} \, \mathbf{P}^{-1} \, \mathbf{P} \, \mathbf{D}_M \, \mathbf{P}^{-1} \nonumber \\
&=& [\mathbf{P} \, \mathbf{D}_{M^\dagger} \, \mathbf{P}^{-1} ]^{-1}  \, \mathbf{P} \, \mathbf{D}_M \, \mathbf{P}^{-1} \nonumber \\
&=& [\mathbf{M}^\dagger]^{-1} \, \mathbf{M} \nonumber 
\end{eqnarray} 
so that one also have $\mathbf{S} = \{\mathbf{I}+i\mathbf{K}\} \, \{\mathbf{I}-i\mathbf{K}\}^{-1} = \{\mathbf{I}-i\mathbf{K}\}^{-1} \, \{\mathbf{I}+i\mathbf{K}\}$. Because both matrices commute, we can safely write the expression as:
\begin{eqnarray}
\mathbf{S} = \frac{\mathbf{I}+i\mathbf{K}}{\mathbf{I}-i\mathbf{K}} \nonumber .
\end{eqnarray} 
We also know that:
\begin{eqnarray}
\mathbf{N}^{ K} &\equiv & \mathbf{A} = -i \, (\mathbf{A'} - \mathbf{B'}) \nonumber \\
&=& -i \, (\mathbf{I} - \mathbf{B'} \, \mathbf{A'}^{-1} ) \, \mathbf{A'} \nonumber \\
&=& -i \, (\mathbf{I} + \mathbf{S} ) \, \mathbf{N}^S  \nonumber, 
\end{eqnarray} 
and inversely:
\begin{eqnarray}
\mathbf{N}^{S} &\equiv & i/2 \ (\mathbf{A} + i \, \mathbf{B})  \nonumber \\
&=& i/2 \ (\mathbf{I} + i \, \mathbf{B} \, \mathbf{A}^{-1}) \, \mathbf{A} 
\nonumber \\
&=& i/2 \ (\mathbf{I} - i \, \mathbf{K}) \, \mathbf{N}^K  .
\nonumber
\end{eqnarray} 
If we use the forms in Eq.~\eqref{Kasymform2} or Eq.~\eqref{Sasymform2}, by developing we can find:
\begin{eqnarray}
\mathbf{F}^K &=& \{ \mathbf{F}^{(1)} - \mathbf{F}^{(2)} \,  \mathbf{K} \}  \nonumber \\
&=& \{ \mathbf{F}^{-} - \mathbf{F}^{+} \, \{ \mathbf{I} + i \mathbf{K}\} \, \{\mathbf{I}-
i\mathbf{K}\}^{-1} \} \, \{- \{\mathbf{I}-i\mathbf{K}\} \} /2i  \nonumber \\
&=& \{ \mathbf{F}^{-} - \mathbf{F}^{+} \,  \mathbf{S} \} \, \{- \{\mathbf{I}-i\mathbf{K}\} \} /2i \nonumber \\
&=& \mathbf{F}^{S} \, \{- \{\mathbf{I}-i\mathbf{K}\} \} /2i \nonumber \\
&=& \mathbf{F}^{S} \, \mathbf{N}^{S} \, (\mathbf{N}^{K})^{-1} \nonumber,
\end{eqnarray} 
so that we check that $ \mathbf{F}^K \, \mathbf{N}^{K} = \mathbf{F}^{S} \, \mathbf{N}^{S} = \mathbf{F}$.

\subsection*{Proof 4}

\begin{eqnarray}
\mathbf{S}^t = 
\bigg[ \frac{\mathbf{I}+i\mathbf{K}}{\mathbf{I}-i\mathbf{K}}\bigg]^t = 
\frac{\mathbf{I}+i\mathbf{K}^t}{\mathbf{I}-i\mathbf{K}^t} = \mathbf{S} \nonumber
\end{eqnarray}
so that $\mathbf{S}$ is symmetric.
\begin{eqnarray}
\mathbf{S}^\dagger = 
\bigg[ \frac{\mathbf{I}+i\mathbf{K}}{\mathbf{I}-i\mathbf{K}}\bigg]^\dagger = 
\frac{\mathbf{I}-i\mathbf{K}^\dagger}{\mathbf{I}+i\mathbf{K}^\dagger} = \mathbf{S}^{-1} \nonumber
\end{eqnarray}
so that $\mathbf{S}$ is unitary.


\begin{thebibliography}{127}

\bibitem{Chu_RMP_70_685_1998}
S.~Chu,
\newblock {\em Nobel Lecture: The manipulation of neutral particles},
\newblock Rev. Mod. Phys. {\bf 70}, 685 (1998).

\bibitem{Cohen-Tannoudji_RMP_70_707_1998}
C.~N. Cohen-Tannoudji,
\newblock {\em Nobel Lecture: Manipulating atoms with photons},
\newblock Rev. Mod. Phys. {\bf 70}, 707 (1998).

\bibitem{Phillips_RMP_70_721_1998}
W.~D. Phillips,
\newblock {\em Nobel Lecture: Laser cooling and trapping of neutral atoms},
\newblock Rev. Mod. Phys. {\bf 70}, 721 (1998).

\bibitem{Cornell_RMP_74_875_2002}
E.~A. Cornell and C.~E. Wieman,
\newblock {\em Nobel Lecture: Bose-Einstein condensation in a dilute gas, the
  first 70 years and some recent experiments},
\newblock Rev. Mod. Phys. {\bf 74}, 875 (2002).

\bibitem{Ketterle_RMP_74_1131_2002}
W.~Ketterle,
\newblock {\em Nobel lecture: When atoms behave as waves: Bose--Einstein
  condensation and the atom laser},
\newblock Rev. Mod. Phys. {\bf 74}, 1131 (2002).

\bibitem{Lewenstein_AP_56_243_2007}
M.~Lewenstein, A.~Sanpera, V.~Ahufinger, B.~Damski, A.~Sen(De), and U.~Sen,
\newblock {\em Ultracold atomic gases in optical lattices: mimicking condensed
  matter physics and beyond},
\newblock Adv. Phys. {\bf 56}, 243 (2007).

\bibitem{Bloch_RMP_80_885_2008}
I.~Bloch, J.~Dalibard, and W.~Zwerger,
\newblock {\em Many-body physics with ultracold gases},
\newblock Rev. Mod. Phys. {\bf 80}, 885 (2008).

\bibitem{Baranov_PRep_464_71_2008}
M.~Baranov,
\newblock {\em Theoretical progress in many-body physics with ultracold dipolar
  gases},
\newblock Phys. Rep. {\bf 464}, 71 (2008).

\bibitem{Shuman_N_467_820_2010}
E.~S. Shuman, J.~F. Barry, and D.~DeMille,
\newblock {\em Laser cooling of a diatomic molecule},
\newblock Nature {\bf 467}, 820 (2010).

\bibitem{Schnell_ACIE_48_6010_2009}
M.~Schnell and G.~Meijer,
\newblock {\em Cold Molecules: Preparation, applications, and challenges},
\newblock Angew. Chem. Int. Ed. {\bf 48}, 6010 (2009).

\bibitem{Dulieu_RPP_72_086401_2009}
O.~Dulieu and C.~Gabbanini,
\newblock {\em The formation and interactions of cold and ultracold molecules:
  new challenges for interdisciplinary physics},
\newblock Rep. Prog. Phys. {\bf 72}, 086401 (2009).

\bibitem{Hutzler_CR_112_4803_2012}
N.~R. Hutzler, H.-I. Lu, and J.~M. Doyle,
\newblock {\em The Buffer Gas Beam: An intense, cold, and slow source for atoms
  and molecules},
\newblock Chem. Rev. {\bf 112}, 4803 (2012).

\bibitem{VanDeMeerakker_CR_112_4828_2012}
S.~Y.~T. van~de Meerakker, H.~L. Bethlem, N.~Vanhaecke, and G.~Meijer,
\newblock {\em Manipulation and control of molecular beams},
\newblock Chem. Rev. {\bf 112}, 4828 (2012).

\bibitem{Narevicius_CR_112_4879_2012}
E.~Narevicius and M.~G. Raizen,
\newblock {\em Toward cold chemistry with magnetically decelerated supersonic
  beams},
\newblock Chem. Rev. {\bf 112}, 4879 (2012).

\bibitem{Zeppenfeld_N_491_570_2012}
M.~{Zeppenfeld}, B.~G.~U. {Englert}, R.~{Gl{\"o}ckner}, A.~{Prehn},
  M.~{Mielenz}, C.~{Sommer}, L.~D. {van Buuren}, M.~{Motsch}, and G.~{Rempe},
\newblock {\em Sisyphus cooling of electrically trapped polyatomic molecules},
\newblock Nature {\bf 491}, 570 (2012).

\bibitem{Thorsheim_PRL_58_2420_1987}
H.~R. Thorsheim, J.~Weiner, and P.~S. Julienne,
\newblock {\em Laser-induced photoassociation of ultracold sodium atoms},
\newblock Phys. Rev. Lett. {\bf 58}, 2420 (1987).

\bibitem{Fioretti_PRL_80_4402_1998}
A.~Fioretti, D.~Comparat, A.~Crubellier, O.~Dulieu, F.~Masnou-Seeuws, and
  P.~Pillet,
\newblock {\em Formation of cold $Cs_{2}$ molecules through photoassociation},
\newblock Phys. Rev. Lett. {\bf 80}, 4402 (1998).

\bibitem{Weiner_RMP_71_1_1999}
J.~Weiner, V.~S. Bagnato, S.~Zilio, and P.~S. Julienne,
\newblock {\em Experiments and theory in cold and ultracold collisions},
\newblock Rev. Mod. Phys. {\bf 71}, 1 (1999).

\bibitem{Jones_RMP_78_483_2006}
K.~M. Jones, E.~Tiesinga, P.~D. Lett, and P.~S. Julienne,
\newblock {\em Ultracold photoassociation spectroscopy: long-range molecules
  and atomic scattering},
\newblock Rev. Mod. Phys. {\bf 78}, 483 (2006).

\bibitem{Ulmanis_CR_112_4890_2012}
J.~Ulmanis, J.~Deiglmayr, M.~Repp, R.~Wester, and M.~Weidem{\"u}ller,
\newblock {\em Ultracold molecules formed by photoassociation: heteronuclear
  dimers, inelastic collisions, and interactions with ultrashort laser pulses},
\newblock Chem. Rev. {\bf 112}, 4890 (2012).

\bibitem{Kohler_RMP_78_1311_2006}
T.~K\"ohler, K.~G\'oral, and P.~S. Julienne,
\newblock {\em Production of cold molecules via magnetically tunable Feshbach
  resonances},
\newblock Rev. Mod. Phys. {\bf 78}, 1311 (2006).

\bibitem{Chin_RMP_82_1225_2010}
C.~Chin, R.~Grimm, P.~Julienne, and E.~Tiesinga,
\newblock {\em Feshbach resonances in ultracold gases},
\newblock Rev. Mod. Phys. {\bf 82}, 1225 (2010).

\bibitem{Bergmann_RMP_70_1003_1998}
K.~Bergmann, H.~Theuer, and B.~W. Shore,
\newblock {\em Coherent population transfer among quantum states of atoms and
  molecules},
\newblock Rev. Mod. Phys. {\bf 70}, 1003 (1998).

\bibitem{Ni_S_322_231_2008}
K.-K. Ni, S.~Ospelkaus, M.~H.~G. de~Miranda, A.~Pe'er, B.~Neyenhuis, J.~J.
  Zirbel, S.~Kotochigova, P.~S. Julienne, D.~S. Jin, and J.~Ye,
\newblock {\em A high phase-space-density gas of polar molecules},
\newblock Science {\bf 322}, 231 (2008).

\bibitem{Danzl_S_321_1062_2008}
J.~G. Danzl, E.~Haller, M.~Gustavsson, M.~J. Mark, R.~Hart, N.~Bouloufa,
  O.~Dulieu, H.~Ritsch, and H.-C. N\"agerl,
\newblock {\em Quantum gas of deeply bound ground state molecules},
\newblock Science {\bf 321}, 1062 (2008).

\bibitem{Koch_CR_112_4928_2012}
C.~P. Koch and M.~Shapiro,
\newblock {\em Coherent control of ultracold photoassociation},
\newblock Chem. Rev. {\bf 112}, 4928 (2012).

\bibitem{Bergmann_JCP_142_170901_2015}
K.~Bergmann, N.~V. Vitanov, and B.~W. Shore,
\newblock {\em Perspective: stimulated Raman adiabatic passage: the status
  after 25 years},
\newblock J. Chem. Phys. {\bf 142}, 170901 (2015).

\bibitem{Krems_IRPC_24_99_2005}
R.~V. Krems,
\newblock {\em Molecules near absolute zero and external field control of
  atomic and molecular dynamics},
\newblock Int. Rev. Phys. Chem. {\bf 24}, 99 (2005).

\bibitem{Krems_PCCP_10_4079_2008}
R.~V. Krems,
\newblock {\em Cold controlled chemistry},
\newblock Phys. Chem. Chem. Phys. {\bf 10}, 4079 (2008).

\bibitem{Quemener_CR_112_4949_2012}
G.~Qu\'em\'ener and P.~S. Julienne,
\newblock {\em Ultracold molecules under control!},
\newblock Chem. Rev. {\bf 112}, 4949 (2012).

\bibitem{Lemeshko_MP_111_1648_2013}
M.~Lemeshko, R.~V. Krems, J.~M. Doyle, and S.~Kais,
\newblock {\em Manipulation of molecules with electromagnetic fields},
\newblock Mol. Phys. {\bf 111}, 1648 (2013).

\bibitem{Carr_NJP_11_055049_2009}
L.~D. Carr, D.~DeMille, R.~V. Krems, and J.~Ye,
\newblock {\em Cold and ultracold molecules: science, technology and
  applications},
\newblock New J. Phys. {\bf 11}, 055049 (2009).

\bibitem{Micheli_PRA_76_043604_2007}
A.~Micheli, G.~Pupillo, H.~P. B\"uchler, and P.~Zoller,
\newblock {\em Cold polar molecules in two-dimensional traps: tailoring
  interactions with external fields for novel quantum phases},
\newblock Phys. Rev. A {\bf 76}, 043604 (2007).

\bibitem{Gorshkov_PRL_107_115301_2011}
A.~V. Gorshkov, S.~R. Manmana, G.~Chen, J.~Ye, E.~Demler, M.~D. Lukin, and
  A.~M. Rey,
\newblock {\em Tunable superfluidity and quantum magnetism with ultracold polar
  molecules},
\newblock Phys. Rev. Lett. {\bf 107}, 115301 (2011).

\bibitem{Baranov_CR_112_5012_2012}
M.~A. Baranov, M.~Dalmonte, G.~Pupillo, and P.~Zoller,
\newblock {\em Condensed matter theory of dipolar quantum gases},
\newblock Chem. Rev. {\bf 112}, 5012 (2012).

\bibitem{Wall_BookChapter_2014}
M.~L. Wall, K.~R.~A. Hazzard, and A.-M. Rey,
\newblock {\em Quantum magnetism with ultracold molecules},
\newblock Chapter 1 in From atomic to mesoscale: the role of quantum coherence
  in systems of various complexities. Edited by S. A. Malinovskaya, I.
  Novikova, World Scientific Publishing Co {\bf 1406}, 4758 (2014).

\bibitem{DeMille_PRL_88_067901_2002}
D.~DeMille,
\newblock {\em Quantum computation with trapped polar molecules},
\newblock Phys. Rev. Lett. {\bf 88}, 067901 (2002).

\bibitem{Yelin_PRA_74_050301_2006}
S.~F. Yelin, K.~Kirby, and R.~C\^ot\'e,
\newblock {\em Schemes for robust quantum computation with polar molecules},
\newblock Phys. Rev. A {\bf 74}, 050301 (2006).

\bibitem{Karra_JCP_144_094301_2016}
M.~Karra, K.~Sharma, B.~Friedrich, S.~Kais, and D.~Herschbach,
\newblock {\em Prospects for quantum computing with an array of ultracold polar
  paramagnetic molecules},
\newblock J. Chem. Phys. {\bf 144}, 094301 (2016).

\bibitem{Hinds_PS_1997_34_1997}
E.~A. Hinds,
\newblock {\em Testing time reversal symmetry using molecules},
\newblock Phys. Scr. {\bf 1997}, 34 (1997).

\bibitem{Tarbutt_BookChapter_2009}
M.~R. {Tarbutt}, J.~J. {Hudson}, B.~E. {Sauer}, and E.~A. {Hinds},
\newblock {\em Preparation and manipulation of molecules for fundamental
  physics tests},
\newblock Chapter 15 in Cold molecules: theory, experiments, applications.
  Edited by R. Krems, B. Friedrich, B. and W. C. Stwalley, CRC Press, 69
  (2009).

\bibitem{Gonzalez-Martinez_PRA_90_052716_2014}
M.~L. Gonz{\'a}lez-Mart{\'\i}nez, O.~Dulieu, P.~Larr{\'e}garay, and L.~Bonnet,
\newblock {\em Statistical product distributions for ultracold reactions in
  external fields},
\newblock Phys. Rev. A {\bf 90}, 052716 (2014).

\bibitem{Tscherbul_PRL_115_023201_2015}
T.~V. Tscherbul and R.~V. Krems,
\newblock {\em Tuning bimolecular chemical reactions by electric fields},
\newblock Phys. Rev. Lett. {\bf 115}, 023201 (2015).

\bibitem{Weck_IRPC_25_283_2006}
P.~F. Weck and N.~Balakrishnan,
\newblock {\em Importance of long-range interactions in chemical reactions at
  cold and ultracold temperatures},
\newblock Int. Rev. Phys. Chem. {\bf 25}, 283 (2006).

\bibitem{Hutson_IRPC_26_1_2007}
J.~M. Hutson and P.~Sold{\'a}n,
\newblock {\em Molecular collisions in ultracold atomic gases},
\newblock Int. Rev. Phys. Chem. {\bf 26}, 1 (2007).

\bibitem{Quemener_BookChapter_2009}
G.~Qu\'em\'ener, N.~Balakrishnan, and A. Dalgarno,
\newblock {\em Inelastic collisions and chemical reactions of molecules at ultracold temperatures},
\newblock Chapter 3 in Cold molecules: theory, experiments, applications.
  Edited by R. Krems, B. Friedrich, B. and W. C. Stwalley, CRC Press, 3
  (2009).

\bibitem{Brandsen_Joachain_Book_2003}
B.~Brandsen and C.~Joachain,
\newblock {\em Physics of atoms and molecules},
\newblock Addison-Wesley, 2003.

\bibitem{Cohen-Tannoudji_Book_1997}
C.~Cohen-Tannoudji, B.~Diu, and F.~Lalo{\"e},
\newblock {\em M{\'e}canique quantique},
\newblock Hermann, 1997.

\bibitem{Friedrich_Book_2005}
H. Friedrich,
\newblock {\em Theoretical atomic physics, third edition},
\newblock Springer, 2005.

\bibitem{Landau_Book_1958}
L.~D. Landau and L.~M. Lifshitz,
\newblock {\em Quantum mechanics (non-relativistic theory)},
\newblock Butterworth Heinemann, 1958.

\bibitem{Child_Book_1996}
M.~S. Child,
\newblock {\em Molecular collision theory},
\newblock Dover Publications, 1996.

\bibitem{Atkins_Friedman_Book_2005}
P. W. Atkins and R. S. Friedman,
\newblock {\em Molecular quantum mechanics},
\newblock Oxford University Press, 2005.

\bibitem{Launay_Book_2000}
J.-M. Launay,
\newblock {\em Collisions mol{\'e}culaires, cours du DEA Physique, option
  ``physique atomique et mol{\'e}culaire", Universit\'e de Rennes 1}.

\bibitem{Quemener_PRA_83_012705_2011}
G.~Qu\'em\'ener and J.~L. Bohn,
\newblock {\em Dynamics of ultracold molecules in confined geometry and
  electric field},
\newblock Phys. Rev. A {\bf 83}, 012705 (2011).

\bibitem{Grishkevich_PRA_84_062710_2011}
S. Grishkevich, S. Sala and A. Saenz,
\newblock {\em Theoretical description of two ultracold atoms in finite three-dimensional optical lattices using realistic interatomic interaction potentials},
\newblock Phys. Rev. A {\bf 84}, 062710 (2011).

\bibitem{Whitten_JMP_9_1103_1968}
R.~C. Whitten and F.~T. Smith,
\newblock {\em Symmetric representation for three-body problems. II. Motion in
  space},
\newblock J. Math. Phys. {\bf 9}, 1103 (1968).

\bibitem{Johnson_JCP_79_1916_1983}
B.~R. Johnson,
\newblock {\em The quantum dynamics of three particles in hyperspherical
  coordinates},
\newblock J. Chem. Phys. {\bf 79}, 1916 (1983).

\bibitem{Pack_JCP_87_3888_1987}
R.~T. Pack and G.~A. Parker,
\newblock {\em Quantum reactive scattering in three dimensions using
  hyperspherical (APH) coordinates. Theory},
\newblock J. Chem. Phys. {\bf 87}, 3888 (1987).

\bibitem{Launay_CPL_163_178_1989}
J.~M. Launay and M. Le Dourneuf,
\newblock {\em Hyperspherical close-coupling calculation of integral cross
  sections for the reaction H+H{$_2$} $\to$ H{$_2$}+H},
\newblock Chem. Phys. Lett. {\bf 163}, 178 (1989).

\bibitem{Rittenhouse_JPBAMOP_44_172001_2011}
S.~T. Rittenhouse, J.~von Stecher, J.~P. D’Incao, N.~P. Mehta, and C.~H.
  Greene,
\newblock {\em The hyperspherical four-fermion problem},
\newblock J. Phys. B: At. Mol. Opt. Phys. {\bf 44}, 172001 (2011).

\bibitem{Curtiss_JCP_21_2045_1953}
C.~F. Curtiss,
\newblock {\em The quantum mechanics of collisions between diatomic molecules},
\newblock J. Chem. Phys. {\bf 21}, 2045 (1953).

\bibitem{Takayanagi_PTP_11_557_1954}
K.~Takayanagi,
\newblock {\em The theory of collisions between two diatomic molecules},
\newblock Prog. Theor. Phys. {\bf 11}, 557 (1954).

\bibitem{Arthurs_PRS_256_540_1960}
A.~M. Arthurs and A.~Dalgarno,
\newblock {\em The theory of scattering by a rigid rotator},
\newblock Proc. Roy. Soc. {\bf 256}, 540 (1960).

\bibitem{Pack_JCP_60_633_1974}
R.~T. Pack,
\newblock {\em Space-fixed vs body-fixed axes in atom-diatomic molecule
  scattering. Sudden approximations},
\newblock J. Chem. Phys. {\bf 60}, 633 (1974).

\bibitem{Green_JCP_62_2271_1975}
S.~Green,
\newblock {\em Rotational excitation in H{$_2$}-H{$_2$} collisions:
  close-coupling calculations},
\newblock J. Chem. Phys. {\bf 62}, 2271 (1975).

\bibitem{Alexander_JCP_66_2166_1977}
M.~H. Alexander and A.~E. DePristo,
\newblock {\em Symmetry considerations in the quantum treatment of collisions
  between two diatomic molecules},
\newblock J. Chem. Phys. {\bf 66}, 2166 (1977).

\bibitem{Launay_JPBAMOP_9_1823_1976}
J.~M. Launay,
\newblock {\em Body-fixed formulation of rotational excitation: exact and
  centrifugal decoupling results for CO-He},
\newblock J. Phys. B: At. Mol. Opt. Phys. {\bf 9}, 1823 (1976).

\bibitem{Heil_JCP_68_2562_1978}
T.~G. Heil, S.~Green, and D.~J. Kouri,
\newblock {\em The coupled states approximation for scattering of two diatoms},
\newblock J. Chem. Phys. {\bf 68}, 2562 (1978).

\bibitem{Takayanagi_AAMP_1_149_1965}
K.~Takayanagi,
\newblock {\em The production of rotational and vibrational transitions in
  encounters between molecules},
\newblock Adv. At. Mol. Phys. {\bf 1}, 149 (1965).

\bibitem{Zarur_JCP_60_2057_1974}
G.~Zarur and H.~Rabitz,
\newblock {\em Effective potential formulation of molecule-molecule collisions
  with application to H{$_2$}-H{$_2$}},
\newblock J. Chem. Phys. {\bf 60}, 2057 (1974).

\bibitem{Quemener_PRA_88_012706_2013}
G.~Qu\'em\'ener and J.~L. Bohn,
\newblock {\em Ultracold molecular collisions in combined electric and magnetic
  fields},
\newblock Phys. Rev. A {\bf 88}, 012706 (2013).

\bibitem{Tscherbul_JCP_133_184104_2010}
T.~V. Tscherbul and A.~Dalgarno,
\newblock {\em Quantum theory of molecular collisions in a magnetic field:
  Efficient calculations based on the total angular momentum representation},
\newblock J. Chem. Phys. {\bf 133}, 184104 (2010).

\bibitem{Tscherbul_PRA_85_052710_2012}
T.~V. Tscherbul,
\newblock {\em Total-angular-momentum representation for atom-molecule
  collisions in electric fields},
\newblock Phys. Rev. A {\bf 85}, 052710 (2012).

\bibitem{Johnson_JCP_13_445_1973}
B.~R. Johnson,
\newblock {\em The multichannel log-derivative method for scattering
  calculations},
\newblock J. Comp. Phys. {\bf 13}, 445 (1973).

\bibitem{Johnson_JCP_69_4678_1978}
B.~R. Johnson,
\newblock {\em The renormalized Numerov method applied to calculating bound
  states of the coupled-channel Schr{\"o}dinger equation},
\newblock J. Chem. Phys. {\bf 69}, 4678 (1978).

\bibitem{Manolopoulos_JCP_85_6425_1986}
D.~E. Manolopoulos,
\newblock {\em An improved log derivative method for inelastic scattering},
\newblock J. Chem. Phys. {\bf 85}, 6425 (1986).

\bibitem{Stone_Book_1996}
A.~J. Stone,
\newblock {\em The theory of intermolecular forces},
\newblock Oxford University Press, 1996.

\bibitem{Hutson_CPC_84_1_1994}
J.~M. Hutson,
\newblock {\em Coupled channel methods for solving the bound-state
  Schr{\"o}dinger equation},
\newblock Comput. Phys. Commun. {\bf 84}, 1 (1994).

\bibitem{Abramowitz_Stegun_Book_1964}
M. Abramowitz and I. Stegun,
\newblock {\em Handbook of mathematical functions with formulas, graphs, and mathematical tables},
\newblock United States Department of Commerce, National Bureau of Standards, 1964.

\bibitem{Burke_PhDThesis_1999}
J.~P. {Burke, Jr.},
\newblock {\em Theoretical investigation of cold alkali atom collisions},
\newblock PhD thesis, University of Colorado, Boulder (USA), 1999.

\bibitem{Tscherbul_NJP_11_055021_2009}
T.~V. Tscherbul, Y.~V. Suleimanov, V.~Aquilanti, and R.~V. Krems,
\newblock {\em Magnetic field modification of ultracold molecule molecule
  collisions},
\newblock New J. Phys. {\bf 11}, 055021 (2009).

\bibitem{Pethick_Smith_Book_2001}
C. J. Pethick and H. Smith,
\newblock {\em Bose--Einstein condensation in dilute gases},
\newblock Cambridge University Press, 2001.

\bibitem{Pitaevskii_Stringari_Book_2003}
L. P. Pitaevskii and S. Stringari,
\newblock {\em Bose--Einstein condensation},
\newblock Oxford: Clarendon Press, 2003.

\bibitem{Sadeghpour_JPBAMOP_33_93_2000}
H.~R. Sadeghpour, J.~L. Bohn, M.~J. Cavagnero, B.~D. Esry, I.~I. Fabrikant,
  J.~H. Macek, and A.~R.~P. Rau,
\newblock {\em Collisions near threshold in atomic and molecular physics},
\newblock J. Phys. B: At. Mol. Opt. Phys. {\bf 33}, 93 (2000).

\bibitem{Hutson_BookChapter_2009}
J.~M. Hutson,
\newblock {\em {Theory of cold atomic and moleculer collisions}},
\newblock Chapter 1 in Cold molecules: theory, experiments, applications.
  Edited by R. Krems, B. Friedrich, B. and W. C. Stwalley, CRC Press, 3
  (2009).

\bibitem{Wigner_PR_73_1002_1948}
E.~P. Wigner,
\newblock {\em On the behavior of cross sections near thresholds},
\newblock Phys. Rev. {\bf 73}, 1002 (1948).

\bibitem{Ospelkaus_PRL_104_030402_2010}
S.~Ospelkaus, K.-K. Ni, G.~Qu\'em\'ener, B.~Neyenhuis, D.~Wang, M.~H.~G.
  de~Miranda, J.~L. Bohn, J.~Ye, and D.~S. Jin,
\newblock {\em Controlling the hyperfine state of rovibronic ground-state polar
  molecules},
\newblock Phys. Rev. Lett. {\bf 104}, 030402 (2010).

\bibitem{Zuchowski_PRA_81_060703_2010}
P.~S. \ifmmode~\dot{Z}\else \.{Z}\fi{}uchowski and J.~M. Hutson,
\newblock {\em Reactions of ultracold alkali-metal dimers},
\newblock Phys. Rev. A {\bf 81}, 060703 (2010).

\bibitem{Byrd_PRA_82_010502_2010}
J.~N. Byrd, J.~A. Montgomery, and R.~C\^ot\'e,
\newblock {\em Structure and thermochemistry of $K_2$Rb, KRb$_2$, and
  $K_2$Rb$_2$},
\newblock Phys. Rev. A {\bf 82}, 010502 (2010).

\bibitem{Meyer_PRA_82_042707_2010}
E.~R. Meyer and J.~L. Bohn,
\newblock {\em Product-state control of bi-alkali-metal chemical reactions},
\newblock Phys. Rev. A {\bf 82}, 042707 (2010).

\bibitem{Byrd_PRA_86_032711_2012}
J.~N. Byrd, J.~A. Montgomery, and R.~C\^ot\'e,
\newblock {\em Long-range forces between polar alkali-metal diatoms aligned by
  external electric fields},
\newblock Phys. Rev. A {\bf 86}, 032711 (2012).

\bibitem{Wang_NJP_17_035015_2015}
G.~Wang and G.~Qu{\'e}m{\'e}ner,
\newblock {\em Tuning ultracold collisions of excited rotational dipolar
  molecules},
\newblock New J. Phys. {\bf 17}, 035015 (2015).

\bibitem{Kotochigova_NJP_12_073041_2010}
S.~Kotochigova,
\newblock {\em Dispersion interactions and reactive collisions of ultracold
  polar molecules},
\newblock New J. Phys. {\bf 12}, 073041 (2010).

\bibitem{Lepers_PRA_88_032709_2013}
M.~Lepers, R.~Vexiau, M.~Aymar, N.~Bouloufa-Maafa, and O.~Dulieu,
\newblock {\em Long-range interactions between polar alkali-metal diatoms in
  external electric fields},
\newblock Phys. Rev. A {\bf 88}, 032709 (2013).

\bibitem{Zuchowski_PRA_87_022706_2013}
P.~S. Zuchowski, M.~Kosicki, M.~Kodrycka, and P.~Sold\'an,
\newblock {\em Van der Waals coefficients for systems with ultracold polar
  alkali-metal molecules},
\newblock Phys. Rev. A {\bf 87}, 022706 (2013).

\bibitem{Idziaszek_PRA_82_020703_2010}
Z.~Idziaszek, G.~Qu\'em\'ener, J.~L. Bohn, and P.~S. Julienne,
\newblock {\em Simple quantum model of ultracold polar molecule collisions},
\newblock Phys. Rev. A {\bf 82}, 020703 (2010).

\bibitem{Idziaszek_PRL_104_113202_2010}
Z.~Idziaszek and P.~S. Julienne,
\newblock {\em Universal Rate Constants for reactive collisions of ultracold
  molecules},
\newblock Phys. Rev. Lett. {\bf 104}, 113202 (2010).

\bibitem{Bishof_PRA_84_052716_2011}
M.~Bishof, M.~J. Martin, M.~D. Swallows, C.~Benko, Y.~Lin, G.~Qu\'em\'ener,
  A.~M. Rey, and J.~Ye,
\newblock {\em Inelastic collisions and density-dependent excitation
  suppression in a ${}^{87}$Sr optical lattice clock},
\newblock Phys. Rev. A {\bf 84}, 052716 (2011).

\bibitem{Ludlow_PRA_84_052724_2011}
A.~D. Ludlow, N.~D. Lemke, J.~A. Sherman, C.~W. Oates, G.~Qu\'em\'ener, J.~von
  Stecher, and A.~M. Rey,
\newblock {\em Cold-collision-shift cancellation and inelastic scattering in a
  Yb optical lattice clock},
\newblock Phys. Rev. A {\bf 84}, 052724 (2011).

\bibitem{Jachymski_PRL_110_213202_2013}
K. Jachymski, M. Micha\l{}, P. S. Julienne, and Z. Idziaszek,
\newblock {\em Quantum theory of reactive collisions for $1/{r}^{n}$ potentials},
\newblock Phys. Rev. Lett. {\bf 110}, 213202 (2013).

\bibitem{Mayle_PRA_85_062712_2012}
M.~Mayle, B.~P. Ruzic, and J.~L. Bohn,
\newblock {\em Statistical aspects of ultracold resonant scattering},
\newblock Phys. Rev. A {\bf 85}, 062712 (2012).

\bibitem{Mayle_PRA_87_012709_2013}
M.~Mayle, G.~Qu\'em\'ener, B.~P. Ruzic, and J.~L. Bohn,
\newblock {\em Scattering of ultracold molecules in the highly resonant
  regime},
\newblock Phys. Rev. A {\bf 87}, 012709 (2013).

\bibitem{Takekoshi_PRL_113_205301_2014}
T. Takekoshi, L. Reichs\"ollner, A. Schindewolf, J. M. Hutson, C. R. Le Sueur, O. Dulieu, F. Ferlaino, R. Grimm, and H.-C. N\"agerl,
\newblock {\em Ultracold dense samples of dipolar RbCs molecules in the rovibrational and hyperfine ground state},
\newblock Phys. Rev. Lett. {\bf 113}, 205301 (2014).

\bibitem{Park_PRL_114_205302_2015}
J. W. Park, S. A. Will, and M. W. Zwierlein,
\newblock {\em Ultracold dipolar gas of fermionic $^{23}\mathrm{Na}^{40}\mathrm{K}$ molecules in their absolute ground state},
\newblock Phys. Rev. Lett. {\bf 114}, 205302 (2015).

\bibitem{Guo_PRL_116_205303_2016}
M. Guo, B. Zhu, B. Lu, X. Ye, F. Wang, R. Vexiau, N. Bouloufa-Maafa, G. Qu\'em\'ener, O. Dulieu, and D. Wang, 
\newblock {\em Creation of an ultracold gas of ground-state dipolar $^{23}\mathrm{Na}^{87}\mathrm{Rb}$ molecules},
\newblock Phys. Rev. Lett. {\bf 116}, 205303 (2016).

\bibitem{Aikawa_PRL_105_203001_2010}
K.~Aikawa, D.~Akamatsu, M.~Hayashi, K.~Oasa, J.~Kobayashi, P.~Naidon,
  T.~Kishimoto, M.~Ueda, and S.~Inouye,
\newblock {\em Coherent Transfer of Photoassociated Molecules into the
  Rovibrational Ground State},
\newblock Phys. Rev. Lett. {\bf 105}, 203001 (2010).

\bibitem{Bohn_BookChapter_2009}
J. L. Bohn,
\newblock {\em Electric dipoles at ultralow temperatures},
\newblock Chapter 2 in Cold molecules: theory, experiments, applications.
  Edited by R. Krems, B. Friedrich, B. and W. C. Stwalley, CRC Press, 3
  (2009).

\bibitem{Aymar_JCP_122_204302_2005}
M. Aymar and O. Dulieu, 
\newblock {\em Calculation of accurate permanent dipole moments of the lowest $^1,3 Sigma ^+$ states of heteronuclear alkali dimers using extended basis sets},
\newblock J. Chem. Phys. {\bf 122}, 204302 (2005).

\bibitem{Ni_N_464_1324_2010}
K.-K. Ni, S.~Ospelkaus, D.~Wang, G.~Qu\'em\'ener, B.~Neyenhuis, M.~H.~G.
  de~Miranda, J.~L. Bohn, D.~S. Jin, and J.~Ye,
\newblock {\em Dipolar collisions of polar molecules in the quantum regime},
\newblock Nature {\bf 464}, 1324 (2010).

\bibitem{Quemener_PRA_81_022702_2010}
G.~Qu\'em\'ener and J.~L. Bohn,
\newblock {\em Strong dependence of ultracold chemical rates on electric dipole
  moments},
\newblock Phys. Rev. A {\bf 81}, 022702 (2010).

\bibitem{Quemener_PRA_84_062703_2011}
G.~Qu\'em\'ener, J.~L. Bohn, A.~Petrov, and S.~Kotochigova,
\newblock {\em Universalities in ultracold reactions of alkali-metal polar
  molecules},
\newblock Phys. Rev. A {\bf 84}, 062703 (2011).

\bibitem{Bohn_NJP_11_055039_2009}
J.~L. Bohn, M.~Cavagnero, and C.~Ticknor,
\newblock {\em Quasi-universal dipolar scattering in cold and ultracold gases},
\newblock New J. Phys. {\bf 11}, 055039 (2009).

\bibitem{Gao_PRA_78_012702_2008}
B.~Gao,
\newblock {\em General form of the quantum-defect theory for $-{}1/r^{\alpha{}}$ type of potentials with $\alpha{}\&gt;2$},
\newblock Phys. Rev. A {\bf 78}, 012702 (2008).

\bibitem{Langevin_ACP_5_245_1905}
P.~Langevin,
\newblock {\em A fundamental formula of kinetic theory},
\newblock Ann. Chim. Phys. {\bf 5}, 245 (1905).

\bibitem{Gao_PRL_105_263203_2010}
B.~Gao,
\newblock {\em Universal model for exoergic bimolecular reactions and inelastic
  processes},
\newblock Phys. Rev. Lett. {\bf 105}, 263203 (2010).

\bibitem{Avdeenkov_PRA_73_022707_2006}
A.~V. Avdeenkov, M.~Kajita, and J.~L. Bohn,
\newblock {\em Suppression of inelastic collisions of polar $^{1}\Sigma$ state
  molecules in an electrostatic field},
\newblock Phys. Rev. A {\bf 73}, 022707 (2006).

\bibitem{Quemener_PRA_93_012704_2016}
G.~Qu\'em\'ener and J.~L. Bohn,
\newblock {\em Shielding $^{2}\mathrm{\ensuremath{\Sigma}}$ ultracold dipolar
  molecular collisions with electric fields},
\newblock Phys. Rev. A {\bf 93}, 012704 (2016).

\bibitem{DeMiranda_NP_7_502_2011}
M.~H.~G. de~Miranda, A.~Chotia, B.~Neyenhuis, D.~Wang, G.~Qu\'em\'ener,
  S.~Ospelkaus, J.~Bohn, J. L.~Ye, and D.~S. Jin,
\newblock {\em Controlling the quantum stereodynamics of ultracold bimolecular
  reactions},
\newblock Nature Physics {\bf 7}, 502 (2011).

\bibitem{Frisch_PRL_115_203201_2015}
A.~Frisch, M.~Mark, K.~Aikawa, S.~Baier, R.~Grimm, A.~Petrov, S.~Kotochigova,
  G.~Qu\'em\'ener, M.~Lepers, O.~Dulieu, and F.~Ferlaino,
\newblock {\em Ultracold dipolar molecules composed of strongly magnetic
  atoms},
\newblock Phys. Rev. Lett. {\bf 115}, 203201 (2015).

\bibitem{Quemener_PRA_92_042706_2015}
G.~Qu\'em\'ener, M.~Lepers, and O.~Dulieu,
\newblock {\em Dynamics of ultracold dipolar particles in a confined geometry
  and tilted fields},
\newblock Phys. Rev. A {\bf 92}, 042706 (2015).

\bibitem{Gorshkov_PRL_101_073201_2008}
A.~V. Gorshkov, P.~Rabl, G.~Pupillo, A.~Micheli, P.~Zoller, M.~D. Lukin, and
  H.~P. B\"uchler,
\newblock {\em Suppression of inelastic collisions between polar molecules with
  a repulsive shield},
\newblock Phys. Rev. Lett. {\bf 101}, 073201 (2008).

\bibitem{Alyabyshev_PRA_80_033419_2009}
S.~V. Alyabyshev and R.~V. Krems,
\newblock {\em Controlling collisional spin relaxation of cold molecules with
  microwave laser fields},
\newblock Phys. Rev. A {\bf 80}, 033419 (2009).

\bibitem{Avdeenkov_PRA_86_022707_2012}
A.~V. Avdeenkov,
\newblock {\em Dipolar collisions of ultracold polar molecules in a microwave
  field},
\newblock Phys. Rev. A {\bf 86}, 022707 (2012).

\bibitem{Ticknor_PRL_105_013201_2010}
C.~Ticknor and S.~T. Rittenhouse,
\newblock {\em Three body recombination of ultracold dipoles to weakly bound
  dimers},
\newblock Phys. Rev. Lett. {\bf 105}, 013201 (2010).

\bibitem{Wang_PRL_106_233201_2011}
Y.~Wang, J.~P. D'Incao, and C.~H. Greene,
\newblock {\em Efimov effect for three interacting bosonic dipoles},
\newblock Phys. Rev. Lett. {\bf 106}, 233201 (2011).

\bibitem{Wang_PRL_107_233201_2011}
Y.~Wang, J.~P. D'Incao, and C.~H. Greene,
\newblock {\em Universal three-body physics for fermionic dipoles},
\newblock Phys. Rev. Lett. {\bf 107}, 233201 (2011).

\bibitem{Lepers_JPBAMOP_49_014004_2016}
M.~Lepers, G.~Qu\'em\'ener, E.~Luc-Koenig, and O.~Dulieu,
\newblock {\em Four-body long-range interactions between ultracold weakly-bound
  diatomic molecules},
\newblock J. Phys. B: At. Mol. Opt. Phys. {\bf 49}, 014004 (2016).

\end{thebibliography}
\end{document}